\documentclass[12pt,epsf]{article}
\usepackage{verbatim}
\usepackage{anyfontsize}
\usepackage{dsfont}
\usepackage{tikz}
\usepackage{sidecap}
\sidecaptionvpos{figure}{t}
\usepackage{subfigure}

\usepackage[english]{babel}
\usepackage{enumitem}  

\usepackage{amssymb,amsmath}
\usepackage{graphicx, xcolor, varwidth}
\usepackage{setspace}
\usepackage[permil]{overpic}
\usepackage{cite}
\usepackage{mathtools}

\DeclareSymbolFont{matha}{OML}{txmi}{m}{it}
\DeclareMathSymbol{v}{\mathord}{matha}{118}

\colorlet{darkblue}{blue!70!black}
\colorlet{darkgreen}{green!70!black}

\usepackage[colorlinks=true,urlcolor=darkblue,linktocpage=true,linkcolor=darkblue,citecolor=darkblue]{hyperref}

\numberwithin{equation}{section}
\DeclareMathSymbol{v}{\mathord}{matha}{118}
\newcommand{\be}{\begin{equation}}
\newcommand{\ee}{\end{equation}}
\newcommand{\bea}{\begin{eqnarray}}
\newcommand{\eea}{\end{eqnarray}}
\newcommand{\bear}{\begin{eqnarray}}
\newcommand{\eear}{\end{eqnarray}}
\newcommand{\beas}{\begin{eqnarray*}}

\newcommand{\eeas}{\end{eqnarray*}}
\newcommand{\ba}{\begin{array}}
\newcommand{\ea}{\end{array}}

\def\ba#1\ea{\begin{align}#1\end{align}}
\def\bs#1\es{\begin{split}#1\end{split}}

\newcommand{\ep}{\epsilon}

\newcommand{\pd}[2][1]{\ifnum#1=1 \frac{\partial}{\partial {#2}} \else
  \frac{\partial^#1}{\partial {#2}^{#1}}\fi}
\newcommand{\dpd}[2][1]{\ifnum#1=1 \dfrac{\partial}{\partial {#2}} \else
  \frac{\partial^#1}{\partial {#2}^{#1}}\fi}
\newcommand{\td}[2][1]{\ifnum#1=1 \frac{d}{d{#2}} \else
  \frac{d^#1}{d{#2}^{#1}}\fi}
\renewcommand{\d}{\partial}

\newcommand{\Deltag}{\Delta_{\text{gap}}}

\newcommand{\e}{\varepsilon}

\newcommand{\tn}{\tilde{n}}

\renewcommand{\d}{\partial}

\renewcommand{\(}{\left(}
\renewcommand{\)}{\right)}

\newcommand{\nbox}{{\,\lower0.9pt\vbox{\hrule \hbox{\vrule height 0.2 cm \hskip 0.19 cm \vrule height 0.2 cm}\hrule}\,}}

\newcommand{\E}{{\mathcal E}}
\def\O{{\cal O}}

\newcommand{\half}{\tfrac{1}{2}}

\newcommand{\N}{{\cal N}}


\newcommand{\pv}{\vec{\mathbf{p}}}
\newcommand{\Lv}{\vec{\mathbf{L}}}
\newcommand{\Xv}{\vec{\mathbf{x}}_3}
\newcommand{\Tv}{\vec{\mathbf{T}}}
\newcommand{\kv}{\vec{\mathbf{k}}}

\begin{document}
\begin{spacing}{1.3}
\begin{titlepage}

\begin{center}
{\Large \bf 

A Conformal Collider for Holographic CFTs

}

\vspace*{6mm}

Nima Afkhami-Jeddi,$^*$ Sandipan Kundu,$^{\dagger}$ and Amirhossein Tajdini$^*$

\vspace*{6mm}

\textit{$^*$Department of Physics, Cornell University, Ithaca, New York, USA\\}

\vspace{3mm}

\textit{$^\dagger$Department of Physics and Astronomy, Johns Hopkins University,
Baltimore, Maryland, USA\\}

\vspace{6mm}

{\tt \small na382@cornell.edu, kundu@jhu.edu, at734@cornell.edu}

\vspace*{6mm}
\end{center}

\begin{abstract}

We develop a formalism to study the implications of causality on OPE coefficients in conformal field theories with large central charge and a sparse spectrum of higher spin operators.  The formalism has the interpretation of a new conformal collider-type experiment for these class of CFTs and hence it has the advantage of requiring knowledge only about  CFT three-point functions.  This is accomplished by considering the holographic null energy operator which was introduced in \cite{Afkhami-Jeddi:2017rmx} as a generalization of the averaged null energy operator. Analyticity properties of correlators in the Regge limit imply that the holographic null energy operator is a positive operator in a subspace of the total CFT Hilbert space. Utilizing this positivity condition, we derive bounds on three-point functions $\langle TO_1O_2\rangle$ of the stress tensor with various operators for CFTs with large central charge and a sparse spectrum. After imposing these constraints, we also find that the operator product expansions of all primary operators in the Regge limit have certain universal properties. All of these results are consistent with the expectation that CFTs in this class, irrespective of their microscopic details, admit universal gravity-like holographic dual descriptions. Furthermore, this connection enables us to constrain various inflationary observables such as the amplitude of chiral gravity waves, non-gaussanity of gravity waves and tensor-to-scalar ratio.

\end{abstract}

\end{titlepage}
\end{spacing}

\vskip 1cm
\setcounter{tocdepth}{2}  
\tableofcontents

\begin{spacing}{1.3}

\section{Introduction}

In conformal field theory (CFT), causality of four-point functions places nontrivial constraints on CFT three-point couplings. In particular, causality in the lightcone limit leads to constraints \cite{Hartman:2015lfa, Hartman:2016dxc, Hofman:2016awc} which are identical to the bounds obtained from the conformal collider experiment \cite{Hofman:2008ar}. Of course, this is not a coincidence. In fact, the proof of the averaged null energy condition (ANEC) $\int T_{uu}du \ge 0$ from causality \cite{Hartman:2016lgu} made it apparent that for generic CFTs, the conformal collider set-up provides an efficient tool for deriving causality constraints.

The conformal collider set-up is a simple yet powerful thought experiment that was introduced by Hofman and Maldacena \cite{Hofman:2008ar}. In this set-up, the CFT is prepared in an excited state by creating a localized excitation which couples to some operator $\O$ (with or without spin) of the CFT. This excitation propagates outwards and the response of the CFT is measured by a distant calorimeter. The calorimeter effectively measures the averaged null energy flux $\langle \int T_{uu}du \rangle$ far away from the region where the excitation was created and hence the calorimeter readings should be non-negative. This gives rise to constraints on the three-point function $\langle \O T\O\rangle$, where  $T$ is the stress tensor operator. Recently, the conformal collider set-up was extended to study interference effects, leading to new bounds on OPE coefficients \cite{Cordova:2017zej,Meltzer:2017rtf}.\footnote{Similar method was also used by \cite{Chowdhury:2017vel} to constrain parity violating CFTs in $d=3$.} All of these causality constraints are valid for every CFT in $d\ge3$, however, additional assumptions about the CFT can lead to stronger constraints. In particular, similar logic in certain class of CFTs can shed  light on how gravity emerges from CFT.

\subsubsection*{Holographic CFTs}
The low energy behavior of gravitons, in any sensible theory of quantum gravity, is described by the Einstein-Hilbert action plus higher derivative correction terms. However, these higher derivative terms can lead to causality-violating propagation in nontrivial backgrounds  \cite{Brigante:2008gz,Hofman:2009ug,Camanho:2009vw}. Requiring the theory to be causal in shockwave states, as shown by Camanho, Edelstein, Maldacena, and Zhiboedov \cite{Camanho:2014apa} (CEMZ),  does impose strong constraints on gravitational three-point interactions. For example, causality dictates that the graviton three-point coupling should be universal in quantum gravity \cite{Camanho:2014apa} -- a claim consistent with constraints obtained from unitarity and analyticity \cite{Bellazzini:2015cra}. Furthermore, the AdS/CFT correspondence  \cite{Maldacena:1997re, Witten:1998qj,Gubser:1998bc} immediately suggests that in any CFT with a holographic dual description, certain three-point functions (for example  $\langle TTT\rangle$) must also have specific structures. 

Over the past several years, it has become clear that a large class of CFTs, with or without supersymmetry, exhibits gravity-like behavior \cite{Strominger:1997eq,Keller:2011xi,Hartman:2013mia,Hartman:2014oaa,Fitzpatrick:2014vua,Perlmutter:2016pkf,Cornalba:2006xk,Cornalba:2006xm,Cornalba:2007zb,Heemskerk:2009pn,Mack:2009gy,Mack:2009mi,Fitzpatrick:2010zm,Heemskerk:2010ty,Fitzpatrick:2011hu,Fitzpatrick:2011ia,
ElShowk:2011ag,Komargodski:2012ek,Fitzpatrick:2012yx,Fitzpatrick:2012cg,Goncalves:2014rfa,Hijano:2015zsa,Alday:2016htq,Costa:2014kfa,Alday:2014tsa,Caron-Huot:2017vep}. More recently, the CEMZ causality constraints have been derived from the CFT side for dimension $d\ge 3$ \cite{Afkhami-Jeddi:2016ntf,Kulaxizi:2017ixa,Costa:2017twz,Afkhami-Jeddi:2017rmx,Meltzer:2017rtf}, under the assumptions:
\begin{itemize}
\item{The central charge $c_T$ is large\footnote{$c_T$ is the coefficient of the stress tensor two-point function (see equation (\ref{cT})). For gauge theories, the large $c_T$ limit is equivalent to the large-$N$ limit.}: $c_T\gg 1$}
\item{A sparse spectrum: the lightest single trace operator with spin $\ell >2$ has dimension
\be
\Deltag\gg 1\ . \nonumber
\ee
}
\end{itemize}
All of these observations indicate that CFTs in this class, irrespective of their microscopic details, admit a universal gravity-like holographic dual description at low energies. Furthermore, this connection provides us with a powerful tool to constrain gravitational interactions by studying CFTs with a large central charge and a sparse spectrum. In this paper, we intend to adopt this point of view. First, for CFTs in this universality class (henceforth denoted {\it holographic CFTs}), we will derive general constraints on CFT three-point functions from causality. In light of the AdS/CFT correspondence, these CFT causality constraints translate into constraints on the low energy gravitational effective action from UV consistency.

The CEMZ causality constraints for CFTs with large central charge and a sparse spectrum were first derived in \cite{Afkhami-Jeddi:2016ntf} from causality of  the four-point function $\langle \psi \psi T_{\alpha \beta}T_{\gamma \delta}\rangle$ in the Regge limit, where $\psi$ is a heavy scalar operator. The derivation heavily relied on the fact that the stress tensor operators in the correlator were smeared in a specific way that projected out $[TT]$ double trace contributions to the Regge correlator. 
The same constraints were also derived in \cite{Kulaxizi:2017ixa,Costa:2017twz} by imposing unitarity on a differently smeared correlator $\langle \psi \psi T_{\alpha \beta}T_{\gamma \delta}\rangle$ in the Regge limit. Moreover, this approach was recently extended to study a mixed system of four-point functions in the Regge limit yielding new bounds on the OPE coefficients of low spin operators in holographic CFTs\cite{Meltzer:2017rtf}. From the dual gravity perspective, all of these set-ups are probing local high energy scattering deep in the bulk. However, the actual CFT analysis involves computations of CFT four-point functions of spinning operators using the conformal Regge theory \cite{Costa:2012cb}, which is technically challenging even in the large central charge limit. One might hope that in the Regge limit causality of CFT four-point functions can be translated to some holographic energy condition which is a generalization of the averaged null energy condition for holographic CFTs. Such an energy condition was recently derived in \cite{Afkhami-Jeddi:2017rmx}. In this paper, we will exploit this energy condition to design a new experiment, similar to the conformal collider experiment of  \cite{Hofman:2008ar}, for holographic CFTs which will allow us to bypass the conformal Regge theory.

\subsubsection*{Holographic null energy condition}

In the Regge limit, causality dictates that the {\it shockwave operator} $\int h_{uu}du$ must be non-negative for CFTs with large central charge and a sparse spectrum \cite{Afkhami-Jeddi:2017rmx}. This immediately allows us to imagine an ``AdS collider" where the boundary CFT is again prepared in the Hofman-Maldacena state $|HM\rangle$. But now the measuring device is in the bulk and measures  $\langle HM| \int h_{uu}du|HM \rangle \ge 0$ (see figure \ref{ads_coll}). It is obvious that this set-up will reproduce all of the causality constraints, however, both technically and conceptually this is not very satisfying for several reasons. First, this correlator should be computed using Witten diagrams which is difficult when the state $|HM\rangle$ is prepared using spinning operators. Second, in the CFT language, this set-up is not illuminating because the operator $\int h_{uu}du$ has a complicated decomposition into CFT operators which consists of  the stress tensor and an infinite tower of double trace operators. 

In this paper, we consider the stress tensor component of the shockwave operator \cite{Afkhami-Jeddi:2017rmx}
\be\nonumber
 \E_{r}(v)=\int_{-\infty}^{+\infty}  du' \int_{\vec{x}^2 \le r^2} d^{d-2}\vec{x}\left( 1-\frac{ \vec{x}^2}{ r^2}\right) T_{uu} \left(u' ,v , i \vec{x} \right)\ ,
\ee
which we will refer to as the {\it holographic null energy operator}.\footnote{$u$ and $v$ are the null coordinates.} Causality of CFT four-point function in the Regge limit \cite{Afkhami-Jeddi:2017rmx} implies that the expectation value of the holographic null energy operator is positive in a large subspace of the total Hilbert space of holographic CFTs. Note that this operator is the averaged null energy operator smeared over a finite sphere along the imaginary transverse directions. Of course, the positivity of the holographic null energy operator is not implied by the ANEC because of the imaginary transverse directions. In fact, this operator, in general, is not positive. 

A key ingredient of the positivity argument is that there exists a class of states $|\Psi\rangle$ which projects out certain double trace contributions to $\int h_{uu}du$. This is an extension of the observations made in \cite{Afkhami-Jeddi:2017rmx}.\footnote{We should note that in this paper we will not provide a general technical proof of the observation made in \cite{Afkhami-Jeddi:2017rmx} about double trace contributions. However, we will argue that the same statement about double trace contributions is true in general.} These states, as we will show,  are equivalent to the Hofman-Maldacena state $|HM\rangle$ which will allow us to introduce a new formalism to study causality constraints. Our formalism can be interpreted as a new collider-type experiment for holographic CFTs (see figure \ref{coll_fig}). Consider a CFT with large central charge and a sparse spectrum in $d$-dimensions. The CFT is prepared in the excited state $|HM\rangle$ by inserting a spinning operator $O$ near the origin and an instrument measures the holographic null energy far away from the excitation:
\be\nonumber
\E(\rho)=\lim_{R\rightarrow \infty} R^2 \langle HM|\E_{r=\sqrt{\rho}R}(R) |HM\rangle\ .
\ee
The holographic null energy condition implies that $\E(\rho)$ is a positive function for $0< \rho < 1$. The parameter $\rho$ is a measure of the angular size of our measuring device at the origin and the parameter $\rho$ can be tuned by changing the size of the device. In the gravity language, $\rho$ plays the role of the bulk direction. In particular, $\rho \rightarrow 0$ represents the lightcone limit (AdS boundary) and hence in this limit, this set-up is equivalent to the original conformal collider experiment. On the other hand, we are interested in probing high energy scattering deep in the bulk of the dual geometry which corresponds to the limit $\rho\rightarrow 1$.

Our conformal collider set-up has several advantages over previous methods \cite{Afkhami-Jeddi:2016ntf,Kulaxizi:2017ixa,Costa:2017twz,Afkhami-Jeddi:2017rmx,Meltzer:2017rtf}. First, we do not need to compute  conformal Regge amplitudes. In our set-up, all of the constraints are directly obtained from CFT three-point functions which are fixed by conformal symmetry up to a few constant coefficients -- a simplification which enables us to derive constraints in a more systematic way. Finally, our approach connects causality constraints in the Regge limit with the holographic null energy condition. This is reminiscent of the ANEC which relates causality in the lightcone limit with entanglement. So, the appearance of the holographic null energy condition perhaps is an indication of some deeper connection between boundary entanglement and bulk locality. Moreover, the recent generalization of the ANEC to continuous spin \cite{Kravchuk:2018htv} suggests that there might also be a generalization of the holographic null energy condition to continuous spin.

\subsubsection*{Summary of results}

The formalism that we developed in this paper efficiently computes the expectation value of $\E_{r}$ in states $|\Psi\rangle$, constructed by inserting spinning operators.\footnote{This formalism can easily be adapted to computing the contribution of any conformal multiplet to the Regge limit of four-point correlation functions.} This formalism involves performing certain integrals over CFT three-point functions which are fixed up to OPE coefficients by symmetries. Furthermore, we decompose the results into independent bounds corresponding to representations under spatial rotations. The inequalities following from these bounds lead to surprising equalities among the various OPE coefficients involving spinning operators and the stress-tensor. 

The first set of constraints are obtained by considering expectation values in states constructed from a single low spin operator ($\ell\le 2$). The second set of constraints follows from the interference effects in our collider. In particular, positivity of the holographic null energy operator in states created by superposition of smeared local operators $O_1$ and $O_2$ leads to a bound on the off-diagonal matrix elements of the operator $\E$:
\be\nonumber
|\E_{O_1 O_2}(\rho)|^2 \le \E_{O_1 O_1}(\rho)\E_{O_2 O_2}(\rho)\  .
\ee
Let us now summarize the resulting constraints  for all single trace low spin ($\ell \le 2$) operators in a holographic CFT (in $d\ge 3$).
\begin{itemize}
\item{All three-point functions of the form $\langle T OO \rangle$ are completely fixed by the two-point function $\langle OO \rangle$. }
\item{All three-point functions $\langle T O_1 O_2 \rangle$, where $O_1$ and $O_2$ are different operators, are suppressed by $\Deltag$.}\footnote{There is a caveat. Our argument does not necessarily hold if scaling dimensions of $O_1$ and $O_2$ coincide with the scaling dimension of double-trace operators (at leading order in $c_T$). For more discussion see  \cite{Cordova:2017zej,Meltzer:2017rtf}.}
\end{itemize}
These constraints encompasses, and generalizes,  all known causality constraints as obtained in \cite{Afkhami-Jeddi:2016ntf,Kulaxizi:2017ixa,Costa:2017twz,Afkhami-Jeddi:2017rmx,Meltzer:2017rtf} by studying various four-point functions in holographic CFTs. Moreover, after imposing these causality constraints, we find that the expectation value of the holographic null energy operator is universal and it is completely determined by the lightcone limit result. This observation suggests the following conclusion about the operator product expansions in holographic CFTs:
\begin{itemize}
\item{The operator product expansion of any two smeared primary single trace operators (with or without spin) in the Regge limit is given by a  {\it  universal shockwave operator}:
\be\nonumber
\Psi^*[O_1] \Psi[O_2] \approx \langle \Psi^*[O_1] \Psi[O_2] \rangle-2 i E_{O_1 O_2} \int_0^\infty dt\ t^2 h_{z+t\ z+t} \  ,
\ee
where, $E_{O_1 O_2}$ is the matrix element of the total energy operator. The operators $O_1$ and $O_2$ are smeared in such a way that they can create states which belong to the class $|\Psi\rangle$ (see section \ref{section_ope}). On the right hand side, the spherical shockwave operator is written as an integral of the metric perturbation over a null geodesic: $z=t$ (where $z$ is the bulk direction and $t$ is the Lorentzian time) in AdS$_{d+1}$ for $d\ge 3$.}
\end{itemize}

In the gravity language, the above CFT constraints translate into the statement that all higher derivative interactions in the low energy effective action must be suppressed by the new physics scale. Furthermore, in agreement with the proposal made by Meltzer and Perlmutter in \cite{Meltzer:2017rtf}, we find that in $d\ge 4$ CFT dual of a bulk derivative is $1/\Deltag$. However, we also notice that in $d=3$ there is a logarithmic violation of this simple relationship between the bulk derivative and $\Deltag$.

As a simple example of the above bounds, we derive ``$a\approx c$" type relations between conformal  trace anomalies in $d=6$. In $d=6$, there are four Weyl anomaly coefficients $a_6,c_1,c_2,c_3$, however, three of them $(c_1,c_2,c_3)$ are determined by the stress tensor three-point function $\langle TTT\rangle$. Our bounds immediately imply that the anomaly coefficients must satisfy  $c_1=4c_2=-12c_3$. These relations between $c_1, c_2, c_3$ are exactly what is expected for $(2,0)$ supersymmetric theories, both holographic and non-holographic \cite{Beccaria:2017dmw}. This is reminiscent of the Ooguri-Vafa conjecture \cite{Ooguri:2016pdq} which states that holographic duality with low energy description in term of the Einstein gravity coupled to a finite number of matter fields exists only for supersymmetric theories.

Finally, as a new application of the holographic null energy condition, we constrain various inflationary observables such as the amplitude of chiral gravity waves, nongaussanity of gravity waves and tensor-to-scalar ratio. Our argument parallels the argument made by Cordova, Maldacena, and Turiaci in \cite{Cordova:2017zej}. The bounds on higher curvature interactions in AdS$_4$ strongly suggests that these higher curvature terms should also be suppressed by the scale of new physics in the effective action in de Sitter space. Hence, any effect that arises from these higher curvature terms must be vanishingly small. For example, in $(3+1)-$dimensional gravity all parity odd interactions appear in higher derivative order. Therefore, all inflationary observables that violate parity including chiral gravity waves and parity odd graviton nongaussanity, must be suppressed by the scale of new physics. Furthermore, any detection of these effects  in future experiments will imply the presence of an infinite tower of new particles with spins $\ell >2$ and masses comparable to the Hubble scale.

\subsubsection*{Outline}
The rest of the paper is organized as follows. In section \ref{section_hnec}, we discuss the conformal collider set-up for holographic CFTs and review the holographic null energy condition. Then in section \ref{section_ope}, we summarize our causality constraints as a statement about Regge OPE of smeared operators. In this section, we also propose a relation that connects the Regge limit with the lightcone limit for holographic CFTs. In section \ref{section_int}, we present a systematic approach of calculating the expectation value of the holographic null energy operators in states created by smeared operators. This section mainly contains technical details, so it can be safely skipped by casual readers. In sections \ref{section_hd} and \ref{section_intf}, we derive explicit constraints on CFT three-point functions for $d\ge 4$. The $d=3$ case is more subtle and hence we treat it separately in section \ref{section_3d}. In section \ref{section_cosmo}, we discuss the cosmological implications of our CFT bounds. Finally, we end with concluding remarks in section \ref{section_final}.

\section{Causality and conformal collider physics}\label{section_hnec}

\begin{figure}
\begin{center}
\includegraphics[width=0.75\textwidth]{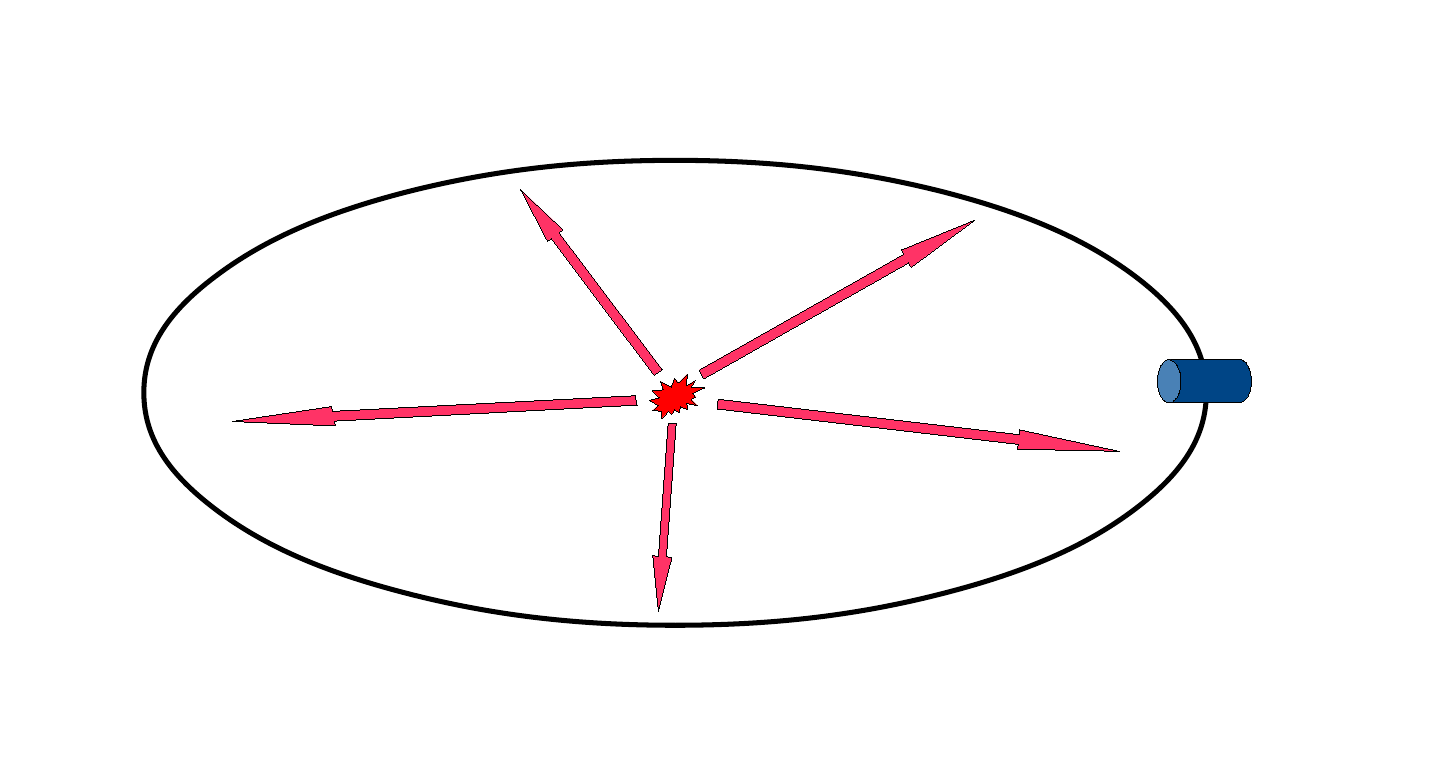}
\caption{Conformal collider experiment: A localized excitation is created in a holographic CFT and an instrument which is shown in blue, measures the holographic null energy $\E_r$ far away from the excitation.}\label{coll_fig}
\end{center}
\end{figure}


In the lightcone limit, causality dictates that the averaged null energy operator $\int T_{uu}du$ should be non-negative  \cite{Hartman:2016lgu}.\footnote{The averaged null energy condition for interacting quantum field theories in Minkowski spacetime was first derived in \cite{Faulkner:2016mzt} from monotonicity of relative entropy.} The ANEC immediately leads to positivity of all CFT three-point functions which have the form: $\langle O|\int T_{uu}du| O\rangle\ge 0$. On the other hand, for CFTs with large central charge and a sparse spectrum, causality of four-point functions in the Regge limit leads to stronger constraints. However, all of these causality conditions involve computations of CFT four-point functions of spinning operators using the conformal Regge theory \cite{Costa:2012cb}. The causality of CFT four-point functions even in the Regge limit can be translated to positivity of certain (holographic) energy operator \cite{Afkhami-Jeddi:2017rmx}. In this section, with the help of that positivity condition, we  develop a new conformal collider set-up enabling us to derive causality bounds directly from three-point functions.

\subsection{A collider for holographic CFTs}
We will use the following convention for points $x \in \mathbb{R}^{1,d-1}$:
\be
x = (t,x^1,\vec{x})\equiv (u,v,\vec{x})\ , \qquad \text{where},  \qquad u=t-x^1\ , \qquad v=t+x^1\ .
\ee
Let us now define the holographic null energy operator:
 \be\label{hne}
 \E_{r}(v)=\int_{-\infty}^{+\infty}  du' \int_{\vec{x}^2 \le r^2} d^{d-2}\vec{x}\left( 1-\frac{ \vec{x}^2}{ r^2}\right) T_{uu} \left(u' ,v , i \vec{x} \right)\ .
 \ee
The holographic null energy operator is a generalization of the averaged null energy operator which was first introduced in \cite{Afkhami-Jeddi:2017rmx}.\footnote{Also see \cite{Czech:2016xec,deBoer:2016pqk} for a connection between the holographic null energy operator and the modular Hamiltonian.} In particular, in the limit $r\rightarrow 0$, this operator is equivalent to the averaged null energy operator. The kernel in (\ref{hne}) is positive and hence one might expect that the operator $ \E_{r}(v)$ should also be positive. However, this is not true because the stress tensor is also integrated over imaginary transverse coordinates and in general $\int du' T_{uu} \left(u' ,v , i \vec{x} \right)$ can have either sign.

Let us now carry out a collider physics thought experiment similar to \cite{Hofman:2008ar} but with a holographic CFT in $d$-dimensions where $d\ge 3$ (see figure \ref{coll_fig}). We prepare the CFT in an excited state by inserting a spinning operator $O$ near the origin\footnote{$O$ is not necessarily a primary operator. In fact $O$ can  be a linear combination of various operators with different spins. Also note that in equation (\ref{psi}), $\epsilon.O\equiv \epsilon_{\mu \nu...}O^{\mu\nu...}$.}:
\be\label{psi}
|\Psi\rangle=\int dy^1 d^{d-2}\vec{y}\, \epsilon.O(-i\delta,y^1,\vec{y})|0 \rangle\ ,
\ee
where, $\epsilon$ is the polarization of the operator $O$ and $\delta>0$. Similarly,
\be
\langle \Psi|=\int dy^1 d^{d-2}\vec{y}\, \langle 0| \epsilon^*.O(i\delta,y^1,\vec{y})\ .
\ee
The state $|\Psi\rangle$ is equivalent to the Hofman-Maldacena state of the original conformal collider experiment \cite{Hofman:2008ar}. Now we imagine an instrument that measures the holographic null energy $\E_r(v)$ far away from the excitation:
\be\label{collider}
\E(\rho)=\lim_{B\rightarrow \infty}\langle \Psi| \E_{\sqrt{\rho}B}(B) |\Psi  \rangle\ ,
\ee
where, $0< \rho < 1$. The parameter $\rho$ is a measure of the size of the measuring device which we can tune. The measuring device is placed at a distance $B$ away from the excitation and the angular size of the device is roughly $\rho^{\frac{d-2}{2}}$.  A priori it is not obvious that the measured value $\E(\rho)$ has to be positive. However, later in this section, by using the positivity conditions of \cite{Afkhami-Jeddi:2017rmx}, we will show that for CFTs with large central charge and a sparse spectrum in $d\ge 3$:
\be\label{hnec}
\E(\rho) \ge 0\ , \qquad  0< \rho < 1
\ee
for a class of states that has the form (\ref{psi}). This inequality will play an important role in this paper and we will refer to this as holographic null energy condition. In the limit $\rho\rightarrow 0$, the holographic null energy operator becomes $\int du' T_{uu}(u')$ and  $\E(\rho) \ge 0$ is true for any CFT. In this limit, the positivity of $\E(\rho)$ reproduces the conformal collider bounds of \cite{Hofman:2008ar,Hartman:2016lgu, Cordova:2017zej,Meltzer:2017rtf,Chowdhury:2017vel}. Note that the wavepacket  of \cite{Hofman:2008ar} is implemented here by the order of limits. We first perform the $u'$-integral in (\ref{collider}) and then take the limit $B\rightarrow \infty$. The same trick was used in \cite{Hartman:2016lgu} to derive conformal collider bounds directly from a Rindler reflection symmetric set-up.

This conformal collider set-up is equivalent to the set-up used in \cite{Afkhami-Jeddi:2016ntf,Afkhami-Jeddi:2017rmx}, however, now we do not need to compute a four-point function. For example, in $d=4$, if we take $O$ to be the stress tensor  and choose the polarization $\epsilon^\mu=(-i,-i,i\lambda, \lambda)$, as we demonstrate in appendix \ref{sec:pol}, each power of $\lambda$ should individually satisfy (\ref{hnec}). In particular, in the limit $\rho\rightarrow 1$, we recover $a=c$ from (\ref{hnec}).

Before we proceed, let us rewrite (\ref{collider}) in a more familiar form. The Hofman-Maldacena state of the original conformal collider experiment \cite{Hofman:2008ar} is given by
\be\label{HM}
|HM\rangle = \int dt dy^1 d^{d-2}\vec{y} e^{-(t^2+{(y^1)}^2+\vec{y}^2)/D}e^{-i \omega t} \epsilon^{\mu \nu ...} O_{\mu \nu...}(t,y^1,\vec{y})\ , \qquad \omega D\gg 1\ .
\ee
Then (\ref{hnec}) immediately implies that
\be
\lim_{R\rightarrow \infty} R^2 \langle HM|\E_{r=\sqrt{\rho}R}(R) |HM\rangle \ge 0\ .
\ee

\subsection{Holographic null energy condition}
It was shown in our previous paper \cite{Afkhami-Jeddi:2017rmx} that causality of CFT four-point functions in the Regge limit implies positivity of certain smeared CFT three-point functions. First,  we review and further explore that positivity condition. Then, we derive (\ref{hnec}) as a simple consequence.

\subsubsection{Regge limit and OPE of heavy scalars}

\begin{figure}
\begin{center}
\includegraphics[width=0.75\textwidth]{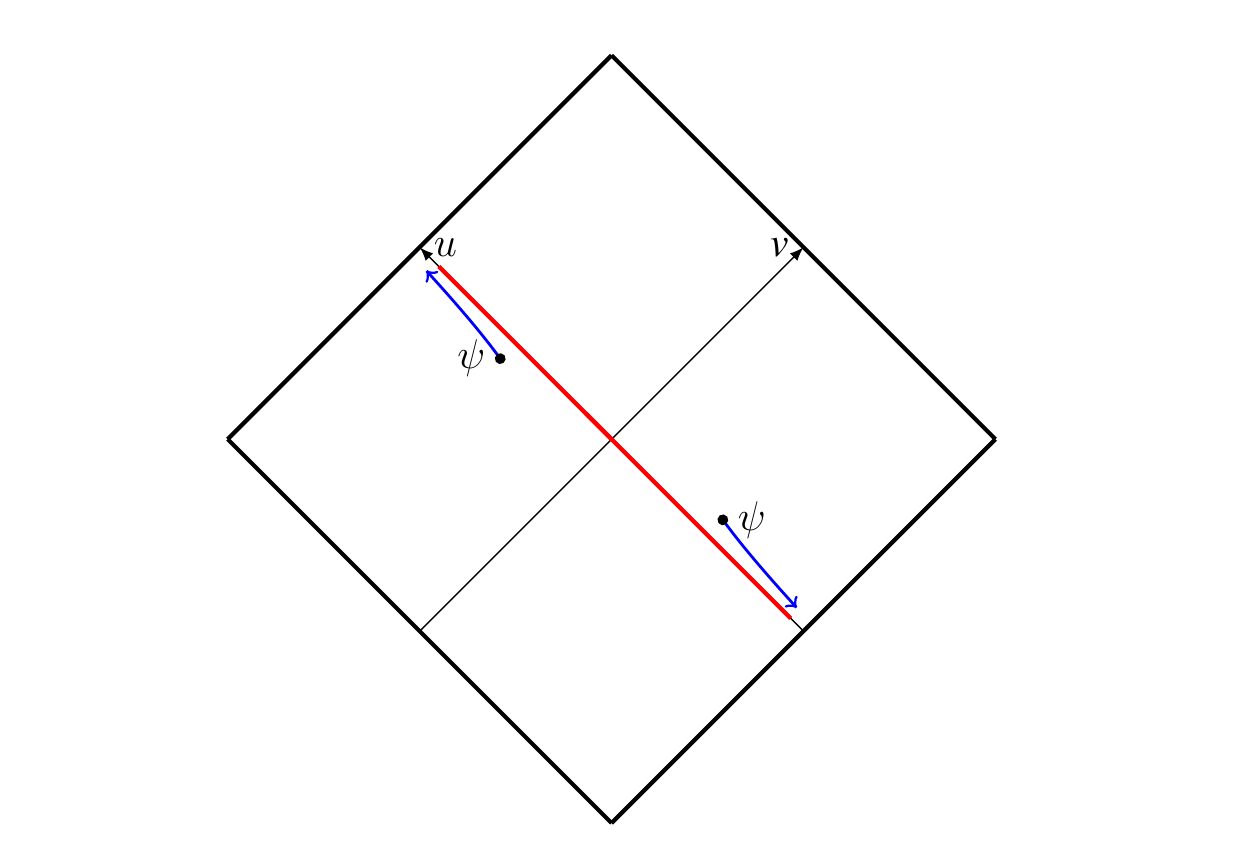}
\end{center}
\caption{\label{regge} \small In the Regge limit the leading correction to the $\psi\psi$ OPE is the graviton $h_{uu}$ integrated over the red line.}
\end{figure}

We start with a discussion on the Regge OPE of heavy operators in the holographic limit. Let us consider a real scalar primary $\psi$ in a $d-$dimensional CFT with $\Delta_\psi \gg 1$. In general, one can replace any two nearby operators by their OPE. For example, $\psi(u,v)\psi(-u,-v)$ can be written as\footnote{Whenever we drop some spacetime coordinates, those coordinates are set to zero.}
\be
\psi(u,v)\psi(-u,-v)=\sum_p C_p (u,v; \partial_u, \partial_v) \O_p(0,0)\ ,
\ee
where, the sum is over all primaries. In a generic CFT, the lightcone and the Regge limits of a correlator are controlled by different sets of operators. In the standard lightcone limit $v\rightarrow 0$ (with $u$ fixed), the above OPE can be organized as an expansion in twist $\tau_p=\Delta_p-\ell_p$ ($\Delta$ is scaling dimension and $\ell$ is spin) which leads to a simple lightcone OPE \cite{Hartman:2016lgu}. On the other hand, the Regge limit is obtained by taking (see figure \ref{regge})
\be\label{rg}
v\rightarrow 0\ , \qquad u\rightarrow \infty\ , \qquad uv=\text{fixed}\ .
\ee
Unlike the lightcone limit, the Regge limit gets significant contributions from high spin exchanges. Even when the central charge $c_T$ (defined in (\ref{cT})) is large, complication arises because an infinite tower of double trace operators become relevant in the Regge limit. However, under the additional assumption that the spectrum of single trace operators with $\ell >2$ is sparse, simplification emerges and the Regge OPE can  be written as \cite{Afkhami-Jeddi:2017rmx}
\be\label{reggeope1}
\frac{\psi(u,v)\psi(-u,-v)}{\langle \psi(u,v)\psi(-u,-v)\rangle }= 1-\frac{\Delta_\psi u}{2} \int_{-\infty}^{\infty} du' h_{uu}(u',v'=0,\vec{x}'=0,z'=\sqrt{-uv})+\cdots\ ,
\ee
where,  $c_T\gg \Delta_\psi \gg 1$ and dots are $\O(u^0,\Delta_\psi^0, 1/c_T^2)$ terms. $h_{uu}$ in the above equation is the bulk metric perturbation in AdS$_{d+1}$ (where $z$ is the bulk coordinate) which is integrated over a null geodesic. In the gravity language, contributions of an infinite tower of primary operators translate into a single term because the dominant contribution to the four-point function comes from the Witten diagram with a single graviton exchange. Hence, the right hand side of (\ref{reggeope1}) should be thought of as a  CFT operator written in terms of the bulk metric. In particular, $\int du h_{uu}$ contains the stress tensor $T_{\mu \nu}$ as well as all double trace operators $[O_1 O_2]$ built from the light operators in theory, e.g., $[TT]$ double trace operators. The stress tensor contribution of $\int du h_{uu}$ can be computed using the HKLL prescription for $h_{uu}$\cite{Kabat:2012hp}.

Causality of the Regge correlator dictates that the operator  $\int du h_{uu}$ has to be positive \cite{Afkhami-Jeddi:2017rmx} and hence any three-point function which has the form $\langle O|\int h_{uu}du|O\rangle$ must be positive as well. From the  CFT perspective, this positivity condition both technically and conceptually is not very useful. However, we will show that for specific states, only the stress tensor contribution of $\int du h_{uu}$ is important which will lead us to the holographic null energy condition. Before we proceed, let us note that the contribution of the single trace stress tensor and its derivatives to the Regge OPE (\ref{reggeope1}) can be written in terms of the holographic null energy operator \cite{Afkhami-Jeddi:2017rmx}
\begin{align}\label{regope}
\frac{\psi(u,v) \psi(-u,-v)|_T}{\langle \psi(u,v) \psi(-u,-v) \rangle }  =- \frac{\Delta_\psi 2^d \pi^{\half - d}\Gamma(\frac{d+2}{2})\Gamma(\frac{d+3}{2})u}{c_T (d-1)}
  \E_{r=\sqrt{-uv}}(0)\ ,
\end{align}
where, $\E_r(v)$ is defined in (\ref{hne}).

\subsubsection{Positivity}
\begin{figure}
\begin{center}
\includegraphics[width=0.80\textwidth]{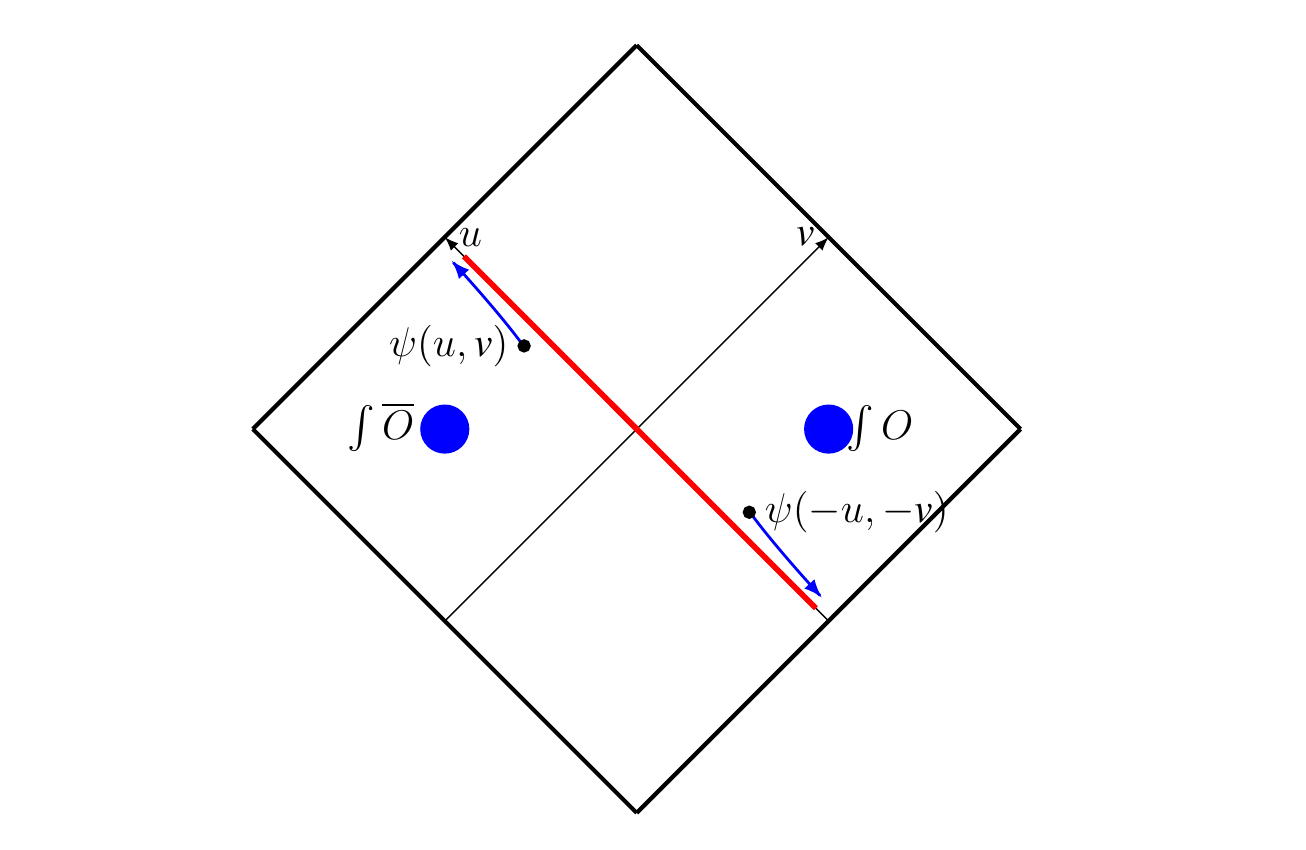}
\end{center}
\caption{\label{config} \small Kinematics for the derivation of the holographic null energy condition. Operators $O$s are smeared over some regions in a Rindler reflection symmetric way.}
\end{figure}
Consider a Rindler reflection symmetric four-point function
\be\label{maing}
G = \frac{\langle \overline{\varepsilon.O(B)} \psi(u, v)\psi( -u, - v) \varepsilon.O(B)\rangle}{\langle \overline{\varepsilon.O(B)}\varepsilon.O(B)\rangle \langle  \psi(u, v)\psi( -u, - v) \rangle} \ ,
\ee
in the regime (\ref{rg}), as shown in figure \ref{config}. $\varepsilon.O(B)$ is an arbitrary operator with or without spin (not necessarily a primary operator) smeared over some region:
\be\label{ob}
\varepsilon.O(B)=\int d\tau d^{d-2}\vec{y}\, \varepsilon.O(t=i(B+\tau),y^1=\delta,\vec{y})\ ,
\ee
where, $\delta>0$ and $\varepsilon$ is the polarization (when $O$ is a spinning operator). Operator $ \overline{\varepsilon.O}$ is the Rindler reflection of the operator $O$ (see \cite{Hartman:2016lgu} for a detailed discussion): 
\be\label{obbar}
\overline{\varepsilon.O(B)}=\int d\tau d^{d-2}\vec{y}\, \overline{\varepsilon}.O^\dagger(t=i(B+\tau),y^1=-\delta,\vec{y})\ ,
\ee
where, the Hermitian conjugate on the right-hand side does not act on the coordinates. $\overline{\varepsilon}$ is the Rindler reflection of the polarization $\varepsilon$:
\be
\overline{\varepsilon^{\mu \nu \cdots}} \equiv (-1)^{P}(\varepsilon^{\mu \nu \cdots})^*
\ee
where $P$ is the number of $t$-indices plus $y^1$-indices. 

Following \cite{Afkhami-Jeddi:2016ntf}, let us define
\be\label{coordslB}
u=\frac{1}{\sigma}\ , \qquad v=-\sigma B^2 \rho
\ee
with $B>0, \sigma>0$ and $0< \rho <1$. The Regge limit is obtained by taking $\sigma \rightarrow 0$ with $\rho, B$ fixed. Now using the OPE \eqref{reggeope1}, we obatin
\be\label{gcor}
G\equiv 1+\delta G = 1 - \frac{\Delta_\psi}{2\sigma \N}   \langle  \overline{\varepsilon.O(B)} \, \int du' h_{uu}(u',z=B \sqrt{\rho}) \, \varepsilon.O(B) \rangle 
\ee
with $\N = \langle \overline{\varepsilon.O(B)}\varepsilon.O(B)\rangle> 0 $. The null line integral in the above expression is computed by choosing appropriate contour. We can now repeat the arguments of \cite{Hartman:2016lgu,Afkhami-Jeddi:2016ntf} which tells us that the boundary CFT will be causal if and only if
\be\label{condition}
\text{Im} (\delta G) \le 0\ ,
\ee
which is precisely the chaos bound of \cite{Maldacena:2015waa}. Since, $\delta G$ as obtained from (\ref{gcor}) is purely imaginary, therefore the last inequality is equivalent to 
\be\label{tyrion}
i     \langle  \overline{\varepsilon.O(B)} \, \int du' h_{uu}(u',z=B \sqrt{\rho}) \, \varepsilon.O(B) \rangle  \ge 0\ 
\ee
for any operator $O$. After we perform a rotation by $\pi/2$ in the Euclidean $\tau-x^1$ plane, this is precisely the statement that the shockwave operator $\int du h_{uu}$ is positive  \cite{Afkhami-Jeddi:2017rmx}. This is a CFT version of the a bulk causality condition proposed by Engelhardt and Fischetti in \cite{Engelhardt:2016aoo}. They showed that asymptotically AdS spacetimes satisfy boundary causality if and only if metric perturbations satisfy $\int du h_{uu}\ge 0$. This requirement is weaker than the bulk null energy condition which was the starting point of the Gao-Wald theorem \cite{Gao:2000ga}.

\begin{figure}[h]

\begin{center}
\includegraphics[width=0.70\textwidth]{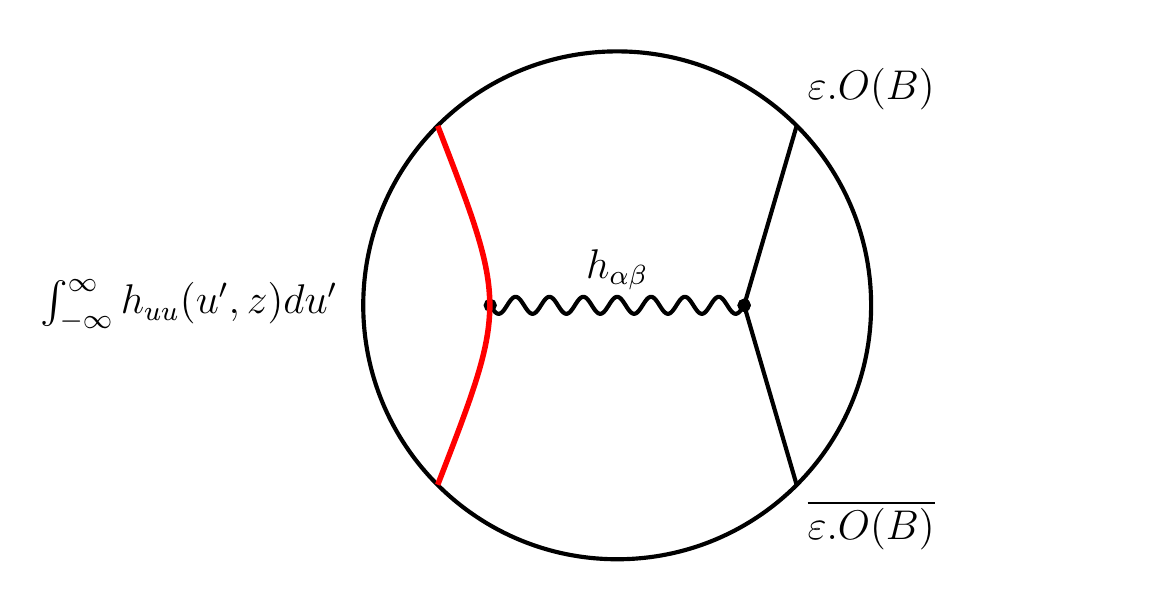}
\end{center}
\caption{ \small The Witten diagram for the correlator $\langle  \overline{\varepsilon.O(B)} \, \int du' h_{uu}(u',z=B \sqrt{\rho}) \, \varepsilon.O(B) \rangle $.}\label{ads_coll}
\end{figure}

The three-point function (\ref{tyrion}) can  be computed using the Witten diagram \ref{ads_coll}, however, we want a three-point function that can be computed directly in CFT. It was also shown in \cite{Afkhami-Jeddi:2017rmx} that in the limit $B\rightarrow \infty$, the smearing projects out $[OO]$ double-trace contributions in the correlator (\ref{maing}) when $O$ is a scalar operator or a spin-1 conserved current or the stress tensor $T$. There are plenty of evidences which suggest that the same statement is true for any operator $O$. We will not attempt to prove this statement, instead  we conjecture that the smearing projects out $[OO]$ double-trace contributions for any operator $O$. The intuition comes from gravity. In the Witten diagram \ref{ads_coll}, the smearing puts the field dual to the operator $O$ onto a geodesic, converting the Witten diagram \ref{ads_coll} into a geodesic Witten diagram  which receives contribution only from the stress tensor exchange \cite{Hijano:2015zsa}. As an immediate consequence, we can replace $\int du' h_{uu}$ in (\ref{tyrion}) by the single trace stress tensor contribution (\ref{regope}), yielding 
 \be\label{snec}
- i\ \lim_{B\rightarrow \infty} \langle\overline{\varepsilon.O(B)} \,   \E_{r=\sqrt{\rho}B}(0) \, \varepsilon.O(B) \rangle \ge 0\ .
 \ee
This is a statement about CFT three-point function which allows us to circumvent the computation of four-point functions.\footnote{Let us note that if there are non-conserved spin-2 single trace primaries in the theory, they can also contribute to the four-point function in the Regge limit and hence equation (\ref{snec}) will not be true. For now, we assume that if they are present, they do not contribute as an exchange operator. However, as will discuss in section \ref{section_intf}, this assumption is not entirely necessary.} Let us note that our conjecture about the double trace operator is simply a technical fact about the smearing that we performed. Later in the paper, we will justify our conjecture about double trace operators by demonstrating that the inequality (\ref{snec}) reproduces all known causality constraints for holographic CFTs. This is a non-trivial check of the conjecture, however, one can perform a more direct check by utilizing the conformal Regge theory. It is not difficult to show case by case   that $\delta G$ as obtained from (\ref{maing})  receives contribution only from the stress tensor exchange.  But we admit that it will be nice to have a more general technical proof.

Let us make few comments regarding the regime of validity for the inequality (\ref{snec}).
\begin{itemize}
\item {The inequality is true for any $0< \rho < 1$ for CFTs in $d\ge 3$ with large central charge and a sparse spectrum. In particular, in the limit $\rho\rightarrow 1$, (\ref{snec}) probes scattering at a point deep in the interior of AdS, similar to \cite{Camanho:2014apa, Afkhami-Jeddi:2017rmx}. 
}

\item{The limit $\rho\rightarrow 0$ corresponds to the lightcone limit and in this limit, the inequality is true for any interacting CFT in $d\ge 3$.  Furthermore, in this limit, the inequality (\ref{snec}) is equivalent to the conformal collider set-up of \cite{Hofman:2008ar} and hence it yields optimal bounds. 
}
\end{itemize}

We will use (\ref{snec}) to derive constraints for holographic CFTs. So, let us rewrite (\ref{snec}) in a more explicit form that we will use in later sections:
\begin{align}\label{snec2}
-i   \int d\tau d^{d-2}\vec{y}  \lim_{B\rightarrow \infty} \langle \overline{\varepsilon}.O^\dagger(iB,-\delta,\vec{0})\,   \E_{r=\sqrt{\rho}B}(0)\, \varepsilon.O(i(B+\tau),\delta,\vec{y}) \rangle \ge 0\ .
\end{align}
We want to stress that in the above expression, order of limits is important. We perform the $u'$ integral first and then take the large $B$ limit. Also note that we are only smearing one of the operators because the other smearing integral will only give an overall volume factor. This is a consequence of the large $B$ limit and this volume factor is the same factor that appears in the smeared two-point function.

The inequality (\ref{snec2}) is not yet an expectation value of the holographic null energy operator in a state which has the form (\ref{psi}). However we can rewrite the inequality (\ref{snec2}) as an expectation value. First, we perform a rotation $R$ in (\ref{snec2}) that rotates by $\pi/2$ in the Euclidean $\tau-x^1$ plane where $\tau=it$ (see appendix A of \cite{Hartman:2016lgu}). Then we perform a translation along $x^1$-direction by $B$. This procedure converts (\ref{snec2}) into an expectation value\footnote{We should also transform the polarization tensor accordingly (see \cite{Hartman:2016lgu}). In particular, polarizations $\epsilon^{\mu\nu ...}$ (as used in equation (\ref{psi})) and $\varepsilon^{\mu\nu ...}$ (which has been used throughout the paper whenever we have a Rindler reflection symmetric set-up) are related in the following way:
\be
\epsilon^{\mu \nu ...}=\left( \Lambda^\mu_\alpha \Lambda^\nu_\beta ...\right) \varepsilon^{\alpha \beta ...}\ , \qquad  \Lambda^\mu_\alpha =\begin{pmatrix}
  0 & -i & 0 \\
  -i & 0 & 0 \\
  0 & 0 & \mathds{1}
 \end{pmatrix}\ .
\ee
Note that if $\overline{\varepsilon}_1=\varepsilon_2$, then $\epsilon_1^\dagger =\epsilon_2$.
}
:
\be
\lim_{B\rightarrow \infty} \langle \Psi| \E_{\sqrt{\rho}B}(B)|\Psi\rangle \ge 0\ ,
\ee
where, $|\Psi\rangle$ is a class of states which has the form (\ref{psi}). This concludes the proof of the holographic null energy condition.
\subsection{Corrections from higher spin operators}\label{gap_crctn}

The holographic null energy condition is exact strictly in the $\Deltag\rightarrow \infty$ limit. Therefore, all of the constraints obtained from the holographic null energy condition in the limit $\rho\rightarrow 1$ will receive corrections from higher spin operators above the gap. A finite number of such operators will violate causality/chaos bound and hence this scenario is ruled out. However, it is expected that an infinite tower of new higher spin operators with $\Delta >\Deltag$  starts contributing as we approach the limit $\rho\rightarrow 1$. Let us now estimate the correction to the causality constraints if we include these higher spin operators with $\Delta>\Deltag$, where,
\be
\sqrt{c_T} \gg \Deltag\gg 1\ .
\ee
We consider a single higher spin operator with spin $\ell$ and dimension $\Delta=\Deltag$ and generalize the argument of our previous paper \cite{Afkhami-Jeddi:2016ntf}. Contribution of this operator to (\ref{maing}) in the limit $\rho\rightarrow 1$ is given by \cite{Afkhami-Jeddi:2016ntf}
\be\label{crctn}
\delta G \sim \frac{i}{\sigma^{\ell-1}}\frac{e^{-s\Deltag/2}}{s^a}\ , \qquad s=1-\rho\ ,
\ee
where,  $a$ is a positive number and we have assumed that $\Deltag\gg \ell$. Therefore, these higher spin operators becomes relevant in the strict limit of $s\rightarrow 0$. On the other hand, we can safely ignore these operators when $s \gtrsim 1/\Deltag$.\footnote{We should note that $\delta G$ has large numerical factors. Here, similar to \cite{Afkhami-Jeddi:2016ntf}, we are making an additional assumption that OPE coefficients which appear in $\delta G$ are small enough to cancel these large numerical factors.} So, we can trust the causality condition (\ref{snec}) as well as the collider bound (\ref{hnec}) only in the regime $1/\Deltag \lesssim s <1$ and the strongest constraints can be obtained by setting $s \sim 1/\Deltag$.

Let us now schematically write 
\be
\text{Im} \lim_{B\rightarrow \infty} \langle\overline{\varepsilon.O(B)} \,   \E_{r=\sqrt{\rho}B}(0) \, \varepsilon.O(B) \rangle \sim \sum_{n} (\pm) \frac{t_n}{(1-\rho)^n} + \frac{c_0}{(1-\rho)^{d-3}}+\cdots\ ,
\ee
where, the sum is over terms which change sign for different polarizations and hence in the absence of the higher spin operators causality condition leads to $t_n=0$. On the other hand, we will show in the rest of the paper that after imposing the causality constraints the leading non-vanishing term in the limit $\rho\rightarrow 1$ goes as  $\frac{c_0}{(1-\rho)^{d-3}}$, where $c_0$ is positive.\footnote{In $d=3$, the leading nonzero term goes as $-c_0 \ln(1-\rho)$ and hence the $\Deltag$-correction is given by
\be
\Big|\frac{t_n}{c_0}\Big|\lesssim \frac{\ln \Deltag}{\Deltag^{n}}\ .
\ee
} Now, setting $\rho\sim 1-1/\Deltag$, from the causality/chaos bound (\ref{snec}), we obtain
\be\label{gap}
\Big|\frac{t_n}{c_0}\Big|\lesssim \frac{1}{\Deltag^{n-d+3}}\ .
\ee

\section{Universality of the smeared Regge OPE}\label{section_ope}
In the rest of the paper, we will derive constraints using the conformal collider for the holographic null energy operator. In this section, we summarize the results as a statement about the Regge OPE of smeared single trace operators with low spin. Causality of the Regge correlators suggests that the operator product expansion of any two   smeared primary  operators (with or without spin) of CFTs with large central charge and a sparse spectrum should approach a universal form in the Regge limit. 

Let us consider two arbitrary primary single trace low spin operators $O_1$ and $O_2$ ($\ell \le 2$). We now smear the operators following (\ref{psi}):
\begin{align}
&\Psi^*[O_1]=\int dy^1 d^{d-2}\vec{y}\, \epsilon_1^*.O_1(i\delta, y^1, \vec{y})\ ,\\
&\Psi[O_2]=\int dy^1 d^{d-2}\vec{y}\, \epsilon_2.O_2(-i\delta, y^1, \vec{y})\ ,
\end{align}
where, $\epsilon_1$ and $\epsilon_2$ are polarizations of operators $O_1$ and $O_2$, respectively (when they have spins). We then perform the rescaling $\delta=\sigma \delta$, $y^1=\sigma y^1$, and $\vec{y}=\sigma \vec{y}$ and take the limit $\sigma\rightarrow 0$. In this limit, we claim that chaos/causality bounds guarantee  that the OPE of $\Psi^*[O_1]$ and $\Psi[O_2]$ (up to order $1/c_T$) is given by a universal operator $\mathbf{H}$:
\be\label{ope}
\Psi^*[O_1] \Psi[O_2]=\langle \Psi^*[O_1] \Psi[O_2] \rangle +\langle \Psi^*[O_1]  \E_{lc}\, \Psi[O_2]\rangle \mathbf{H} +\cdots\ ,
\ee
where, dots represent terms which are suppressed by either the large gap limit or the large $c_T$ limit or the Regge limit. And $\E_{lc}$ is the lightcone limit of the operator (\ref{hne}):
\be
\E_{lc}\equiv \int du' T_{uu}(u',v=1) \sim \lim_{r\rightarrow 0}\frac{\E_{r}(v=1)}{r^{d-2}} \ .
\ee
This OPE holds if all other operator insertions are finite distance away. In general, $\mathbf{H}$ is a complicated operator which contains the stress tensor and an infinite set of double trace operators. However, the important point is that the same operator $\mathbf{H}$ appears in the OPE of all operators and does not depend on the polarizations. Only the coefficient of $\mathbf{H}$ depends on $O_1$ and $O_2$. This coefficient can be chosen to be the contribution in a regular conformal collider experiment which is determined by the lightcone limit. Also note that when $O_1$ and $O_2$ are different operators the first term in (\ref{ope}) vanishes, however, the second term can still be nonzero.

When $O_1$ and $O_2$ are scalar operators, (\ref{ope}) is a simple consequence of the smeared Regge OPE of \cite{Afkhami-Jeddi:2017rmx}. Moreover, we are also claiming that the OPE (\ref{ope}) holds in the Regge limit even when $O_1$ and $O_2$ are spinning operators. However, for spinning operators, the OPE (\ref{ope}) is true only after we first impose chaos/causality constraints that we obtained from the holographic null energy condition 
\be
\lim_{B\rightarrow \infty}\langle (c_1^*\Psi^*[O_1]+c_2^*\Psi^*[O_2]) \E_{\sqrt{\rho}B}(B) (c_1 \Psi[O_1]+c_2\Psi[O_2])  \rangle\ge 0
\ee
for arbitrary $c_1$ and $c_2$. For scalar operators, the Regge correlator is trivially causal. Since the same operator $\mathbf{H}$ appears in the OPE (\ref{ope}) of all operators, it is obvious that the equation (\ref{ope}) is a sufficient condition that makes all of the Regge correlators causal. In this paper, we will not explicitly prove that (\ref{ope}) is a necessary condition. Rather, in the rest of the paper, we will show that (\ref{ope}) is true for various spinning operators. Note that a hint of this property of the Regge OPE was present even in our previous paper \cite{Afkhami-Jeddi:2017rmx}.

For heavy scalar operators, the smearing integrals in (\ref{ope}) can be ignored because they only produce overall volume factors. Hence, for a heavy scalar $\O_H$, with $1\ll \Delta_{H} \ll \sqrt{c_T}$ the OPE (\ref{ope}) is very simple. Therefore, the Regge OPE of any two smeared primary operators is determined by the OPE of two heavy scalar operators, in particular 
\be\label{opH}
\mathbf{H}=\frac{\O_H(i\delta) \O_H(-i\delta)-\langle \O_H(i\delta) \O_H(-i\delta) \rangle}{\langle \O_H(i\delta) \E_{lc}\, \O_H(-i\delta) \rangle}\ .
\ee

Let us now consider correlator of the holographic null energy operator with two arbitrary smeared operators $\Psi^*[O_1]$ and $\Psi[O_2]$. The equation (\ref{ope}) predicts that after imposing all of the causality conditions the correlator $\langle \Psi^*[O_1] \E_{r}(v) \Psi[O_2]\rangle$ can be written as a product of the lightcone answer and a correlator of the holographic null energy operator with heavy scalars. In particular, if we define
\be\label{def_f}
f_{O_1 O_2}(\rho)\equiv \lim_{B\rightarrow \infty}\langle \Psi^*[O_1] \E_{\sqrt{\rho} B}(B) \Psi[O_2] \rangle 
\ee
then it can be easily shown that equations (\ref{ope}) and (\ref{opH}) imply
\be
f_{O_1 O_2}(\rho)=\frac{f_{O_1 O_2}(\rho\rightarrow 0) f_{\O_H \O_H}(\rho)}{f_{\O_H \O_H}(\rho\rightarrow 0)}+\cdots\ ,
\ee
where, dots represent terms suppressed by $\Deltag$. We can further simplify by computing the scalar part of the above equation, yielding
\be\label{opecheck}
f_{O_1 O_2}(\rho)=\lim_{\rho_0\rightarrow 0} f_{O_1 O_2}(\rho_0) \left(\frac{\rho}{\rho_0}\right) ^{\frac{d-2}{2}} \, _2F_1\left(\frac{d-2}{2},d-1;\frac{d+2}{2};\rho \right)+\cdots\ .
\ee
Broadly speaking, this equation relates UV (Regge limit) with IR (lightcone limit). It is rather remarkable that for holographic CFTs the Regge limit is completely determined by the lightcone limit. In the following sections, we will check the OPE (\ref{ope}) by demonstrating that the above relation holds for various operators with or without spin. 

\subsection{Gravity interpretation} 
The Regge OPE (\ref{ope}) has a nice gravity interpretation. The operator $\mathbf{H}$ is a complicated CFT operator, however, when written in terms of the bulk metric it has a simple expression. In particular, in the gravity language the Regge OPE (\ref{ope}) can be rewritten as\footnote{$h_{z+t\ z+t}$ is defined in the usual way: $h_{z+t\ z+t}=\frac{1}{4}(h_{tt}+2h_{tz}+h_{zz})$.}
\be\label{ope2}
\Psi^*[O_1] \Psi[O_2]=\langle \Psi^*[O_1] \Psi[O_2] \rangle -2 i E_{O_1 O_2} \int_0^\infty t^2 h_{z+t\ z+t}(z=t, t) dt \ ,
\ee
where, $E_{O_1 O_2}$ is the matrix element of the total energy operator $\langle \Psi^*[O_1] \mathbf{E} \Psi[O_2] \rangle$. On the right hand side the operator $\mathbf{H}$ is now written as the bulk metric perturbation integrated over a null geodesic $z=t, y^1=0, \vec{y}=\vec{0}$ in AdS$_{d+1}$. Therefore, $\mathbf{H}$ is a {\it shockwave operator} that creates a spherical shockwave in AdS.

The OPE (\ref{ope2}) has been derived by starting from the planar shockwave operator of \cite{Afkhami-Jeddi:2017rmx}. In the gravity language, the OPE of heavy scalars $\O_H(i\delta) \O_H(-i\delta)$ can be obtained from  the Regge OPE of \cite{Afkhami-Jeddi:2017rmx} by performing the following change of coordinates:
\be
u \rightarrow \frac{z_0^2}{u}\ , \qquad v \rightarrow -v+\frac{\vec{y}^2}{u}+\frac{z^2}{u}\ , \qquad \vec{y}\rightarrow \frac{z_0 \vec{y}}{u}\ , \qquad z\rightarrow \frac{z z_0}{u}\ , 
\ee
where, $z_0$ is the position of the planar shockwave operator in \cite{Afkhami-Jeddi:2017rmx}. On the boundary this change of coordinates acts as a conformal transformation. On the other hand, in the bulk this change of coordinate converts the planar shockwave operator into the spherical shockwave operator. Now the universality of the Regge OPE immediately implies that the same spherical shockwave operator will also appear in (\ref{ope2}). 

It is important to note that it is not surprising that the smeared Regge OPE can be expressed as an integral over a geodesic. After all, this has already been shown in \cite{Afkhami-Jeddi:2017rmx} for light scalar operators. Moreover, our conjecture about double trace contributions  implies the same for any primary single trace operator. However, the non-trivial consequence of the HNEC is the appearance of the same spherical shockwave operator in the OPE (\ref{ope2}) for all single trace operators. This universality of the Regge OPE can be interpreted as the CFT version of the equivalence principle in the bulk.

The form of the OPE (\ref{ope2}) is fixed by the conformal symmetry and causality of the boundary CFT and in the dual gravity language, it has an interesting consequence. First, consider a single light operator $O_1$ with spin $\ell \le 2$. The OPE (\ref{ope2}) implies that one can create a spherical shockwave in the bulk by inserting the smeared operator $\Psi[O_1]$. The resulting shockwave has an energy $\sim E_{O_1 O_1}$ and the bulk metric is identical to the shockwave geometry obtained from an infinitely boosted AdS-Schwarzschild black hole \cite{Horowitz:1999gf}. Furthermore, the form of the OPE (\ref{ope2}) also dictates that this process of creating bulk shockwaves obeys a simple {\it superposition principle}. Consider an operator $\O$ which is a linear combination of several low spin operators 
\be
\O= c_1 O_1+c_2 O_2+c_3 O_3 +\cdots\  .
\ee
The smeared operator $\Psi[\O]$ again creates a spherical shockwave in the bulk but now with an energy $\sim E_{\O \O}$. Therefore, causality of four-point functions of the boundary CFT translates into a shockwave superposition principle in the bulk.


 \section{Nitty-gritty of doing the integrals}\label{section_int}
The aim of the rest of the paper is to derive constraints by evaluating (\ref{snec}) for various spinning operators. So, in this section we present a systematic approach of calculating (\ref{snec}). As an example, we will explicitly show the computation of (\ref{snec}) for scalars which can be easily generalized for spinning operators. Then, we briefly sketch the calculation for the spinning case. This section consists of technical details, so casual readers can skip this section.

 Let us now introduce the notation: 
 \be\label{eo1o2}
 \E_{O_1 O_2}(\rho)\equiv - i \ \lim_{B\rightarrow \infty} \langle\overline{\varepsilon_1.O_1(B)} \,   \E_{r=\sqrt{\rho}B}(0) \, \varepsilon_2.O_2(B) \rangle\ ,
 \ee
 where, $\varepsilon.O(B)$ and $\overline{\varepsilon.O(B)}$ are defined in (\ref{ob}) and (\ref{obbar}) respectively. $ \E_{O_1 O_2}(\rho)$ is a positive function when $O_1$ and $O_2$ are the same operators and $\varepsilon_1=\varepsilon_2$. This positivity is equivalent to the holographic null energy condition (\ref{hnec}): 
 \be
 \E_{O O}(\rho) = \E(\rho) \ge 0\ .
 \ee
The function  $ \E_{O_1 O_2}(\rho)$ is also closely related to $f_{O_1 O_2}(\rho)$ as defined in (\ref{def_f}). However, there is a key difference: $ \E_{O_1 O_2}(\rho)=f_{O_1 O_2}(\rho)$  only after we impose causality constraints on $ \E_{O_1 O_2}(\rho)$.
 
 Let us now evaluate $ \E_{O_1 O_2}(\rho)$: 
 \begin{align}\label{masterint}
\E_{O_1 O_2}(\rho)&=-\frac{i B^{d-2}}{\rho}\lim_{B\rightarrow \infty}\int_{-\infty}^\infty d\tilde{u} \int_{\vec{x}_3^2\leq \rho} d^{d-2}\vec{x}_3\int d\tau d^{d-2}\vec{y}\, (\rho-\vec{x}_3^2)\notag\\
&\hspace{.5in}\times \langle \overline{\varepsilon_1}.O_1(iB,-\delta,0) T_{uu}(\tilde{u},0,iB\vec{x}_3) \varepsilon_2.O_2(i(B+\tau),\delta,\vec{y})\rangle\ ,
 \end{align} 
where, we have rescaled $\vec{x}_3$ to $B \vec{x}_3$ so that the bounds of integration becomes $\vec{x}_3^2\leq \rho$.\footnote{For the sake of clarity let us again note that positions of operators $O_1$ and $O_2$ in (\ref{masterint}) are labelled by $(t,x^1,\vec{x})$. Whereas, position of the stress tensor operator in (\ref{masterint}) is labelled by $(u,v,\vec{x})$.} Note that we are only smearing one of the operators because the other smearing integral will only give an overall volume factor. So, the computation of $\E_{O_1 O_2}(\rho)$ is reduced to performing certain integrals over a CFT 3-point function whose form is fixed by conformal invariance up to constant OPE coefficients.

 \subsubsection*{Order of limits:}
 The expression (\ref{masterint}) is evaluated by first performing the $\tilde{u}$-integral using an appropriate contour. Then we take the $B\rightarrow \infty$ limit, yielding a relatively simple expression. To perform the smearing integrals, it is convenient to package $\tau$ and $\vec{y}$ together in a $(d-1)$-dimensional vector $\vec{\mathbf{k}}$. The resulting expression can be written covariantly by decomposing the $d$-dimensional vectors $x_i$ and polarization vectors $\varepsilon_i$ into scalars and $(d-1)$-vectors under rotations in $(\tau$-$\vec{y})$-space that is $\mathbb{R}^{d-1}$. The smearing integrals can then be performed in a covariant way using familiar techniques used in Feynman diagram computations. Finally we perform the $(d-2)$-dimensional integral over $\vec{x}_3$. Note that we have exchanged the order in which we perform integrations.
 
 The advantage of this method is that the spin and scaling dimension of the external and exchanged operators as well as the space-time dimensions are simply constant parameters in the integrand and the integrals can in principle be performed for arbitrary values resulting in general expressions as functions of these parameters.

\subsection{Scalar operators}
As a demonstration of the formalism in action we will now compute (\ref{masterint}) for scalar operators. The three point function of interest in this case is 
entirely fixed by conformal invariance \cite{Osborn:1993cr}
\be\label{OP}
\langle  \O(x_1)\O(x_2) T_{\mu\nu}(x_3)\rangle=\frac{C_{\O\O T} I_{\mu \nu}}{x_{23}^{d-2} x_{12}^{2\Delta_\O+2-d}x_{13}^{d-2}}\ ,
\ee
where,
\be
x_{IJ}=|x_I-x_J|\ , \qquad I^{\mu\nu}=\left(\frac{x_{13}^\mu}{x_{13}^2}-\frac{x_{23}^\mu}{x_{23}^2}\right)\left(\frac{x_{13}^\nu}{x_{13}^2}-\frac{x_{23}^\nu}{x_{23}^2}\right)-\frac{x_{12}^2}{x_{13}^2x_{23}^2}\frac{\eta^{\mu \nu}}{d}\ .
\ee
The OPE coefficient $C_{\O\O T}$ is fixed by the Ward identity
\be
C_{\O\O T} = - \Delta_\O \frac{ \Gamma(d/2) d}{2\pi^{d/2}(d-1)}\ . 
\ee
We therefore wish to compute 
 \begin{align}\
\E_{\O\O}(\rho)=-\frac{i C_{\O\O T} B^{d-2}}{\rho} \int_{-\infty}^\infty d\tilde{u} \int_{\vec{x}_3^2\leq \rho} d^{d-2}\vec{x}_3\int d\tau d^{d-2}\vec{y} \frac{( \rho -\vec{x}_3^2) I_{uu}}{x_{23}^{d-2} x_{12}^{2\Delta_\O+2-d}x_{13}^{d-2}}\ 
\end{align} 
in the large $B$ limit, where points $x_1$, $x_2$ and $x_3$ are given by (\ref{masterint}). 

\subsubsection*{Performing the $\tilde{u}$-integral:}
In our coordinates, we find that the factors in the denominator have the form
\begin{align}
x_{13}^2= c_1 \tilde{u}+c_2,\qquad x_{23}^2= c_3 \tilde{u}+c_4\ ,
\end{align}
where $c_i$'s are $\tilde{u}$-independent complex constants and the numerator will in general be a finite degree polynomial $P(\tilde{u})$ in $\tilde{u}$. If we perform the $\tilde{u}$-integral with the usual $i\epsilon$-prescription, then the $\tilde{u}$-contour does not enclose any poles (or branch cuts) and the integral vanishes. Instead, we need to follow a prescription similar to the prescription of \cite{Afkhami-Jeddi:2017rmx} to obtain the operator ordering of (\ref{masterint}). Whenever the holographic null energy operator appears inside a correlator, we define the $\tilde{u}$-integral with the $\tilde{u}$-contour such that the $\tilde{u}$-integral in (\ref{masterint}) is determined by the residue at the pole due to  the operator $ O_1$ (in the presence of branch cuts the integral is determined by the integral of the discontinuity across the branch cut due to the operator $O_1$). This contour can be motivated in many different ways. In equation (\ref{masterint}), both the stress tensor and the operator $O_2$ are smeared over some region. To give a physical interpretation of the contour, consider centers of these smeared operators:
 \begin{align}
\int_{-\infty}^\infty d\tilde{u}  \langle \overline{\varepsilon_1}.O_1(iB,-\delta) T_{uu}(\tilde{u},0) \varepsilon_2.O_2(iB,\delta)\rangle\ .
 \end{align}
In general this $\tilde{u}$-integral has branch cut singularities at $u=i B \pm \delta$. And the above contour is equivalent to the prescription of analytic continuation of \cite{Afkhami-Jeddi:2017rmx}. Another way to understand this choice of contour is to perform a $\pi/2$ rotation in the Euclidean $\tau-x^1$ plane and start with (\ref{collider}) instead of (\ref{masterint}). Now if we consider the centers of the smeared operators, the choice of contour for $\tilde{u}$-integral is obvious. 

To summarize, effectively the $\tilde{u}$-integral in (\ref{masterint}) is given by the contour:
\be\label{ucontour}
\begin{gathered}\includegraphics[width=0.46\textwidth]{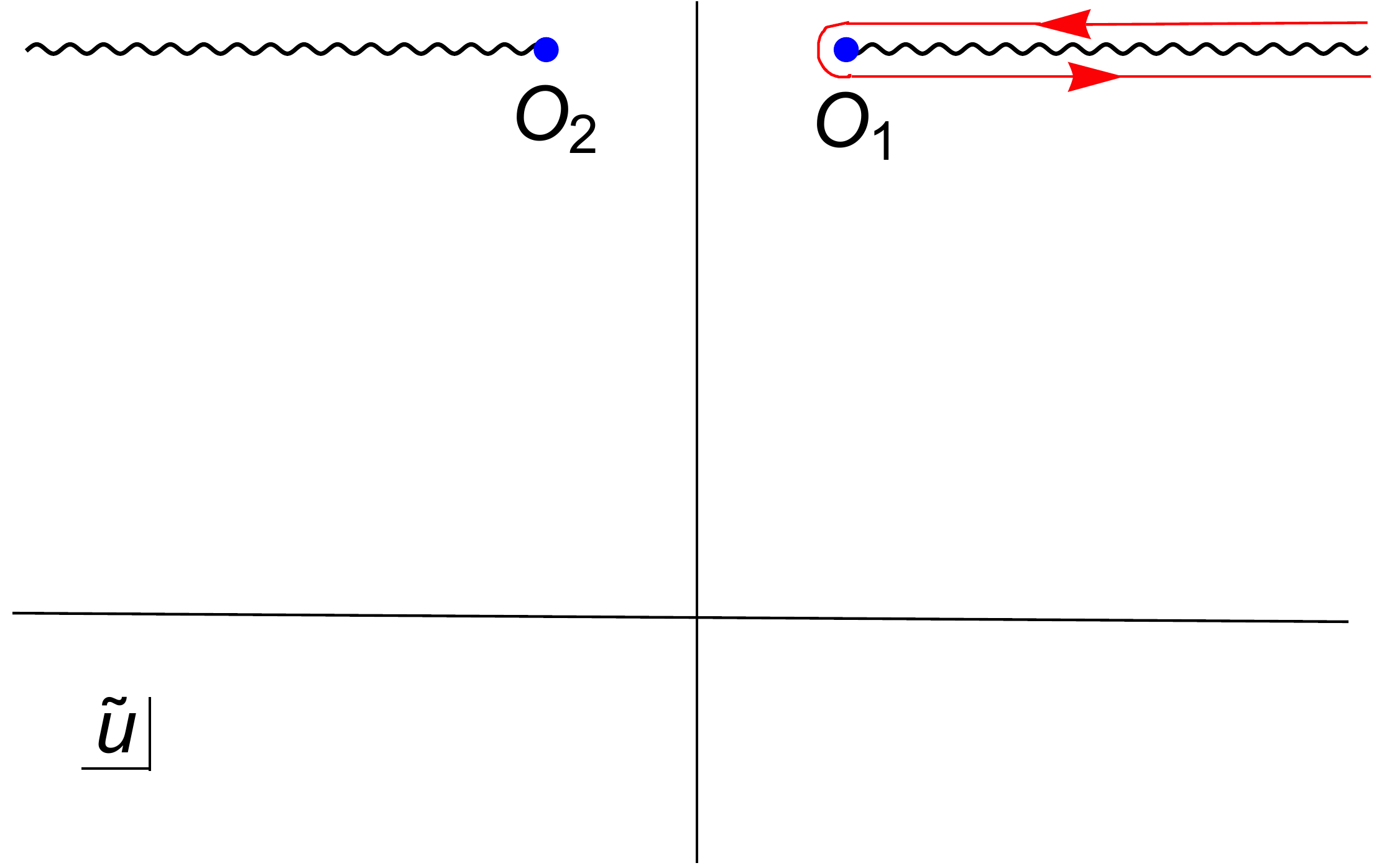}\end{gathered}
\ee

Let us now use this contour to perform integrals of the form:
\begin{align}
\int_{-\infty}^{+\infty} d\tilde{u} \frac{P(\tilde{u})}{(c_1 \tilde{u}+c_2)^{a_1}(c_3 \tilde{u}+c_4)^{a_2}}\equiv \int_\gamma d\tilde{u} \frac{P(\tilde{u})}{(c_1 \tilde{u}+c_2)^{a_1}(c_3 \tilde{u}+c_4)^{a_2}}\ ,
\end{align}
where $P(u)$ is a polynomial in $u$. These integrals can be easily evaluated by  using the identity 
\begin{align}
\int_\gamma d\tilde{u}\frac{1}{(\tilde{u} +c_2)^{p_1}(c_4-\tilde{u} )^{p_2}}= \frac{2\pi i}{(c_4+c_2)^{p_1+p_2-1}}\frac{ \Gamma(p_1+p_2-1)}{\Gamma(p_1)\Gamma(p_2)}\ ,
\end{align}
where, $p_1$ and $p_2$ are positive numbers with $p_1+p_2 >1$. So, now performing the $\tilde{u}$-integral and taking the large-$B$ limit we find, \footnote{Naively it seems that $\pv$ integral is divergent near $\pv\rightarrow 0$. However $\pv$ is a complex valued vector and the integration region is shifted in the imaginary direction. In practice this means that the integration must be performed by analytic continuation using appropriate choice of contours to ensure convergence as described in appendix \ref{smintegrals}.}
\begin{align}\label{eqn123}
\E_{\O\O}(\rho)=
\frac{\pi   2^{2 d-3} \Gamma (d+1)C_{\O\O T}}{\rho  \Gamma \left(\frac{d}{2}+1\right)^2}
\int_{\vec{x}_3^2\leq \rho} d^{d-2}\vec{x}_3\int d^{d-1}\pv \frac{(\Xv\cdot\Xv-\rho)(1-\Xv\cdot \Xv)^{1-d}}{(\pv^2+\pv\cdot\Lv)^{1 - d/2 + \Delta_\O}(\pv\cdot\Lv)^{d-1}},
\end{align}
where we have made a change of variables from $(\tau,\vec{y})$ to $\pv$ and defined the following $(d-1)$-dimensional vectors running over time and $d-2$ transverse coordinates $(\tau, \vec{y})$ 
\begin{align}\label{defp}\
& \kv=(\tau,\vec{y})\ ,\notag\\
& \pv=\kv-\frac{\Lv}{2}\ , \notag\\
& \Lv=\frac{8 \delta  \Xv+4 i \delta  (\Xv\cdot \Xv+1) \Tv}{\Xv\cdot \Xv-1}\ , \notag\\
& \Tv=(1,\vec{0})\ ,\notag\\
& \Xv=(0,\vec{x}_3)\ .
\end{align}
Before we proceed, let us note that if one starts with (\ref{collider}) instead of (\ref{masterint}), the $\tilde{u}$-integral should be performed in a similar way. After taking the large-$B$ limit, one ends up with exactly (\ref{eqn123}) and hence the rest of the calculation is identical.
\subsubsection*{Performing the $\pv$-integral:}

It turns out that even in the most general correlation function, the smearing integrals reduce to the form
\begin{align}\label{pintegral}
\int d^{d-1}\pv \frac{\prod_i\pv.\vec{\mathbf{v}}_i}{(\pv^2+\pv\cdot\Lv)^{p_1}(\pv\cdot\Lv)^{p_2}},
\end{align}
where $\vec{\mathbf{v}}_i$ are constant vectors. These integrals have closed form expressions in the most general case and the relevant results are summarized in appendix \ref{smintegrals}. In this example, performing the smearing integrals yields\footnote{From now on we set $\delta=1$ for simplicity. In the final expression one can restore $\delta$ back by dimensional analysis.}
\begin{align}
\E_{\O\O}(\rho)=
\frac{\pi ^{d/2} \Gamma \left(\frac{d+1}{2}\right) 2^{2 \left(d-\Delta _{\mathcal{O}}-\frac{3}{2}\right)} \Gamma \left(-\frac{d}{2}+\Delta _{\mathcal{O}}+\frac{3}{2}\right)C_{\O\O T}}{\rho  \Gamma \left(\frac{d}{2}+1\right) \Gamma \left(\Delta _{\mathcal{O}}+1\right)}
\int_{\vec{x}_3^2\leq \rho} d^{d-2}\vec{x}_3\frac{\vec{x}_3\cdot \vec{x}_3-\rho}{(1-\vec{x}_3\cdot \vec{x}_3)^{d-1}}.
\end{align}

\subsubsection*{Performing the $\vec{x}_3$ integral:}

The most general integrals of the kind that appeared in our last expression, after going to the radial coordinate, can be done using
\begin{align}
\int_0^{\sqrt{\rho}} dx \left(1-x^2\right)^a x^b\left(x^2-\rho\right)^c=\frac{\rho ^{\frac{b+1}{2}} \Gamma \left(\frac{b+1}{2}\right) (-\rho )^c \Gamma (c+1) \, _2F_1\left(-a,\frac{b+1}{2};\frac{b+3}{2}+c;\rho \right)}{2 \Gamma \left(\frac{b+3}{2}+c\right)}\ ,
\end{align}
where, $b,c >-1$ and $0<\rho<1$. Using this identity we finally obtain
\begin{align}
\E_{\O\O}(\rho)=
-\frac{ \pi ^{d-1} \rho ^{\frac{d}{2}-1} C_{\O\O T}\Gamma \left(\frac{d+1}{2}\right) \, _2F_1\left(\frac{d}{2}-1,d-1;\frac{d}{2}+1;\rho \right) 4^{d-\Delta _{\mathcal{O}}-\frac{3}{2}} \Gamma \left(-\frac{d}{2}+\Delta _{\mathcal{O}}+\frac{3}{2}\right)}{\Gamma \left(\frac{d}{2}+1\right)^2 \Gamma \left(\Delta _{\mathcal{O}}+1\right)}.
\end{align}
For scalars, the causality condition $\E_{\O\O}(\rho) \ge 0$ is already satisfied because of the Ward identity. Note that $\E_{\O\O}(\rho)$ satisfies the relation (\ref{opecheck}) which is the first check of the UV/IR connection.\footnote{Let us recall that for scalars $\E_{\O\O}(\rho)=f_{\O\O}(\rho)$.} As described in the previous section the lightcone limit is obtained by taking $\rho\rightarrow 0$:
\begin{align}
\E_{\O\O}(\rho)=-\frac{\pi ^{d-1} \rho ^{\frac{d}{2}-1} \Gamma \left(\frac{d+1}{2}\right) 4^{d-\Delta _{\mathcal{O}}-\frac{3}{2}}C_{\O\O T} \Gamma \left(-\frac{d}{2}+\Delta _{\mathcal{O}}+\frac{3}{2}\right)}{\Gamma \left(\frac{d}{2}+1\right)^2 \Gamma \left(\Delta _{\mathcal{O}}+1\right)}+\O(\rho^{d/2}).
\end{align}

The ``bulk-point" limit\footnote{This bulk point limit is similar to the bulk point limit studied in \cite{Maldacena:2015iua}, however, it is not exactly the same. Our bulk point limit is in fact the limit discussed in \cite{Afkhami-Jeddi:2016ntf} in the context of causality.} is obtained by taking the limit $\rho\rightarrow 1$ and in $d\ge 4$, we obtain:
\begin{align}
\E_{\O\O}(\rho)=-\frac{ d \pi ^{d-1} \Gamma \left(\frac{d+1}{2}\right) 4^{d-\Delta _{\mathcal{O}}-\frac{5}{2}}C_{\O\O T} \Gamma \left(-\frac{d}{2}+\Delta _{\mathcal{O}}+\frac{3}{2}\right)}{(d-3) \Gamma \left(\frac{d}{2}+1\right)^2 \Gamma \left(\Delta _{\mathcal{O}}+1\right) (1-\rho )^{d-3}} +\O(1-\rho)^{4-d}.
\end{align}
In $d=3$, there is a logarithmic divergence in the limit $\rho\rightarrow 1$
\be
\E_{\O\O}(\rho)=\frac{4^{\frac{5}{2}-\Delta_\O}\pi C_{\O\O T} }{3 \Delta_\O }\ln (1-\rho)+\O(1)\ .
\ee

\subsection{Spinning operators}
It was shown in \cite{Osborn:1993cr,Costa:2011mg} that the most general 3-point functions of symmetric traceless spinning operators in a CFT can be written as a sum over certain elementary spinning structures:
\begin{align}\label{general3pt}
 \langle \Phi_1\Phi_2\Phi_3\rangle=\sum_{\{n_{23},n_{13},n_{12}\}}C^{\Phi_1\Phi_2\Phi_3}_{n_{23},n_{13},n_{12}}\frac{V_{1,23}^{\ell_1-n_{12}-n_{13}}V_{2,13}^{\ell_2-n_{12}-n_{23}}V_{3,12}^{\ell_3-n_{13}-n_{23}}H_{12}^{n_{12}}H_{13}^{n_{13}}H_{23}^{n_{23}}}{(x_{12}^2)^{\frac{1}{2}(h_1+h_2-h_3)}(x_{13}^2)^{\frac{1}{2}(h_1+h_3-h_2)}(x_{23}^2)^{\frac{1}{2}(h_2+h_3-h_1)}},
 \end{align} 
where $C^{\Phi_1\Phi_2\Phi_3}_{n_{23},n_{13},n_{12}}$ are constant coefficients and $h_i \equiv \Delta_i+\ell_i $. The structures are given by
 \begin{align}\label{HV}
 H_{ij}&\equiv x_{ij}^2\e_i\cdot\e_j- 2 (x_{ij}\cdot\e_i)(x_{ij}\cdot\e_j),\qquad V_{i,jk}\equiv \frac{x_{ij}^2 x_{ik}\cdot \e_i-x_{ik}^2 x_{ij}\cdot \e_i}{x_{jk}^2}\ ,
 \end{align}
 where, $x_{ij}^\mu=(x_i-x_j)^\mu$ and $\e_i$ is a null polarization vector contracted with spinning indices of $\Phi_i$ in the following way:
 \be
 \left(\e^{\mu}\e^\nu \cdots\right)\Phi_{\mu \nu \cdots}\equiv \e.\Phi\ .
 \ee 
 For a traceless symmetric tensor, one can easily convert the null polarization $\e^{\mu}\e^\nu \cdots$ into an arbitrary polarization tensor $\e^{\mu \nu \cdots}$ by using projection operators \cite{Costa:2011mg}.

The sum in (\ref{general3pt}) is over all triplets of non-negative integers $\{n_{12},n_{13},n_{23}\}$ satisfying
 \begin{align}
	\ell_1-n_{12}-n_{13} \geq 0\ , \qquad 	\ell_2-n_{12}-n_{23} \geq 0\ , \qquad 	\ell_3-n_{13}-n_{23} \geq 0\ .
 \end{align}
For a general correlation function, the coefficients $C^{\Phi_1\Phi_2\Phi_3}_{n_{23},n_{13},n_{12}}$ are all independent parameters, however imposing conservation equations or Ward identities will impose relations amongst these coefficients.

From equation (\ref{general3pt}), we see that the most general integrals in $\E_{O_1 O_2}(\rho)$ are of the form
 \begin{align}
 \int d\tilde{u}  \int_{\vec{x}_3^2 \le \rho}d^{d-2}\vec{x}_3\int d\tau d^{d-2}\vec{y}\frac{( \rho -\vec{x}_3^2) V_{1,23}^{a_1}V_{2,13}^{a_2}V_{3,12}^{a_3}H_{12}^{b_1}H_{13}^{b_2}H_{23}^{b_3}}{ (x_{12}^2)^{\frac{1}{2}(h_1+h_2-h_3)}(x_{13}^2)^{\frac{1}{2}(h_1+h_3-h_2)}(x_{23}^2)^{\frac{1}{2}(h_2+h_3-h_1)}}\ ,
 \end{align}
where the exponents in the numerator, $a_i$ and $b_i$, are positive integers. Polarizations are given by (in $d\ge 4$)\footnote{We will treat the $d=3$ case separately.}
\begin{align}\label{polarization}
\e_1^\mu=(1,\xi_1,\vec{\e}_{1,\perp})\ ,\qquad \e_2^\mu=(1,\xi_2,\vec{\e}_{2,\perp})\ , \qquad \e_3^\mu=\frac{1}{2}(1,-1,\vec{0})\ ,
\end{align}
with $\xi_{1,2}=\pm 1$ and $\vec{\e}_{1,\perp}{}^2=\vec{\e}_{2,\perp}{}^2=0$.

\subsubsection*{Angular integrals:}
In the case where the external operators are non-scalars, similar to (\ref{defp}) we also need to introduce $(d-1)$-dimensional vectors made out of the polarization vectors $\e_1^\mu$, $\e_2^\mu$:
\begin{align}
\vec{\boldsymbol{\epsilon}}_{1,\perp}=(0,\vec{\e}_{1,\perp}),\qquad \vec{\boldsymbol{\epsilon}}_{2,\perp}=(0,\vec{\e}_{2,\perp}).
\end{align}
Now after $\pv$-integrals, we will have to perform angular integrals for $\vec{x}_3$ which is of the form 
\begin{align}
\int_{\mathbb{S}^{d-3}} d\hat{\Omega} (\vec{\e}_{1,\perp}\cdot \vec{x}_3)^n(\vec{\e}_{2,\perp}\cdot \vec{x}_3)^m=\frac{\pi ^{\frac{d-2}{2}} 2^{1-n} |\vec{x}_3|^{2 n} \Gamma (n+1)  (\vec{\e}_{1,\perp}\cdot \vec{\e}_{2,\perp})^n\delta_{m,n}}{\Gamma \left(\frac{d-2}{2}+n\right)},
\end{align}
where $d\hat{\Omega}$ is the standard measure on $\mathbb{S}^{d-3}$ and we have used the fact that $\vec{\e}_{2,\perp}{}^2=\vec{\e}_{1,\perp}{}^2=0$.  Rest of the computation is identical to the scalar case and can be efficiently automated in Mathematica.

\section{Bounds on $\langle TTT\rangle$, $\langle JJT\rangle$, and $\langle \O_{\ell=1,2} \O_{\ell=1,2} T\rangle$}\label{section_hd}
In this section, we will use the methods described above to derive constraints in $d\ge 4$. These constraints encompasses, and generalizes,  the constraints obtained in \cite{Afkhami-Jeddi:2016ntf,Kulaxizi:2017ixa,Costa:2017twz,Afkhami-Jeddi:2017rmx,Meltzer:2017rtf} by studying various four-point functions in holographic CFTs. Note that the $d=3$ case is more subtle which we will discuss in a separate section. 
\subsection{$\langle JJT \rangle$}
We start with $\E_{JJ}$ where $J$ is a spin-1 conserved current. The $\langle JJT \rangle$ three-point function is given in Appendix \ref{jjt}. Following our formalism, the leading term in the limit $\rho\rightarrow 1$ is given by 
\begin{align}\label{jjt1}
&\E_{JJ}(\rho)\sim \frac{- 2^{-d-2} \pi ^{d-\frac{1}{2}} \left(d-2 \left(\lambda ^2+1\right)\right) \Gamma \left(\frac{d-1}{2}\right) (4 n_f-n_s)}{\Gamma \left((\frac{d}{2}+1\right)^2 \Gamma \left(\frac{d}{2} \right) (1-\rho )^{d-1}}+\O\left(\frac{1}{(1-\rho)^{2-d}}\right) 
\end{align}
up to some positive overall coefficient. Our choice of polarizations is given in equation (\ref{polarization}) with $\e_2^\mu =\overline{\e}_1^\mu$ and
\be\label{jjt2}
\lambda^2=\frac{1}{2}\overline{\vec{\e}}_{2,\perp}\cdot \vec{\e}_{2,\perp}\ge 0\ .
\ee
 As shown in the Appendix \ref{sec:pol}, given our choice of polarization, different powers of $\lambda^2$ correspond to independent spinning structures and decomposition of $SO(d-1,1)$ to representations under $SO(d-2)$. Therefore positivity of $\E_{JJ}$ implies that coefficients of each powers of $\lambda^2$ must be positive. Hence, from equation (\ref{jjt1}) we obtain
\begin{align}\label{jjt3}
n_s=4 n_f + \O\left(\frac{c_J}{\Deltag^2} \right)=\frac{d(d-2)}{S_d (d-1)}c_J + \O\left(\frac{c_J}{\Deltag^2} \right)\ ,
\end{align}
where, in the last equation we have used the Ward identity (\ref{cj}). The $\Deltag$ correction in the above equation is computed following (\ref{gap}). All subleading contributions to (\ref{jjt1}) are proportional to $n_s\left(2 \lambda ^2+1\right)$, a manifestly positive quantity. Therefore, subleading terms of $\E_{JJ}(\rho)$ do not lead to new constraints. Furthermore, it is obvious from (\ref{jjt3}) that the three-point function $\langle JJT\rangle$ is completely determined by the $\langle JJ \rangle$ two-point function. In fact, this  is a general feature of CFTs with a large central charge and a large gap.

After imposing the constraint (\ref{jjt2}), we can compute $f_{JJ}(\rho)$: 
\begin{align}\label{f_JJ}
f_{\e_1\cdot J\, \e_2\cdot J}(\rho)=\frac{ 2^{-d} \pi ^{d-\frac{1}{2}} \Gamma \left(\frac{d+1}{2}\right)}{  \Gamma \left(\frac{d}{2}+1\right)^2 \Gamma \left(\frac{d}{2}\right)} n_s \left(1+ \vec{\e}_{1,\perp}\cdot \vec{\e}_{2,\perp}\right) \rho ^{\frac{d}{2}-1} \, _2F_1\left(\frac{d}{2}-1,d-1;\frac{d}{2}+1;\rho \right)
\end{align}
which is consistent with the equation (\ref{opecheck}).

In dual gravity language, the three-point function $\langle JJT\rangle$ arises from the following action of a massless gauge field
\ba
  \int d^{d+1}x \sqrt{-g} \;  \left[-F_{\mu \nu} F^{\mu \nu} + \alpha_{AAh}  W_{\mu \nu \alpha \beta} F^{\mu \nu} F^{\alpha \beta}\right], 
\ea
where, $W$ is the Weyl tensor\footnote{The Weyl tensor is given by $W_{\mu \nu \rho \sigma} = R_{\mu \nu \rho \sigma} - \frac{1}{D-2} \left(g_{\mu [\rho} R_{\sigma] \nu}- g_{\nu [\rho}R_{\sigma] \mu} \right)+\frac{1}{(D-1) (D-2)} R g_{\mu [\rho} g_{\sigma ]\nu}$, where $D=d+1$.}. The coefficient $\alpha_{AAh}$ can be written in terms of $n_s$ and $n_f$:
\be
\alpha_{AAh}\sim \frac{n_s-4n_f}{n_s+4(d-2)n_f}\sim \frac{1}{\Deltag^2}\ .
\ee
Hence, $\alpha_{AAh}$ should be suppressed by the scale of new physics. The power dependence of the suppression $\alpha_{AAh} \sim\frac{1}{\Delta^2_{\text{gap}}}$ agrees with the result obtained from causality of the effective field theory in the bulk \cite{Camanho:2014apa}.\footnote{Here we are assuming $R_{AdS}=1$.}

\subsection{$\langle TTT \rangle$}
Let us now consider $\E_{TT}(\rho)$ where $\langle TTT \rangle$ three-point function is given in Appendix \ref{ttt}. Following our formalism, the leading term in the limit $\rho\rightarrow 1$ is given by 
\begin{align}\label{ttt1l}
\E_{TT}(\rho)\sim &\frac{(-1)^d 4^{1-d} \pi ^d \Gamma (d) \left(-8 d \lambda ^2+(d-2) d+8 \lambda ^4\right)}{(d-2)   \Gamma \left(\frac{d}{2}+1\right)^2 \Gamma \left(\frac{d}{2}+2\right) \Gamma \left(\frac{d}{2}\right)(1-\rho)^{d+1}}   \notag\\
&\times   \left((d-2) d^2 (4 \tilde{n}_f-\tilde{n}_s)-64 (d-3) \tilde{n}_v\right)+\O\left(\frac{1}{(1-\rho)^{d}} \right)
\end{align}
up to some overall positive coefficient. Polarizations are given by equation (\ref{polarization}) with $\e_2^\mu =\overline{\e}_1^\mu$  and $\lambda$ is defined in equation (\ref{jjt2}). Positivity of $\E_{TT}$ for all powers of $\lambda$ demands that we must have
\begin{align}
\tilde{n}_v&=\frac{(d-2) d^2 (4 \tilde{n}_f-\tilde{n}_s)}{64 (d-3)}+\O\left(\frac{c_T}{\Deltag^4} \right).
\end{align}
After imposing this condition, the next leading term becomes
\begin{align}
\E_{TT}(\rho)\sim& \frac{(-1)^{d-1} 2^{1-d} (d+1) \pi ^{d-\frac{1}{2}}  \Gamma \left(\frac{d-3}{2}\right)}{(d-1)   \Gamma \left(\frac{d}{2}+1\right) \Gamma \left(\frac{d}{2}\right)^2 (1-\rho)^{d-1}}\left(d^2-4 (d-1) \lambda ^4+2 (d-3) (d-1) \lambda ^2-5 d+6\right)\notag\\
&~~~~~~~~~\times(2 (d-1) \tilde{n}_f-(d-2) \tilde{n}_s)+\O\left(\frac{1}{(1-\rho)^{d}} \right).
\end{align}
Positivity then implies
\begin{align}\label{ttt2} 
\tilde{n}_f&=\frac{(d-2) \tilde{n}_s}{2 (d-1)} +\O\left(\frac{c_T}{\Deltag^2} \right)\ ,\notag\\
\tilde{n}_v&=\frac{(d-2) d^2 \tilde{n}_s}{64 (d-1)}+\O\left(\frac{c_T}{\Deltag^2} \right)\ ,\notag\\
\tilde{n}_s&=\frac{c_T (d-1)}{32 (d-2) (d+1) S_d}+\O\left(\frac{c_T}{\Deltag^2} \right)\ ,
\end{align}
where, we have also used the Ward identity (\ref{ct}) to derive the last equation. After imposing these constraints, the positivity of $\tilde{n}_s$ guarantees that the rest of the terms are always positive and hence no new constraints are obtained from subleading terms. Note that the three-point function $\langle TTT \rangle$ is completely determined by the $\langle TT\rangle$ two-point function. Furthermore, we can now compute our $f_{\e_1\cdot T\, \e_2\cdot T}(\rho)$ function 
\begin{align}\label{f_TT}
f_{\e_1\cdot T\, \e_2\cdot T}(\rho)&=\frac{\left( (d-1) (\vec{\e}_{1,\perp}\cdot \vec{\e}_{2,\perp})^2+ 2 (d-1) \vec{\e}_{1,\perp}\cdot \vec{\e}_{2,\perp}+d-2\right)}{ (d-1)^2 \Gamma \left(\frac{d}{2}-1\right) \Gamma \left(\frac{d}{2}+1\right) \Gamma \left(\frac{d}{2}\right)}    & \nonumber\\
&\times\ \tilde{n}_s\pi ^{d-1/2} 2^{5-d} \Gamma \left(\frac{d+3}{2}\right)\rho ^{\frac{d}{2}-1}  {}_2F_1\left(\frac{d}{2}-1,d-1;\frac{d}{2}+1;\rho \right)&
\end{align}
which is in agreement with the relation (\ref{opecheck}) indicating that the Regge OPE of smeared operators is indeed universal.

On the gravity side, this constrains higher derivative correction terms in the pure gravity action that contribute to three point interactions of gravitons. These higher derivative correction terms can be parametrized as 
\ba
S =M_{pl}^{d-1}\int d^{d+1}x \sqrt{g} \;\left[ R-2\Lambda + \alpha_2 W_{\mu \nu \alpha \beta} W^{\mu \nu \alpha \beta} +\alpha_4 W_{\mu \nu \alpha \beta} W^{\mu \nu \rho \sigma }{W_{\rho \sigma}}^{\alpha \beta}  \right].
\ea
Note that in case of vacuum AdS, Weyl tensor  vanishes. Other terms are encoding the most general form of three-point interaction for gravitons. Coupling constants $\alpha_2$ and $\alpha_4$ are related to the coefficients $\tilde{n}_s$, $\tilde{n}_f$ and $\tilde{n}_v$. In particular, constraints (\ref{ttt2}) translate into 
 $\alpha_2 \lesssim \frac{1}{\Delta_{\text{gap}}^2},\alpha_4 	\lesssim \frac{1}{\Delta_{\text{gap}}^4} $ (for $d\ge 4$) which is in agreement with the expectation from bulk causality \cite{Camanho:2014apa}.

  \subsubsection*{Conformal trace anomaly in 6d}
 In $d=4$, the causality constraints (\ref{ttt2}) can be rewritten as a statement about central charges: $\frac{|a-c|}{c} \lesssim 1/\Deltag^2$. There is a similar relation between trace anomaly coefficients in $d=6$. In particular, the conformal trace anomaly in $d=6$ can be written as \cite{Bonora:1985cq, Deser:1993yx, Bastianelli:2000hi,Boulanger:2007ab}
 \be\label{anomaly}
 \langle T^\mu_\mu\rangle = 2a_6 E_6+ c_1 I_1+c_2 I_2+c_3 I_3\ 
 \ee
 up to total derivative terms which can be removed by adding finite and covariant counter-terms in the effective action. In equation (\ref{anomaly}), $a_6, c_1, c_2, c_3$ are 6d central charges and 
 \begin{align}
 I_1=& W_{\gamma \alpha \beta \delta}W^{\alpha \mu \nu \beta}W_{\mu~~\nu}^{~\gamma \delta}\ , \notag\\
 I_2=& W_{\alpha \beta}^{~~\gamma \delta}W_{\gamma \delta}^{~~\mu \nu}W_{\mu \nu}^{~~\alpha \beta}\ , \notag\\
 I_3=& W_{\alpha \gamma \delta \mu}\left(\nabla^2 \delta^\alpha_\beta+4 R^\alpha_\beta-\frac{6}{5}R \delta^\alpha_\beta \right)W^{\beta \gamma \delta \mu}\ , \notag\\
 E_6=& \frac{1}{384(2\pi)^3}\delta^{\mu_1 \mu_2 \mu_3 \mu_4 \mu_5 \mu_6}_{\nu_1 \nu_2 \nu_3 \nu_4 \nu_5 \nu_6}R^{\nu_1\nu_2}_{~~~~\mu_1 \mu_2}R^{\nu_3\nu_4}_{~~~~\mu_3 \mu_4}R^{\nu_5\nu_6}_{~~~~\mu_5 \mu_6}\ .
 \end{align}
 The $a_6$ coefficient can be determined only from the stress tensor four-point function and hence (\ref{ttt2}) does not constrain $a_6$. However, $c_1, c_2, c_3$ are related to the stress tensor three-point function and hence constraints (\ref{ttt2}) can be translated into constraints on central charges. In particular, using the result of \cite{Hung:2011xb} for Einstein gravity, we can easily show that 
 \begin{align}
 \frac{c_1}{c_3}=-12+\O\left(\frac{1}{\Deltag^2} \right)\ , \qquad 
 \frac{c_2}{c_3}=-3+\O\left(\frac{1}{\Deltag^2} \right)\ .
 \end{align}
Note that the relations between $c_1, c_2, c_3$ are exactly what is expected for $(2,0)$ supersymmetric theories. For these theories, invariants $I_1,I_2,I_3$  can be combined into  a single super-invariant \cite{Beccaria:2015uta,Butter:2016qkx,Butter:2017jqu} which leads to the relation: $c_1=4c_2=-12c_3$ \cite{Beccaria:2017dmw}. This relation between $c_1, c_2, c_3$ was first derived in \cite{ Bastianelli:2000hi} for the free $(2, 0)$  tensor multiplet. On the other hand, the same relation also holds for strongly coupled theories with a supergravity dual \cite{Henningson:1998gx}.

 \subsection{$\langle \O_{\ell=1}\O_{\ell=1} T  \rangle$}
Now we derive bounds on non-conserved spin-1 operators. The three point function $\langle \O_{\ell=1}\O_{\ell=1} T  \rangle$ has five OPE constants $\{C_{0,0,0},C_{0,0,1},C_{0,1,0},C_{1,0,0},C_{1,1,0}\}$. Imposing permutation symmetry and conservation equation reduces this number to three independent coefficients. The leading contribution in the limit $\rho\rightarrow 1$ is 
\begin{align}
\E_{\O\O}(\rho)&\sim -\frac{\pi ^{d-1} (1-\rho )^{1-d} (d-\vec{\e}_\perp\cdot \bar{\vec{\e}}_\perp-2) \Gamma \left(\frac{d+1}{2}\right) 2^{2 d-2 \Delta _{\mathcal{O}}-3} \Gamma \left(-\frac{d}{2}+\Delta _{\mathcal{O}}+\frac{3}{2}\right)}{(d-1) d \sigma  \Gamma \left(\frac{d}{2}\right)^2 \left(d-2 \left(\Delta _{\mathcal{O}}+1\right)\right) \Gamma \left(\Delta _{\mathcal{O}}+2\right)}\notag\\
&\times\left(C_{1,1,0} \left(d^2+d \left(2 \Delta _{\mathcal{O}} \left(\Delta _{\mathcal{O}}+1\right)-1\right)-2 \left(\Delta _{\mathcal{O}} \left(\Delta _{\mathcal{O}}+3\right)+1\right)\right)\right.\notag\\
&\left.-2 (d-1) C_{0,0,1}+C_{0,1,0} \left(4 \Delta _{\mathcal{O}}+2\right)\right) \ ,
\end{align}
where we have used the polarizations $\e=(1,\xi,\vec{\e}_\perp)$ and $\bar{\e}=(-1,-\xi,\bar{\vec{\e}}_\perp)$ with $\xi=\pm 1$. Imposing positivity on the coefficients of powers of $\vec{\e}_\perp\cdot \bar{\vec{\e}}_\perp$ we find
\begin{align}
C_{1,1,0}=\frac{2 (d-1) C_{0,0,1}-2 C_{0,1,0} \left(2 \Delta _O+1\right)}{d^2+d \left(2 \Delta _O \left(\Delta _O+1\right)-1\right)-2 \left(\Delta _O \left(\Delta _O+3\right)+1\right)}+\O\left(\frac{1}{\Deltag^2} \right)\ .
\end{align}
After imposing this condition the next leading term is 
\begin{align}
\E_{\O\O}(\rho)\sim & \frac{ \pi ^{d-1} 4^{d-\Delta _O-1} \xi\left(2 C_{0,0,1} \left((d-1) \Delta _O-1\right)+C_{0,1,0} \left(d-2 \Delta _O^2\right)\right)}{d   \Gamma \left(\frac{d}{2}\right)^2 \left(d^2+d \left(2 \Delta _O \left(\Delta _O+1\right)-1\right)-2 \left(\Delta _O \left(\Delta _O+3\right)+1\right)\right) \Gamma \left(\Delta _O+1\right)}\notag\\
&\times  \frac{1}{(1-\rho )^{d-2}} \Gamma \left(\frac{d+1}{2}\right)  \Gamma \left(-\frac{d}{2}+\Delta _O+\frac{3}{2}\right) \ .
\end{align}
As described previously, the above expression must be positive for $\xi=\pm 1$ resulting in
\begin{align}
C_{0,1,0}=\frac{2 C_{0,0,1} \left(d \Delta _O-\Delta _O-1\right)}{2 \Delta _O^2-d}+\O\left(\frac{1}{\Deltag} \right)\ .
\end{align}
After imposing the condition, the resulting expression  
\begin{align}
&-\frac{ d  \pi ^{d-1} C_{0,0,1}  \Gamma \left(\frac{d+1}{2}\right) 2^{2 d-2 \Delta _O-5} \Gamma \left(-\frac{d}{2}+\Delta _O+\frac{3}{2}\right) \left(d-\Delta _O (\vec{\e}_\perp\cdot \bar{\vec{\e}}_\perp+2)+\vec{\e}_\perp\cdot \bar{\vec{\e}}_\perp\right)}{(1-\rho)^{d-3}(d-3)   \Gamma \left(\frac{d}{2}+1\right)^2 \left(d-2 \Delta _O^2\right) \Gamma \left(\Delta _O\right)}\notag\\
&~~~~~~~~~~~~~~~~~~~~~~~~~~~~~~~~~~~~~~~~~~~~~~~~~~~~~~~~~~~~~~+\O\left(\frac{1}{(1-\rho)^{d-4}}\right)
\end{align}
has only one independent coefficient $C_{0,0,1}$ and is positive if and only if $C_{0,0,1}<0$. 

Finally, imposing causality constraints and conservation equation result in the following relations
\begin{align}
&C_{0,0,1}=\frac{C_{0,0,0} \left(d-2 \Delta _O^2\right)}{d^2-4 d \Delta _O+4 \Delta _O}+\O\left(\frac{1}{\Deltag} \right)\ ,\notag\\ 
&C_{0,1,0}=C_{1,0,0}=-\frac{2 C_{0,0,0} \left(d \Delta _O-\Delta _O-1\right)}{d^2-4 d \Delta _O+4 \Delta _O}+\O\left(\frac{1}{\Deltag} \right)\ ,\notag\\ 
&C_{1,1,0}=\frac{2 C_{0,0,0}}{d^2-4 d \Delta _O+4 \Delta _O}+\O\left(\frac{1}{\Deltag} \right)
\end{align}
and hence there is only one independent coefficient which is related to the two-point function $\langle \O_{\ell=1}\O_{\ell=1}  \rangle$ by the Ward identity. 
Similarly, we can show that after imposing the causality constraints 
\begin{align}
f_{\e_1\cdot \O\, \e_2\cdot \O}(\rho)&=
-\frac{ \Gamma \left(\frac{d+1}{2}\right) \, 2^{2 d-2 \Delta_\O-2} C _{0,0,1} \Gamma \left(-\frac{d}{2}+\Delta_\O+\frac{3}{2}\right) \left(-d+2 \Delta_\O+\left(\Delta_\O-1\right) \vec{\e}_{1,\perp}\cdot \vec{\e}_{2,\perp}\right)}{  \Gamma \left(\frac{d}{2}+1\right)^2 \left(2 \Delta_\O^2-d\right) \Gamma \left(\Delta_\O\right)}\notag\\
&\times  \pi ^{d-1} \rho ^{\frac{d}{2}-1} {}_2F_1\left(\frac{d}{2}-1,d-1;\frac{d}{2}+1;\rho \right)
\end{align}
which is consistent with the equation (\ref{opecheck}).

In the gravity side, the causality constraints imply that the action for a massive spin-1 field in the bulk must have the form
\be
 \int d^{d+1}x \sqrt{-g} \;  \left[-F_{\mu \nu} F^{\mu \nu} + m^2 A_\mu A^\mu+\cdots\right]\ ,
\ee
where dots represent higher derivative terms (for example  $W_{\mu \nu \alpha \beta} F^{\mu \nu} F^{\alpha \beta}$, $A_{\mu} A_\nu R^{\mu \nu} $) which must be suppressed by scale of new physics in the gravity side.

\subsection{$\langle \O_{\ell=2}\O_{\ell=2}T \rangle$}
Similarly, we can find bounds for non-conserved spin-2 operator $\O_{\ell=2}$. For simplicity we quote the results in 4 dimensions but it can be easily generalized in general $d$. We assume that the three-point function $\langle \O_{\ell=2}TT\rangle$ vanishes so that it does not appear as an exchange operator in the Regge four-point function. But the three-point function $\langle \O_{\ell=2}\O_{\ell=2}T \rangle$ is non-vanishing and to begin with it has 11 coupling constants. Permutation symmetry and conservation equation ensure that only 6 of these coefficients are independent. Furthermore, causality demands that only one of these coefficient can be independent. In particular, the leading contribution in the limit $\rho \rightarrow 1$ is given by 
\begin{align}
&\E_{\O\O}(\rho)\sim
-\frac{ \pi ^4 2^{3-4 \Delta_\O} \left(\Delta_\O+1\right) \left(\Delta_\O+2\right) \Gamma \left(2 \Delta_\O-2\right) ((\vec{\e}_\perp\cdot \bar{\vec{\e}}_\perp-8) \vec{\e}_\perp\cdot \bar{\vec{\e}}_\perp+4)}{ (1-\rho)^5\Gamma \left(\Delta_\O+3\right){}^2}\notag\\
&\times\left( 2 C_{0,1,0} \left(2 \Delta_\O+3\right) \left(\Delta_\O \left(\Delta_\O+3\right)+6\right)-24 C_{0,1,1} \left(2 \Delta_\O+3\right)+72 C_{0,0,2}\right.\notag\\
&+36 \left(2 C_{0,2,0}+C_{1,1,0}-6 C_{1,1,1}\right)+\Delta_\O\left(-2 C_{0,2,0} \left(\Delta_\O+1\right) \left(\Delta_\O \left(3 \Delta_\O+19\right)+18\right)\right.\notag\\
&\left.+\left.C_{1,1,0} \left(\Delta_\O \left(\Delta_\O \left(3 \Delta_\O+14\right)+43\right)+60\right)-24 C_{1,1,1} \left(3 \Delta_\O+7\right)\right)\right)\ .
\end{align}

Following the same procedure as for spin 1 and including conservation conditions we find 
\begin{align}
&C_{0,0,1}=\frac{C_{0,0,0} \left(3 \Delta_\O^3-2 \Delta_\O^2-15 \Delta_\O+18\right)}{4 \left(3 \Delta_\O^2-9 \Delta_\O+7\right)}+\O\left(\frac{1}{\Deltag} \right)\ ,\notag\\ 
&C_{0,0,2}=\frac{C_{0,0,0} \left(\Delta_\O^4-5 \Delta_\O^2+8\right)}{16 \left(3 \Delta_\O^2-9 \Delta_\O+7\right)}+\O\left(\frac{1}{\Deltag} \right)\ ,\notag\\ 
&C_{0,1,0}=C_{1,0,0}=\frac{C_{0,0,0} \left(6 \Delta_\O^2-9 \Delta_\O-1\right)}{2 \left(3 \Delta_\O^2-9 \Delta_\O+7\right)}+\O\left(\frac{1}{\Deltag} \right)\ ,\notag\\ 
&C_{0,1,1}=C_{1,0,1}=\frac{C_{0,0,0} \Delta_\O \left(3 \Delta_\O^2+4 \Delta_\O-15\right)}{8 \left(3 \Delta_\O^2-9 \Delta_\O+7\right)}+\O\left(\frac{1}{\Deltag} \right)\ ,\notag\\ 
&C_{0,2,0}=C_{2,0,0}=\frac{C_{0,0,0} \left(3 \Delta_\O^2-2\right)}{4 \left(3 \Delta_\O^2-9 \Delta_\O+7\right)}+\O\left(\frac{1}{\Deltag} \right)\ ,\notag\\ 
&C_{1,1,0}=\frac{C_{0,0,0} \left(3 \Delta_\O^2+1\right)}{2 \left(3 \Delta_\O^2-9 \Delta_\O+7\right)}+\O\left(\frac{1}{\Deltag} \right)\ ,\notag\\ 
&C_{1,1,1}=\frac{C_{0,0,0} \Delta_\O^2}{2 \left(3 \Delta_\O^2-9 \Delta_\O+7\right)}+\O\left(\frac{1}{\Deltag} \right)\ .
\end{align}
Imposing these conditions we find that the subleading term 
\begin{align}
&\E_{\O\O}(\rho)\sim-\frac{3 \pi ^4 4^{-2 \Delta _{\mathcal{O}}-1} \left(\Delta _{\mathcal{O}}-1\right) \Gamma \left(2 \Delta _{\mathcal{O}}-1\right)C_{0,0,0}}{(1-\rho )  \left(3 \left(\Delta _{\mathcal{O}}-3\right) \Delta _{\mathcal{O}}+7\right) \Gamma \left(\Delta _{\mathcal{O}}\right){}^2}\notag\\
&\times\left( 4 \left(\Delta _{\mathcal{O}}-3\right) \left(\Delta _{\mathcal{O}}-2\right)+\left(\Delta _{\mathcal{O}}-1\right) \vec{\e}_\perp\cdot \bar{\vec{\e}}_\perp \left(\Delta _{\mathcal{O}} (\vec{\e}_\perp\cdot \bar{\vec{\e}}_\perp+4)-8\right)
\right)\ ,
\end{align}
 is determined up to one independent coefficient $C _{0,0,0}<0$. This coefficient is related to the coefficient that appears in the two-point function $\langle \O_{\ell=2}\O_{\ell=2}  \rangle$ by the Ward identity. Furthermore, after imposing all of the constraints we find that 
\begin{align}
f_{\e_1\cdot \O\, \e_2\cdot \O}(\rho)&=-\frac{3 \pi ^4 4^{-2 \Delta _{\mathcal{O}}-\frac{1}{2}} \left(\Delta _{\mathcal{O}}-1\right) \Gamma \left(2 \Delta _{\mathcal{O}}-1\right)\rho}{\left(3 \left(\Delta _{\mathcal{O}}-3\right) \Delta _{\mathcal{O}}+7\right) \Gamma \left(\Delta _{\mathcal{O}}\right){}^2(1-\rho)}C _{0,0,0}\notag\\
&\times \left(  4 ( \Delta_\O -3) ( \Delta_\O -2)+( \Delta_\O -1)  \vec{\e}_{1,\perp}\cdot \vec{\e}_{2,\perp} (4  \Delta_\O + \Delta_\O   \vec{\e}_{1,\perp}\cdot \vec{\e}_{2,\perp}-8)\right)
\end{align}
which is consistent with the universality of the Regge OPE of smeared operators. 

In the gravity dual description, there are also 6 possible types of vertices appearing in the on-shell three-point function of 2 massive spin-2 particles with a single graviton. The CFT result shows that the final answer is fixed up to a constant which is in agreement with the gravity result. Furthermore, requiring causality in the bulk \cite{Camanho:2014apa, Hinterbichler:2017qcl} dictates that the three-point function is determined up to a constant corresponding to the minimal coupling between massive spin 2 fields and a graviton. The vertex has the following form
\ba
 \left( (\ep_2 \cdot \ep_3) (\ep_1 \cdot p_2) + (\ep_1 \cdot \ep_3) (\ep_2 \cdot p_3) + (\ep_1 \cdot \ep_2) (\ep_3 \cdot p_1) \right)^2,
\ea
where the momenta are denoted by $p_1, p_2, p_3$, satisfying conservation and on-shell conditions: $p_1^\mu + p_2^\mu + p_3^\mu = 0, p_1^2 = -m^2, p_2^2= -m^2,p_3^2= 0$ and $\ep_i$ denote polarization tensors. For a more complete analysis of vertices and bulk dual, see \cite{Hinterbichler:2017qcl,Bonifacio:2017nnt}.

\section{Bounds from interference effect}\label{section_intf}
In this section, we will leverage the holographic null energy condition to derive bounds on the off-diagonal matrix elements of the operator $\E_r$. To this end we will consider superposition states created by smeared local operators:
\begin{align}
-i \lim_{B\rightarrow \infty}\langle  (\overline{\e_1.O_1(B)}+\overline{\e_2.O_2(B)})\E_{r=\sqrt{\rho}B}(\e_1.O_1(B)+\e_2.O_2(B))\rangle\ge 0
\end{align}
where $O_1$ and $O_2$ are arbitrary operators with or without spin ($\ell_1, \ell_2 \le 2$). This inequality can be expressed as semi-definiteness of the following matrix
\begin{align}\label{matrix}
\left(
\begin{array}{cc}
\E_{O_1 O_1}(\rho) & \E_{O_1 O_2}(\rho) \\ 
\E_{O_1 O_2}(\rho)^* & \E_{O_2 O_2}(\rho)
\end{array} 
\right)\succeq 0\ ,
\end{align}
where, we are using the notation (\ref{eo1o2}). The above condition can also be restated in the following form
\be\label{intf_bound}
|\E_{O_1 O_2}(\rho)|^2 \le \E_{O_1 O_1}(\rho)\E_{O_2 O_2}(\rho)\ , \qquad 0<\rho<1\ .
\ee
This is very similar to the interference effects in conformal collider experiment as studied in \cite{Cordova:2017zej}. In particular, in the limit $\rho\rightarrow 0$, the above relation is equivalent to the interference effects of \cite{Cordova:2017zej}. However, we are interested in the limit $\rho \rightarrow 1$ in which the above inequality imposes stronger constraints on three-point functions $\langle O_1 O_2 T\rangle$. These interference bounds are exactly the same as the bounds obtained in \cite{Meltzer:2017rtf} by studying  mixed system of four-point functions in the Regge limit in holographic CFTs.

As shown in the previous section, in $d\ge 4$ after imposing positivity of $\E_{O_1 O_1}(\rho)$ we have
\begin{align}
\E_{O_1 O_1}(\rho)\sim \O(1-\rho)^{3-d}\ .
\end{align}
Similarly,
\begin{align}
\E_{O_2 O_2}(\rho)\sim \O(1-\rho)^{3-d}\ .
\end{align}
Therefore, $\E_{O_1 O_2}(\rho)$ can not grow faster than $\O(1-\rho)^{3-d}$ in the limit $\rho\rightarrow 1$, or else causality will be violated. However, just from dimensional argument one can show that, in general 
\begin{align}
 \E_{O_1 O_2}(\rho) \sim \frac{1}{(1-\rho)^{-3+d+\ell_1+\ell_2}}\sum_{n=0,1,\cdots}c_n (1-\rho)^n\ 
\end{align}
and hence
\be\label{int_holo}
c_n =\O\left(\frac{1}{\Deltag^{\ell_1+\ell_2-n}} \right) \ , \qquad n=0,1,\cdots, \ell_1+\ell_2-1\ .
\ee
Whereas, $c_{\ell_1+\ell_2}$ is constrained by (\ref{intf_bound}). 

The causality conditions (\ref{int_holo}) are too constraining. In fact, from simple counting, one can argue that constraints (\ref{int_holo}) require all three-point functions of the form $\langle T O_1 O_2\rangle$ to vanish. Constraints (\ref{int_holo}) lead to at least $\ell_1+\ell_2$ linear algebraic equations among the OPE coefficients of $\langle T O_1 O_2\rangle$. However, for low spin operators ($\ell \le 2$), the number of independent OPE coefficients of $\langle T O_1 O_2\rangle$ is always less than $\ell_1+\ell_2$. This immediately suggests
\be
\langle T O_1 O_2\rangle=0\ ,
\ee
when $O_1$ and $O_2$ are different operators. Explicit computations, as we will show next, confirm that this is indeed true when $O_1, O_2$ are single trace primary operators. All our results are consistent with the interference bounds obtained in \cite{Meltzer:2017rtf} by using the conformal Regge theory.

\subsection*{Bound on $\langle T T\psi\rangle$}
As an example, we will obtain bounds on the OPE coefficient $C_{TT\psi}$ of $\langle T T\psi\rangle$ in $d\ge 4$ where $\psi$ is a light scalar operator. The polarization of $T$ is still given by $(1,\xi,\vec{\e}_{\perp})$. Now, from (\ref{matrix}) we have
\begin{align}
\left(
\begin{array}{cc}
\O(1-\rho)^{3-d} & c_0 (1-\rho)^{1-d}+\O(1-\rho)^{2-d} \\ 
c_0 (1-\rho)^{1-d}+\O(1-\rho)^{2-d} & \O(1-\rho)^{3-d}
\end{array} 
\right)\succeq 0.
\end{align}
Positivity of the eigenvalues of this matrix implies 
\begin{align}\label{interference1}
c_0&\sim
\frac{\pi ^{d-1}  \Gamma \left(\frac{d}{2}-\frac{1}{2}\right) 2^{d-\Delta _{\psi }-5} e^{-\frac{1}{2} i \pi  \left(d+\Delta _{\psi }\right)} \Gamma \left(\frac{\Delta _{\psi }}{2}+\frac{3}{2}\right)}{(1-\rho )^{d-1}\Gamma \left(\frac{\Delta _{\psi }}{2}+2\right) \Gamma \left(d-\frac{\Delta _{\psi }}{2}\right) \Gamma \left(\frac{d}{2}+\frac{\Delta _{\psi }}{2}+1\right)} C_{TT\psi}\notag\\
&\sim \O\left(\frac{1}{\Deltag^{2}} \right)\ 
\end{align}
and hence 
\be\label{TTpsi}
\langle T T\psi\rangle \lesssim \O\left(\frac{\sqrt{c_T}}{\Deltag^{2}} \right)
\ee
for all values of $\Delta_\psi$ for which the coefficient in front of $C_{TT\psi}$ does not vanish. Note that the coefficient in front of $C_{TT\psi}$ vanishes when $\Delta_\psi=2d+2n$ which is consistent with the fact that there are double trace stress tensor operators $[TT]_{\ell=0,n}$ which have spin 0. This agrees with the result obtained in \cite{Meltzer:2017rtf}.

In the dual gravity picture, $\langle T T\psi\rangle$ vanishes for a minimally coupled scalar field in AdS. However, in the bulk we can write higher derivative interactions between a scalar and two gravitons which give rise to $\langle T T\psi\rangle$ three-point function. In particular, let us consider the bulk action
\be
S=\frac{M_{Pl}^{d-1}}{2}\int d^{d+1}x \sqrt{g}\left[\left(\nabla \Psi \right)^2 -m^2 \Psi^2\right]+M_{Pl}^{d-1} \alpha_{\Psi hh}\int d^{d+1}x \sqrt{g}\, \Psi W^2\ .
\ee
In $d\ge 4$, the scalar-graviton-graviton vertex of the above action represents the most general bulk interaction which gives rise to the OPE coefficient $C_{TT\psi}$ \cite{Cordova:2017zej}:
\be
\frac{C_{TT\psi}}{\sqrt{c_T}}=\alpha_{\Psi hh} \frac{8 \pi^{d/2}(d-1)\sqrt{2d}}{\sqrt{d+1}\, \Gamma(d/2) \sqrt{f(\Delta_{\psi})}}
\ee
where, the function $f(\Delta)$ is given in \cite{Cordova:2017zej}. Hence, $\alpha_{\Psi hh}$ should be suppressed by the scale of new physics. In particular, the causality constraint  (\ref{TTpsi}) translates into $\alpha_{\Psi hh} \lesssim \frac{1}{\Delta^2_{\text{gap}}}$.\footnote{Note that $\Gamma \left(d-\frac{\Delta _{\psi }}{2}\right)$ in equation (\ref{interference1}) is canceled by $ \sqrt{f(\Delta_{\psi})}$ and hence the constraint $\alpha_{\Psi hh} \lesssim \frac{1}{\Delta^2_{\text{gap}}}$ is valid for any mass of the scalar field $\Psi$.} Of course, this is stronger than the constraint obtained in  \cite{Cordova:2017zej}. In  \cite{Cordova:2017zej}, constraints were obtained by considering interference effects in general CFTs. However, as shown in (\ref{int_holo}), interference effects from the holographic null energy condition lead to stronger constraints.

\subsection*{Bound on $\langle T T\O_{\ell=2}\rangle$}
Let us now obtain bounds on the three-point function $\langle T T\O_{\ell=2}\rangle$. This case is more subtle because a nonzero $\langle T T\O_{\ell=2}\rangle$ implies that the operator $\O_{\ell=2}$ will contribute to a four-point function in the Regge limit as an exchange operator. So, if  $\langle T T\O_{\ell=2}\rangle \neq 0$, the holographic null energy condition is no longer true. However, simplification emerges if we assume that there is at least one heavy scalar in the theory $\psi_H$ for which $\langle \psi_H \psi_H \O_{\ell=2}\rangle=0$. In this case, we can start with the operator $\psi_H$ in (\ref{maing}) and derive the holographic null energy condition even in the presence of $\O_{\ell=2}$.
So, with this additional assumption, we can calculate $\E_{T\O_{\ell=2}}(\rho)$ which is a straight forward generalization of the scalar case. Furthermore, the interference condition (\ref{int_holo}) again leads to
\be
\langle T T\O_{\ell=2}\rangle \lesssim \O\left(\frac{\sqrt{c_T}}{\Deltag} \right)\ .
\ee
Let us note that the above bound is not applicable when the dimension of $\O_{\ell=2}$ satisfies: $\Delta_{\O_{\ell=2}}= 2d+2+2n$. This is consistent with the fact that there are double trace stress tensor operators $[TT]_{\ell=2,n}$ with spin 2. With this caveat, we conclude that the presence of a single heavy scalar operator $\psi_H$ guarantees that the three-point function $\langle T T\O_{\ell=2}\rangle$ is suppressed by the gap for all single trace $\O_{\ell=2}$. An immediate consequence is that the operator $\O_{\ell=2}$ can not contribute as an exchange operator to four-point functions $\langle TT\O \O \rangle$ in the Regge limit for any $\O$. 

Before we proceed, let us also note that we expect that the same conclusion is true even without the presence of $\psi_H$. We believe causality of the four-point function $\langle TTTT \rangle$, requires that $\langle T T\O_{\ell=2}\rangle$ must be suppressed by the gap for all single trace $\O_{\ell=2}$. However, a detailed analysis requires the computation of $\langle TTTT \rangle$ using the conformal Regge theory which we will not attempt in this paper.

\section{Constraints on CFTs in $d=3$}\label{section_3d}

In this section, we will use the holographic null energy condition in $(2+1)$-dimensions to constrain various three-point functions of $(2+1)$-dimensional CFTs. Three-dimensional CFTs are special because of the presence of various parity odd structures. However, we again show that CFTs in $d=3$ with large central charge and a large gap exhibit universal, gravity-like behavior. Furthermore, holography enables us to translate the CFT bounds in to constraints on $(3+1)$-dimensional gravitational interactions. This, as we will discuss in the next section, has important consequences in cosmology. 

There is another aspect of $d=3$ which is different from the higher dimensional case. For $d\ge 4$, we have seen that holographic dual of a bulk derivative is $1/\Deltag$. This observation is consistent with the proposal of \cite{Meltzer:2017rtf}. However, we will show that in $d=3$, this simple relationship between bulk derivative and $\Deltag$ has a logarithmic violation.

\subsection{$\langle TTT\rangle$}

In $(2+1)$ dimensions, $\langle TTT\rangle$ has three tensor structures: two parity even structures with coefficients $\tilde{n}_s$ and $\tilde{n}_f$, and one parity odd structure with coefficient $\tilde{n}_\text{odd}$ (see appendix \ref{3d_odd}). We start with the holographic null energy condition (\ref{snec}) with $O$ being the stress-tensor $T$. In the limit $\rho\rightarrow 1$, the leading contribution to $\E_{TT}(\rho)$ goes as $ \frac{1}{(1-\rho)^4}$, the coefficient of which should always be positive. In particular, 
\begin{align}
\E_{TT}(\rho)|_{\text{Even}}&\sim\frac{32  \pi \left(4 \tilde{n}_f-\tilde{n}_s\right) }{5 (1-\rho )^4  }
\left(e_0^2 \left(\bar{e}_0^2+\bar{e}_2^2\right)-4 e_2 e_0 \bar{e}_0 \bar{e}_2+e_2^2 \left(\bar{e}_0^2+\bar{e}_2^2\right)
\right),\notag\\
\E_{TT}(\rho)|_{\text{Odd}}&\sim  \frac{8 \pi ^2 i \tilde{n}_\text{odd} \left(e_0^2 \bar{e}_0 \bar{e}_2-e_2 e_0 \left(\bar{e}_0^2+\bar{e}_2^2\right)+e_2^2 \bar{e}_0 \bar{e}_2\right)}{15  (1-\rho)^4  },
\end{align}
where we have defined
\begin{align}
\e=(e_0,e_1,e_2)\ , \qquad \bar{\e}=(\bar{e}_0,\bar{e}_1,\bar{e}_2).
\end{align}
The total expression can be conveniently written as
\begin{align}
\E_{TT}(\rho)\sim\frac{8 \pi  \left(- i \pi  \tilde{n}_\text{odd}A B+12  \left(4 \tilde{n}_f-\tilde{n}_s\right)\left(A^2+B^2\right)\right)}{15  (1-\rho )^4  },
\end{align}
where 
\begin{align}
A\equiv|e_0|^2-|e_2|^2\ , \qquad B\equiv e_2 e_0^*-e_0 e_2^*\ .
\end{align}
To find constraints on the coefficients, we first choose
\begin{align}
&\e=(i,\sqrt{-2},1) \Rightarrow (4\tilde{n}_f-\tilde{n}_s)\ge 0,\notag\\
&\e=(0,i,1) \Rightarrow (4\tilde{n}_f-\tilde{n}_s)\le 0,
\end{align}
implying that $\tilde{n}_s=4 \tilde{n}_f$. Imposing this condition we find constraints on the parity odd structure by considering
\begin{align}
\e&=(1+i,\sqrt{-1+2 i},1) \Rightarrow \tilde{n}_\text{odd}\ge 0,\notag\\
\e&=(-1+i,\sqrt{-1-2 i},-1) \Rightarrow \tilde{n}_\text{odd}\le 0,
\end{align}
implying that we must have $\tilde{n}_\text{odd}=0$ to satisfy positivity. Furthermore, after imposing these constraints, one can check that $f_{\e_0\cdot T\, \e_1\cdot T}(\rho)$ is still given by the equation (\ref{f_TT}) with $d=3$. 

Let us now estimate the size of the corrections to the above constrains if we include higher spin operators with large scaling dimensions, but not large
enough to compete with the $c_T$ expansion.  We can repeat the argument of section \ref{gap_crctn} for $d=3$, yielding
\be\label{TTT_3d}
\frac{|\tilde{n}_s-4 \tilde{n}_f|}{c_T} \lesssim \frac{\ln \Deltag}{\Deltag^4}\ , \qquad \frac{|\tilde{n}_\text{odd}|}{c_T}\lesssim \frac{\ln \Deltag}{\Deltag^4}\ .
\ee

On the gravity side, similar to the higher dimensional case, this constrains higher derivative correction terms in the pure gravity action that contribute to three point interactions of gravitons. However, in $(3+1)-$dimensional gravity there are certain crucial differences. First, the four-derivative terms do not contribute to $\langle TTT \rangle$. Second, in $(3+1)-$dimensional gravity, there is a parity odd higher derivative term which gives rise to $\tilde{n}_\text{odd}$. In particular, the higher derivative correction terms can be parametrized as 
\ba\label{gravity_4d}
S =M_{Pl}^{2}\int d^{4}x \sqrt{g} \;\left[ R-2\Lambda +\alpha_4 W_{\mu \nu \alpha \beta} W^{\mu \nu \rho \sigma }{W_{\rho \sigma}}^{\alpha \beta}+ \tilde{\alpha}_4 \tilde{W}_{\mu \nu \alpha \beta} W^{\mu \nu \rho \sigma }{W_{\rho \sigma}}^{\alpha \beta}  \right]\ ,
\ea
where, $\tilde{W}_{\mu \nu \alpha \beta}=\frac{1}{2}\epsilon_{\mu \nu \rho \sigma} W^{\rho \sigma}_{~~\alpha \beta}$. Coupling constants $\alpha_4$ and $\tilde{\alpha}_4$ are related to the coefficients $\tilde{n}_s-4\tilde{n}_f$ and $\tilde{n}_\text{odd}$ respectively.\footnote{In $(3+1)-$dimensional gravity, one can also have another parity violating term in the four-derivative level: $\int d^{4}x \sqrt{g} \tilde{W}W$ which is a total derivative. However,  this term contributes a non-trivial parity violating contact term to the two-point function $\langle TT\rangle$ \cite{Fischler:2015kro}.} Hence, causality constraints translate into $\alpha_4 \sim \frac{\ln \Deltag}{\Deltag^4}, \tilde{\alpha}_4 	\sim \frac{\ln \Deltag}{\Deltag^4} $. It was proposed in \cite{Meltzer:2017rtf} that holographic dual of a bulk derivative is $1/\Deltag$. However, as we see here, for $(3+1)-$dimensional gravity, there is a logarithmic violation.
\subsection{$\langle JJT\rangle$}
Similarly, in $(2+1)$ dimensions $\langle JJT\rangle$ has parity even and odd structures (see appendix \ref{3d_odd}) with the leading terms in the limit $\rho\rightarrow 1$ given by
\begin{align}
\E_{JJ}(\rho)|_{\text{Even}}&\sim -\frac{ \pi  \left(e_0 \bar{e}_0-e_2 \bar{e}_2\right) \left(4 n_f-n_s\right)}{9  (1-\rho )^2 }\ ,\notag\\
\E_{JJ}(\rho)|_{\text{Odd}}&\sim \frac{2 i \pi ^2 n_\text{odd} \left(e_2 \bar{e}_0-e_0 \bar{e}_2\right)}{3 (1-\rho )^2 }\ .
\end{align}
Positivity of $\E_{JJ}(\rho)$ implies the following conditions on the coefficients
\begin{align}
\frac{|n_s-4 n_f|}{c_J} \lesssim \frac{\ln \Deltag}{\Deltag^2}\ ,\qquad \frac{|n_\text{odd}|}{c_J}\lesssim \frac{\ln \Deltag}{\Deltag^2}\ .
\end{align}
After imposing these constraints, one can easily check that our conjectured relation (\ref{opecheck}) is satisfied.

 The three-point function $\langle JJT\rangle$, in dual gravity language, arises from the following 4d-action
\ba
  \int d^{4}x \sqrt{-g} \;  \left[-F_{\mu \nu} F^{\mu \nu} + \alpha_{AAh}  W_{\mu \nu \alpha \beta} F^{\mu \nu} F^{\alpha \beta}+ \tilde{\alpha}_{AAh}  \tilde{W}_{\mu \nu \alpha \beta} F^{\mu \nu} F^{\alpha \beta}\right], 
\ea
where, coefficients $\alpha_{AAh}$ and  $\tilde{\alpha}_{AAh}$ can be written in terms of $n_s$, $n_f$ and $n_\text{odd}$:
\be
\alpha_{AAh}\sim \frac{n_s-4n_f}{n_s+4n_f}\sim \frac{\ln \Deltag}{\Deltag^2}\ , \qquad \tilde{\alpha}_{AAh}\sim \frac{n_\text{odd}}{n_s+4n_f}\sim \frac{\ln \Deltag}{\Deltag^2}\ .
\ee
Appearance of  $\ln \Deltag$ again indicates that the simple relationship between bulk derivative and $\Deltag$ has a logarithmic violation in 3d CFT.
\subsection{$\langle TT \psi\rangle$}
Let us now discuss the three-point function $\langle TT \psi\rangle$ in $d=3$. The analysis is identical to the derivation of causality constraints for $\langle TT \psi\rangle$ in higher dimension using interference effects. So, we will not show the full calculation, instead we only point out the key differences. In $d=3$, conformal invariance also allows for a parity odd structure and the full correlator consists of two structures
\be
\langle TT \psi\rangle= \langle TT \psi\rangle_\text{Even}+\langle TT \psi\rangle_\text{Odd}\ 
\ee
with OPE coefficients $C_{TT\psi}^\text{Even}$ and $C_{TT\psi}^\text{Odd}$ respectively \cite{Cordova:2017zej}. First, we derive causality constraints on the three-point function $\langle TTT \rangle$ which leads to (\ref{TTT_3d}). After imposing these constraints, in the limit $\rho\rightarrow 1$, $\E_{TT}(\rho)\sim \ln(1-\rho)$. On the other hand,  in the limit $\rho\rightarrow 1$, for both even and odd structures $\E_{T\psi}(\rho)\sim \frac{1}{(1-\rho)^2}$. Hence, the interference bound (\ref{intf_bound}) dictates that both $C_{TT\psi}^\text{Even}$ and $C_{TT\psi}^\text{Odd}$ should be suppressed by $\Deltag$:
\be\label{TTpsi_3d}
\frac{C_{TT\psi}^\text{Even}}{\sqrt{c_T}} \lesssim \frac{\ln \Deltag}{\Deltag^2}\ , \qquad \frac{C_{TT\psi}^\text{Odd}}{\sqrt{c_T}} \lesssim \frac{\ln \Deltag}{\Deltag^2}\ .
\ee

Similarly, in the bulk there are two possible vertices between a scalar and two gravitons, one parity even and one parity odd. These interactions can be parametrized as
\be
S=M_{Pl}^{2} \alpha_{\Psi hh}\int d^{4}x \sqrt{g}\, \Psi W^2+M_{Pl}^{2} \tilde{\alpha}_{\Psi hh}\int d^{4}x \sqrt{g}\, \Psi \tilde{W}W\ .
\ee
These interactions were first constrained by Cordova, Maldacena, and Turiaci in \cite{Cordova:2017zej}. Using the averaged null energy condition they showed that in generic CFTs in $d=3$, interference effects impose constraints on the OPE coefficients $C_{TT\psi}^\text{Even}$ and $C_{TT\psi}^\text{Odd}$. These general bounds can be translated into bounds on gravitational interactions  \cite{Cordova:2017zej}
\be
\sqrt{\alpha_{\Psi hh}^2+\tilde{\alpha}_{\Psi hh}^2}\le \frac{1}{12\sqrt{2}}\ .
\ee
However, it is expected that the holographic null energy condition leads to stronger constraints on $ \alpha_{\Psi hh}, \tilde{\alpha}_{\Psi hh}$. In particular, bounds (\ref{TTpsi_3d}) are equivalent to
\be\label{bound_alpha}
\alpha_{\Psi hh}\lesssim \frac{\ln \Deltag}{\Deltag^2}\ , \qquad \tilde{\alpha}_{\Psi hh}\lesssim \frac{\ln \Deltag}{\Deltag^2}\ .
\ee 
In the following section, we will use these constraints to impose bounds on inflationary observables.

\section{Constraining inflationary observables}\label{section_cosmo}

In the last section, we showed that $(2+1)-$dimensional CFTs with large central charge and a sparse spectrum, irrespective of their microscopic details, admit universal holographic dual description at low energies. This connection provides us with a  tool to constrain gravitational interactions in $(3+1)-$dimensions. This has an immediate application in inflation. The period of inflation is an exponential expansion of the universe that powered the epoch of the big bang. On one hand, inflation naturally explains why our universe is flat and homogeneous on the large scale. On the other hand, quantum fluctuations produced during inflation are responsible for the temperature fluctuations of cosmic microwave background (CMB) and the large-scale structures of the universe.

The simplest model of inflation consists of a real scalar field minimally coupled to Einstein gravity. In general, there can be higher derivative interactions which can contribute to various inflationary observables. Therefore, constraints obtained in the previous section can impose bounds on such observables (for example chiral gravity waves, tensor-to-scalar ratio etc.). However, there is a caveat. All of the constraints on gravitational interactions obtained in this paper, strictly speaking, are valid in AdS. Following the philosophy of \cite{Camanho:2014apa,Cordova:2017zej}, we simply assume that the same constraints are also valid in de Sitter after we make the substitution $R_{AdS}\rightarrow 1/H$, where $H$ is the Hubble scale associated with inflation. This is a reasonable assumption but it would be important to have a robust derivation of these de Sitter constraints.

\subsection{Chiral gravity waves}
Chiral gravity waves \cite{Lue:1998mq,Alexander:2004wk} can be produced during inflation from a parity odd higher derivative interaction in the action
\be
M_{Pl}^{2} \int d^{4}x \sqrt{g}\, f_o(\Psi) \tilde{W}W\ , \qquad \Psi=\frac{\phi}{M_{Pl}}\ ,
\ee
where $\phi$ is the inflaton field. In the presence of this term in the action, two-point functions of tensor modes with left handed and right handed circular polarizations are not the same.  The asymmetry $A$ measures the difference between left and right handed polarizations and it is determined by the above parity odd interaction \cite{Cordova:2017zej}
\be
A\equiv \frac{\langle h_Lh_L\rangle-\langle h_R h_R \rangle}{\langle h_Lh_L\rangle+\langle h_R h_R \rangle}=\pm 4\pi \sqrt{2\epsilon}\,  \tilde{\alpha}_{\Psi hh} H^2\ ,
\ee
where, $\epsilon$ is one of the slow-roll parameters of inflation. In the above expression, we have used the fact $\tilde{\alpha}_{\Psi hh}=\frac{\partial f_o(\Psi)}{\partial \Psi}$. So, constraint (\ref{bound_alpha}) strongly suggests that the asymmetry parameter $A$ must be suppressed by the scale of new physics $M$:\footnote{We have identified $\Deltag=M/H$.}
\be
|A|\lesssim 4\pi \sqrt{2\epsilon}\, \left(\frac{H^2}{M^2}\right) \ln \left(\frac{M}{H} \right)\ .
\ee
First of all, note that this is stronger than the bound obtained in \cite{Cordova:2017zej}. Secondly, if the asymmetry parameter $A$ is measured (or in other words it is found to be at least a few percent), then the new physics scale must be $M\sim H$. This scenario necessarily requires the presence of an infinite tower of new particles with spins $\ell >2$ and masses $\sim H$. Therefore, any detection of this parameter in future experiments will serve as an evidence in favor of string theory (or something very similar) with the string scale comparable to the Hubble scale and a very small coupling.

\subsection{Tensor-to-scalar ratio}
Similarly, one can obtain a bound on the ratio $r$ of the amplitudes of tensor fluctuations and scalar fluctuations. In a single-field inflation without any higher derivative couplings, the tensor-to-scalar ratio $r$ obeys a consistency condition \cite{Lidsey:1995np}: $r=-8 n_t$, where $n_t$ is the tensor spectral index. In the presence of the higher derivative interaction 
\be
M_{Pl}^{2} \int d^{4}x \sqrt{g}\, f_e(\Psi) W^2\ ,  
\ee
the consistency condition is violated \cite{Baumann:2015xxa}. In particular, one can show that \cite{Cordova:2017zej}
\be
-\frac{n_t}{r}=\frac{1}{8}\pm \frac{H^2}{\sqrt{2\epsilon}} \alpha_{\Psi hh}\ , \qquad \alpha_{\Psi hh}=\frac{\partial f_e(\Psi)}{\partial \Psi}\ .
\ee
In the above expression we have assumed that the inflaton field has only a canonical kinetic term with two-derivatives.\footnote{In other words, the speed of sound for the inflaton field is $1$.} So far, this is exactly the same as the discussion of \cite{Cordova:2017zej}. But we now derive a stronger bound by using constraint (\ref{bound_alpha}) which suggests that the violation of the consistency relation must be suppressed by $M$
\be
\Big|\frac{n_t}{r}+\frac{1}{8}\Big|\lesssim  \frac{1}{\sqrt{2\epsilon}}\, \left(\frac{H^2}{M^2}\right) \ln \left(\frac{M}{H} \right)\ .
\ee

\subsection{Graviton non-gaussanity}
Let us now consider non-gaussanity of primordial gravitational waves produced during inflation. In Einstein gravity, the three-point function of tensor perturbation goes as
\be
\langle hhh \rangle_E \sim \frac{H^4}{M_{Pl}^4}\ .
\ee
The graviton three-point function (parity preserving part) can also get contributions from $W^3$ term in the gravity action (\ref{gravity_4d}). As shown in  \cite{Camanho:2014apa}, the contribution from this interaction must be suppressed by the scale of new physics:
\be
\frac{\langle hhh \rangle_{W^3}}{\langle hhh \rangle_E} \sim \alpha_4 H^4 \sim \left(\frac{H^4}{M^4}\right) \ln \left(\frac{M}{H} \right)\ .
\ee
Hence, any significant deviation from the Einstein gravity result requires the presence of an infinite tower of new particles with spins $\ell >2$ and masses $\sim H$ \cite{Camanho:2014apa}.

The advantage of studying any parity violating effects during inflation is that these contributions are exactly zero for Einstein gravity. Hence, any detection of parity violation will be a signature of new physics at the Hubble scale. The gravity action in general can have a parity odd term $\tilde{W}W^2$ which is also controlled by the same scale $M$. This term contributes to the  parity odd part of graviton three-point function \cite{Soda:2011am,Shiraishi:2011st,Maldacena:2011nz}. In particular,
\be
\langle h_L h_L h_L\rangle-\langle h_R h_R h_R \rangle \sim \epsilon \left(\frac{H^4}{M_{Pl}^4}\right) \tilde{\alpha}_4 H^4\ .
\ee
Therefore, causality requires that
\be
\frac{\langle h_L h_L h_L\rangle-\langle h_R h_R h_R \rangle}{\langle hhh \rangle}\sim \epsilon \left(\frac{H^4}{M^4}\right) \ln \left(\frac{M}{H} \right)\ .
\ee
This parity violating graviton non-gaussanity will have signatures in the CMB. For example, CMB three-points correlators $\langle TEB\rangle$, $\langle EEB\rangle$, $\langle TTB\rangle$ become nonzero in the presence of the parity violating graviton non-gaussanity. However, one disadvantage of studying the parity violating graviton non-gaussanity is that this contribution is exactly zero in pure de Sitter \cite{Soda:2011am,Maldacena:2011nz}. Hence, for slow-roll inflation this effect is suppressed by the slow-roll parameter $\epsilon$.

We should also note that terms like $f_e(\phi)W^2$ or $f_o(\phi) \tilde{W}W$ in the effective action can also contribute to the graviton three-point function. Both these contributions depend on the details of the inflationary scenario and they can dominate over the contributions from $W^3$ and $\tilde{W}W^2$ \cite{Maldacena:2011nz, Bartolo:2017szm}. However, contributions of $f_e(\phi)W^2$ and $f_o(\phi) \tilde{W}W$ to the graviton three-point function are proportional to $\sqrt{\epsilon} f_e'(\phi)$ and $\sqrt{\epsilon} f_o'(\phi)$ respectively which are bounded by causality as well. So, if these terms are present in the effective action, their contributions to the non-gaussanity of primordial gravitational waves should also be suppressed by $M$ but with a different power
\be
\frac{\langle hhh \rangle_{f_e(\phi)W^2,\ f_o(\phi) \tilde{W}W}}{\langle hhh \rangle_E} \sim \sqrt{\epsilon} \left(\frac{H^2}{M^2}\right) \ln \left(\frac{M}{H} \right)\ .
\ee


\section{Discussion}\label{section_final}
In this paper, we analyzed the implications of causality of correlation functions on CFT data in theories with large $c_T$ and sparse higher spin spectrum. This was accomplished by developing a new formalism that can be interpreted as a collider type experiment in the CFT, set up in such a way to probe scattering processes deep in the bulk interior of the corresponding holographic dual theory. In doing so we consider the {\it holographic null energy operator}, $\E_{r}$ which is a positive operator in a certain subspace of the total CFT Hilbert space. This subspace is spanned by states constructed by acting local operators, smeared with Gaussian wave-packets, on the CFT vacuum. Positivity of this operator was then used to impose bounds on the CFT data.

\subsubsection*{Other representations}
It is worth mentioning that the formalism presented here can easily be adopted to compute the contribution of the holographic null-energy operator to the four-point function of external operators in arbitrary representation including spinors or non-symmetric traceless representations. The only modification required is to compute three-point functions of these operators with the stress-tensor whose form is fixed by conformal symmetry. 

Furthermore with slight modification one may compute the contribution of single-trace exchanged operators other than the stress-tensor. More specifically in \cite{Afkhami-Jeddi:2017rmx} it was shown that in the Regge limit ($ v\rightarrow 0 $ with $uv$ held fixed) the contribution of a spinning operator $X$ (with spin $\ell$ and dimension $\Delta_X$) to the OPE can be written as
\begin{align}
\begin{split}\label{genreg}
& \frac{\psi(u,v) \psi(-u,-v)|_X}{\langle \psi(u,v) \psi(-u,-v) \rangle } =  \pi^{\frac{1 -d}{2}} 2^{\Delta_X}  \frac{\Gamma(\frac{\Delta_X+\ell+1}{2})}{\Gamma(\frac{\Delta_X+\ell}{2})} \frac{\Gamma(\Delta_X -d/2+1)}{\Gamma(\Delta_X-d+2)} \frac{C_{\psi \psi X}}{C_X} \\
&\times \frac{(-u v)^{\frac{d-\ell-\Delta_X}{2}}}{u^{1-\ell}} \int_{-\infty}^{+\infty} d\tilde{u} \int_{\vec{x}^2 \le -u v}d^{d-2}\vec{x} ( -u v -\vec{x}^2)^{\Delta_X-d+1} X_{uu\cdots u}(\tilde{u},0,i\vec{ x})
\end{split}.
\end{align}
This OPE is valid as long as it is evaluated in a correlation function where all other operator insertions are held fixed as we take the Regge limit. 
However, the chaos bound suggests that this contribution does not necessarily dominate in the Regge limit in holographic CFTs.
\subsubsection*{Non-conserved spin-2 exchange}
As previously mentioned, one caveat to our computation is the possibility of competition between the contributions of  non-conserved spin-2 operators with the stress-tensor in the Regge limit. However, using the OPE described above it is possible to explicitly compute the contribution of such an operator to the Regge OPE. Including the contribution of a single non-conserved spin-2 exchange, we find bounds on the OPE coefficients of the stress-tensor as well as the non-conserved spin-2 operator. We expect that some version of the experiment described above, should reproduce the constraints found in \cite{Bonifacio:2017nnt} which resulted from performing a scattering experiment in the bulk. We leave explicit confirmation of this claim to future explorations.

\subsubsection*{Regge OPE of single trace operators}
The operator product expansion of smeared primary operators in the Regge limit, as discussed in section \ref{section_ope}, is universal. When $O_1$ and $O_2$ are different operators, the identity piece in the OPE (\ref{ope2}) does not contribute. Moreover, if  $O_1$ and $O_2$ are single trace operators, then interference effects imply that $\langle T O_1 O_2\rangle =0$. So, for these operators, even the coefficient of the shockwave operator in (\ref{ope2}) vanishes. Hence, for non-identical single trace primary operators the OPE
\be
\Psi^*[O_1] \Psi[O_2]=0+\cdots,
\ee
where, dots represent terms which are suppressed by either the large gap limit or the large $c_T$ limit or the Regge limit.

\subsubsection*{Higher spin ANEC}
Although not pursued in detail here, by taking the lightcone limit of \eqref{genreg}, the same formalism developed here can be used to compute the contribution of the ANEC operator  to correlation functions. Furthermore, this formalism can be easily extended to study the higher spin ANEC \cite{Hartman:2016lgu} which  says
\be
\int du X_{uu\cdots u} \ge 0\ ,
\ee
where, $X$ is the lowest dimension operator with even spin ($\ell \ge 2$). Positivity of these operators holds in the more general class of theories including non-holographic CFTs. A systematic exploration of bounds derived from the positivity of these operators is left to future work.

\subsubsection*{OPE of spinning operators }
It would be interesting to derive the stress tensor contribution to the OPE of spinning operators both in the Regge and the lightcone limits. Using this OPE, an argument similar to the ones used in this paper would lead to new positive spinning null energy conditions. These positivity conditions both conceptually as well as technically, will have important implications. For instance, this will allow us to derive new constraints in a more systematic way. Moreover, based on the analogous constraints obtained in the bulk \cite{Camanho:2014apa}, we expect these positive operators to play an important role in closing the gap in ruling out non-conserved spin-2 exchanges. 
\section*{Acknowledgements}
It is our pleasure to thank Tom Hartman for several helpful discussions as well as comments on a draft. We would also like to thank Ibou Bah, Federico Bonetti, Clay Cordova, Arnab Kundu, Marco Meineri, Masato Nozawa, Eric Perlmutter, and Maresuke Shiraishi for discussions. The work of NAJ and AT is supported by DOE grant DE-SC0014123. SK would like to thank Institut des Hautes Etudes Scientifiques, France for hospitality where some of this work was completed.\\

SK- {\it I dedicate this paper to my teacher and my father Dr. Balai Chand Kundu who  passed away recently. I am eternally grateful to him for his
unconditional love, constant support, and sacrifices. I owe him everything.}

\appendix

\section{Three-point functions of conserved currents}\label{s:currents}
In this appendix we summarize conventions used through out the paper in describing the OPE coefficients appearing in the correlation functions of conserved currents. 
Correlation functions of conserved currents in CFT are derived in \cite{Osborn:1993cr} (see also \cite{Erdmenger:1996yc}) using conformal symmetry. Expression written here can be compared with similar ones written in \cite{Costa:2011dw,Costa:2011mg,Zhiboedov:2012bm,Zhiboedov:2013opa,Li:2015itl,Giombi:2011rz}.

\subsection{$\langle JJT \rangle$}\label{jjt}
Two point function of spin-1 currents is given by
\be
\langle \varepsilon_1 . J(x_1) \varepsilon_2 . J(x_2) \rangle=c_J \frac{H_{12}}{x_{12}^{2d}}\ ,
\ee
where, $H_{12}$ is defined in (\ref{HV}). The three-point function $\langle JJT\rangle$ is given by
\begin{align}\label{jjtbasis}
\langle J(x_1)J(x_2)T(x_3)\rangle = \frac{ \alpha_1 V_{1} V_{2} V_{3}^2 + \alpha_2 H_{12} V_{3}^2 +  \alpha_3 ( H_{23} V_{1}V_{3}+H_{13}V_{2}V_{3} ) + \alpha_5 H_{13}H_{23}}{x_{12}^{d-2}x_{13}^{d-2}x_{23}^{d+2}} \, 
\end{align}
with
\ba
&V_1 = V_{1,23}, \qquad V_2 = V_{2,31}, \qquad V_{3}= V_{3,12}, 
\ea
In the free field basis, this can also be written as 
\be\label{freej}
\langle JJT \rangle = n_s \langle JJT\rangle_{scalar} + n_f \langle JJT \rangle_{fermion}
\ee
where the coefficients are related by \cite{Hartman:2016dxc}
\begin{equation}
\begin{split}
&\alpha_1 = n_s \frac{d-2}{2(d-1)}-8 n_f,~~ \alpha_2= -4 n_f -  \frac{n_s}{2(d-1)} \\
&\alpha_3 = -4n_f - \frac{n_s}{d-1},~~\alpha_5= \frac{n_s}{(d-1)(d-2)}.
\end{split}\end{equation}
The Ward identity relates one combination of $n_s$ and $n_f$ to the two-point function:
\be\label{cj}
c_J  = \frac{S_d}{d}\left(4n_f + \frac{n_s}{d-2}\right) \ ,
\ee
where, $S_d=\frac{2 \pi^{d/2}}{\Gamma(d/2)}$.

\subsection{$\langle TTT\rangle $}\label{ttt}
The central charge $c_T$ is defined as
\be\label{cT}
\langle \varepsilon_1 . T(x_1) \varepsilon_2 . T(x_2) \rangle=c_T \frac{H_{12}^2}{x_{12}^{2(d+2)}}\ ,
\ee
where, $H_{12}$ is given by equation (\ref{HV}).

Three point function $ \langle T(x_1)T(x_2) T(x_3)\rangle$ is fixed by conformal invariance and permutation symmetry
\be \label{TTTcorr}
 \langle T(x_1)T(x_2) T(x_3)\rangle= \frac{\sum_{i=1}^{5}\alpha_i S_i}{ x_{12}^{2+d} x_{13}^{2+d } x_{23}^{2+d} }
\ee
where
\begin{align}
&S_1=  V_1^2 V_2^2 V_3^2 ,~~ S_2= V_1 V_2 V_3 \left( H_{23} V_1+H_{13} V_2+H_{12} V_3\right) \\
& S_3=\left(H_{12} H_{23} V_1 V_3+ H_{13} V_2 \left( H_{23} V_1+ H_{12} V_3  \right)  \right),~~ S_4= H_{12} H_{13} H_{23}\notag\\
&S_5=H_{23}^2 V_1^2+ H_{13}^2 V_2^2 + H_{12}^2 V_3^2.\notag
\end{align}
This three-point function can be translated to the free-field basis
\be\label{freet}
\langle TTT \rangle = \tn_s \langle TTT\rangle_{scalar} + \tn_f \langle TTT \rangle_{fermion}+ \tn_v\langle TTT \rangle_{vector}
\ee
using \cite{Hartman:2016dxc}
\begin{align}
\alpha_1 &= 128 d^2 \tn_f - \frac{8d^2(d-2)^3}{(d-1)^3}\tn_s - 8192\tn_v \\
\alpha_2 &= 64d(d-2)\tn_f + \frac{32(d-2)^2d^2}{(d-1)^3}\tn_s-8192\tn_v\notag\\
\alpha_3 &= -128d\tn_f - \frac{64d^2(d-2)}{(d-1)^3}\tn_s-4096\tn_v\notag\\
\alpha_4 &=\frac{64d^2}{(d-1)^3}\tn_s - \frac{4096}{d-2}\tn_v\notag\\
\alpha_5 &=-64d\tn_f - \frac{16d(d-2)^2}{(d-1)^3}\tn_s-2048\tn_v\ .\notag
\end{align}
Ward identity relates $\tn_s$, $\tn_f$, and $\tn_v$ to the central charge in the following way
\be\label{ct}
c_T = 128S_d\left(\tn_f + \frac{1}{2(d-1)}\tn_s + \frac{16(d-3)}{d(d-2)}\tn_v\right) \ .
\ee

\section{Three-point functions in $d=3$}\label{3d_odd}
\subsection{$\langle JJT \rangle$}
The parity odd part of the correlation functions is given by \cite{Giombi:2011rz}
\begin{align}
\langle J(x_3)J(x_4)T(x_5)\rangle=n_\text{odd}\frac{Q_3^2 S_3^2+2P_5 S_4^2+2P_4S_5^2}{|x_{34}||x_{35}||x_{45}|},
\end{align}
where,
\begin{align}
Q_3&=\frac{2\e_3\cdot x_{35}}{x_{35}^2}-\frac{2\e_3\cdot x_{34}}{x_{34}^2},\notag\\
Q_4&=\frac{2\e_4\cdot x_{43}}{x_{43}^2}-\frac{2\e_4\cdot x_{45}}{x_{45}^2},\notag\\
Q_5&=\frac{2\e_5\cdot x_{54}}{x_{54}^2}-\frac{2\e_5\cdot x_{53}}{x_{53}^2},\notag\\
P_3&=\frac{4 x_{34}\cdot \e_3 x_{34}\cdot \e_4}{\left(x_{34}\cdot x_{34}\right){}^2}-\frac{2 \e_3\cdot \e_4}{x_{34}\cdot x_{34}}\notag\\
P_4&=\frac{4 x_{45}\cdot \e_4 x_{45}\cdot \e_5}{\left(x_{45}\cdot x_{45}\right){}^2}-\frac{2 \e_4\cdot \e_5}{x_{45}\cdot x_{45}}\notag\\
P_5&=\frac{4 x_{53}\cdot \e_3 x_{53}\cdot \e_5}{\left(x_{53}\cdot x_{53}\right){}^2}-\frac{2 \e_5\cdot \e_3}{x_{53}\cdot x_{53}}\notag\\
S_3^2&=-\frac{2 \left(x_{34}^2 \epsilon\left(x_{53},\e_3,\e_4\right)+x_{53}^2 \epsilon\left(x_{34},\e_3,\e_4\right)-2 \epsilon\left(x_{34},x_{53},\e_3\right) \e_4\cdot x_{34}\right)}{|x_{34}|^3 |x_{45}| |x_{53}|}\notag\\
S_4^2&=\frac{2 \left(x_{34}^2 \epsilon\left(x_{45},\e_4,\e_5\right)+x_{45}^2 \epsilon\left(x_{34},\e_4,\e_5\right)-2 \epsilon\left(x_{45},x_{34},\e_4\right) \e_5\cdot x_{45}\right)}{|x_{34}| |x_{45}|^3 |x_{53}|}\notag\\
S_5^2&=-\frac{2 \left(x_{45}^2 \epsilon\left(x_{53},\e_5,\e_3\right)+x_{53}^2 \epsilon\left(x_{45},\e_5,\e_3\right)-2 \epsilon\left(x_{53},x_{45},\e_5\right) \e_3\cdot x_{53}\right)}{|x_{34}| |x_{45}| |x_{53}|^3},
\end{align}
where $\epsilon\left(a,b,c\right)\equiv \epsilon_{\mu\nu\alpha}a^\mu b^\nu c^\alpha$, with $\epsilon_{\mu\nu\alpha}$ denoting the Levi-Civita symbol.
The parity even part is given by \eqref{jjtbasis} with $d=3$.

\subsection{$\langle TTT\rangle$}
The parity odd part of the correlation functions is given by \cite{Giombi:2011rz} 
\begin{align}
\langle T(x_3)T(x_4)T(x_5)\rangle=n_\text{odd}\frac{P_5 Q_5^2 S_5^2+P_3 \left(Q_3^2 S_3^2-5 P_5 S_4^2\right)+P_4 \left(5 P_5 S_3^2+5 P_3 S_5^2-Q_4^2 S_4^2\right)}{|x_{34}| |x_{45}| |x_{53}|},
\end{align}
where the structures are defined in the previous subsection.
The parity even part is given by \eqref{TTTcorr} with $d=3$ and $\tilde{n}_v=0$.
\section{d-dimensional smearing integrals}\label{smintegrals}
We are interested in evaluating integrals of the form
\begin{align}\label{sminta}
\int d^{d-1}\pv \frac{\prod_i\pv.\vec{\mathbf{v}}_i}{(\pv^2+\pv\cdot\Lv)^{p_1}(\pv\cdot\Lv)^{p_2}}.
\end{align}
Let us first define\footnote{note that $p_1,p_2>0$ in all expression appearing in this paper.}
\begin{align}
I_{p_1,p_2}(\Lv)\equiv\int d^{d-1}\pv \frac{1}{(\pv^2+\pv\cdot\Lv)^{p_1}(\pv\cdot\Lv)^{p_2}}.
\end{align}
Using Feynman parametrization we can rewrite this as
\begin{align}
I_{p_1,p_2}(\Lv)=\frac{\Gamma (p_1+p_2)}{\Gamma (p_1) \Gamma (p_2)}\int_0^1 d\alpha\int d^{d-1}\pv \frac{\alpha^{p_1-1} (1-\alpha)^{p_2-1}}{(\pv\cdot \Lv+\alpha\pv\cdot \pv)^{p_1+p_2}}.
\end{align}
The idea is to use derivatives with respect to $\Lv$ to obtain an expression with powers of $\pv$ in the numerator. To this end, let us first define
\begin{align}
K_p(\Lv)&\equiv\int d^{d-1}\pv (\pv\cdot \Lv+\alpha\pv\cdot \pv)^{-p}\notag\\
&=\frac{i^{d-1} \pi ^{\frac{d-1}{2}} (-1)^p 2^{-d+2 p+1} \alpha^{-d+p+1} \Gamma \left(\frac{1-d}{2}+p\right) (\Lv\cdot \Lv)^{\frac{d-1}{2}-p}}{\Gamma (p)}.
\end{align}
Furthermore let us notice that
\begin{align}
\frac{(-1)^{-n} \Gamma (p-n)}{\Gamma (p)}\prod_i^n \frac{\d}{\d L^{\mu_i}}(\pv\cdot \Lv+\alpha \pv\cdot \pv)^{n-p}=(\pv\cdot L+\alpha \pv\cdot \pv)^{-p}\prod_i^n p^{\mu_i}.
\end{align}
Finally we define 
\begin{align}
F^{(n)}_{p_1,p_2}(\Lv)&\equiv \frac{\Gamma (p_1+p_2-n) (-1)^{-n}}{\Gamma (p_1) \Gamma (p_2)}\int_0^1 d\alpha \alpha^{p_1-1} (1-\alpha)^{p_2-1} K_{p_1+p_2-n}(\Lv)\notag\\
&=\frac{i^{d-1} \pi ^{\frac{d-1}{2}}  \Gamma \left(-\frac{d}{2}-n+p_1+p_2+\frac{1}{2}\right) \Gamma (-d-n+2 p_1+p_2+1) (L\cdot L)^{\frac{d-1}{2}+n-p_1-p_2}}{(-1)^{2 n-p_1-p_2} 2^{d-2 (-n+p_1+p_2)-1}\Gamma (p_1) \Gamma (-d-n+2 (p_1+p_2)+1) }.
\end{align}
Using this, we now have a simple way of evaluating integrals:
\begin{align}
\int d^{d-1}\pv \frac{\prod_i^n\pv.\vec{\mathbf{v}}_i}{(\pv^2+\pv\cdot\Lv)^{p_1}(\pv\cdot\Lv)^{p_2}}=\prod_i^n (\vec{\mathbf{v}}_i\cdot\vec{\d}) F^{(n)}_{p_1,p_2}(\Lv),
\end{align}
where $\vec{\d}$ signifies differentiation with respect to $\Lv$.

\section{Polarization vectors}\label{sec:pol}

Throughout this paper, we used a particular null vector \ref{polarization}, to construct the polarization tensors corresponding to the external smeared states. The same null vector was used in \cite{Afkhami-Jeddi:2016ntf} for obtaining $a=c$ bounds in $d=4$. In this appendix we will describe how this choice simplifies the task of extracting positivity conditions from spinning correlators with conserved operator insertions. For the case of non-conserved operators, this is not the most general choice of polarizations and does not necessarily lead to the most optimum bounds. However the bounds obtained using this vector are sufficiently stringent for our purposes.

\subsection*{Conserved operators}
Defining holographic operator $\mathcal{E}_r(v)$ requires choosing a null direction $u$, similar to the conformal collider setup in \cite{Hofman:2008ar}. Let us call this d-dimensional vector $ \hat{u}= (-1, \hat{n})=  (-1, 1, \vec{0}) $ and denote $n^{\mu} = (0,1,\vec{0})$. For most of the following discussion $d \ge 4$ and $d=3$ is considered separately in the paper. \\ We are interested in computing smeared spinning external states,
\ba
{\ep_1}^\star_{\alpha_1 \alpha_2 \cdots \alpha_{s_1}} \langle O_1(\omega)^{\alpha_1 \alpha_2 \cdots \alpha_s} \mathcal{E}_r (\nu) O_2(\omega)^{\beta_1 \beta_2 \cdots \beta_{s_2}} \rangle {\ep_2}_{\beta_1 \beta_2 \cdots \beta_{s_2}},
\ea
where $\star$ denotes complex conjugation.
By smearing external operators, we are preparing states with definite momenta, $\omega^\mu = \omega t^\mu$ along the time direction with $ \; t^2=-1$.  Primary operators considered here are in the symmetric traceless representations, so polarization tensors can be chosen to be symmetric and traceless. Conservation equation implies
\begin{align}
\omega_{\mu_1} \langle O_1(\omega)^{\mu_1 \mu_2 \cdots \mu_{s_1}} \cdots \rangle=0.
\end{align}
Therefore we are free to choose $\ep$ with vanishing time-like components so that we have $\ep=\epsilon_{i_1 \cdots i_{s_1}}$.  

As a first example let us choose external state created by wave-packets of the stress tensor. The expectation value of holographic null energy operator has the following decomposition under $SO(d-1)$ corresponding to spatial rotations :
\ba \label{tensordecomp1}
&\langle \mathcal{E}_r (v) \rangle = \langle 0 | \ep_{ij}^\star  T_{ij}(\omega)   \mathcal{E}_r (v)  \ep_{lk} T_{lk}(\omega) | 0\rangle =  \tilde{t}_0 \ep^\star_{ij} \ep_{ij} + \tilde{t}_2  \ep^\star_{ij} \ep_{il } \hat{n}_j \hat{n}_l + \tilde{t}_4 |\ep_{ij} \hat{n}_i \hat{n}_j|^2.
\ea
Using the positivity of this expectation value for any $\ep_{ij}$, we look for the optimal bounds on coefficients.  Following \cite{Hofman:2008ar}, we further decompose this expression in terms of irreducible representations, i.e.  spin 0, 1, 2 under $SO(d-2)$, corresponding to rotations that leave the spatial part of the null direction $\hat{n}^i$ invariant. More explicitly,  let us parametrize a purely spatial polarization tensor as\footnote{Note that in writing this parametrization, we have chosen $d \ge 4$ as can be seen by the fact that if $d\le 3$, then $e_{ij}=0$ for a traceless tensor.}
\ba\label{eq:decom}
\ep_{ij} = e_{ij} +  b_{( i}\hat{n}_{j)}+ \alpha \(  \hat{n}_i \hat{n}_j -\frac{\delta_{ij}}{d-1}  \),
\ea
  where $e_{ij}$ and $b_i$ satisfy  $ b_i \hat{n}_i =0, e_{i j} \hat{n}^j = 0, e_{ii} = 0$ and $\alpha$ is an arbitrary complex number. \\
Substituting this expression in \eqref{tensordecomp1} we find
\ba
\langle \mathcal{E}_r \rangle =  |\alpha|^2 \left( \tilde{t}_0 \frac{d-2}{d-1} + \tilde{t}_2 \frac{(d-2)^2}{(d-1)^2} + \tilde{t}_4 \frac{(d-2)^2}{(d-1)^2}  \right) + \frac{b_i {b^\star}^i}{2} \left( \tilde{t}_0 + \frac{\tilde{t}_2}{2}  \right) +\tilde{t}_0 e_{ij} {e^\star}^{ij},  
\ea
where each term in this expression corresponds to an irreducible representation. Since these terms do not mix under $SO(d-2)$ rotations, positivity of the holographic null energy operator implies the positivity of each term separately. 

We will now show that the powers of $\lambda^2$ in \eqref{jjt1} and \eqref{ttt1l} are in one to one correspondence with these irreducible representations. To demonstrate this let us consider the following polarization vector,
\ba
&\ep^\mu =  \;  {\hat{v}}^{\mu}+ \ep_\perp^\mu\ , \qquad  \ep_\perp= (0,0, i \lambda,  \lambda,
{\underbrace{ 0, \cdots ,0}_{d-4}} ),  \qquad \nonumber\\
&{\hat{v}} = (1, 1 ,{\underbrace{0,\cdots, 0 }_{d-2}}) ,
\ea
where $ \lambda $ is an arbitrary real number. Contracting this null vector with external operator, $T_{\mu \nu} \ep^\mu \ep^\nu$ we find
\ba
\langle \mathcal{E}_r \rangle = g_0 + g_2 \lambda^2 + g_4 \lambda^4.
\ea
Note that $\ep^\mu \ep^\nu$ is not a purely spatial polarization tensor. Since only the spatial components contribute, we will use the symmetric traceless projector\footnote{Note that the expectation values in states created by smearing conserved operators are unchanged under the action of $\mathcal{Q}$ due to conservation.} $\mathcal{Q}^{\alpha \beta}_{\mu \nu}$ to convert $\ep^\mu \ep^\nu$ into a purely spatial traceless polarization tensor $\mathcal{E}^{\mu \nu}$ :
\ba
&\mathcal{P}_{\mu \nu} = \eta_{\mu \nu} + t_\mu t_\nu \nonumber\\
& \mathcal{Q}_{\mu \nu}^{\alpha \beta} = \frac{1}{2} \left(\mathcal{P}_{\mu}^\alpha \mathcal{P}_{\nu}^\beta +  \mathcal{P}_{\nu}^\alpha \mathcal{P}_{\mu}^\beta \right) - \frac{1}{d-1} \mathcal{P}_{\mu \nu} \mathcal{P}^{\alpha \beta} \qquad {\mathcal{Q}_{\mu}^{\mu}}^{\alpha \beta} = 0\nonumber\\ 
&\mathcal{E}^{\alpha \beta} \equiv Q_{\mu \nu}^{\alpha \beta} \ep^\mu \ep^\nu = \ep_\perp^{( \alpha} \ep_\perp^{\beta )} + \ep_\perp^{ ( \alpha}  \hat{v}^{\beta)} + (\hat{v}\cdot t) \ep_\perp^{ ( \alpha}  t^{\beta) } \nonumber\\
&+ \left(\hat{v}^\alpha + (\hat{v}\cdot t) t^\alpha \right) \left(\hat{v}^\beta + (\hat{v}\cdot t) t^\beta \right) - \frac{(\hat{v}\cdot t)^2}{d-1} (\delta^{\alpha \beta} + t^\alpha t^\beta)  \nonumber \\
&\Rightarrow \quad \mathcal{E}^{i j}=  \ep_\perp^{( i} \ep_\perp^{j )} + \ep_\perp^{( i }\hat{n}^{j)}+ ( \hat{n}^i \hat{n}^j - \frac{\delta_{ij}}{d-1}),
\ea 
which has the form of the decomposition in \ref{eq:decom}. Furthermore, $\ep_\perp^{( i} \ep_\perp^{j )} , \ep_\perp^{( i }\hat{n}^{j)}$ satisfy the same conditions as $e^{ij}$ and $b^{( i} n^{j )}$. In addition, any expression involving  $\ep_\perp$ is multiplied with a power of $\lambda$. Therefore $ \ep_\perp^{( i }\hat{n}^{j)}$ and $\ep_\perp^{( i} \ep_\perp^{j )}$ are multiplied with $\lambda $ and $\lambda^2$ respectively. This implies that each powers of $\lambda^2$ are in one-to-one correspondence with irreducible representations under $SO(d-2)$ rotations and $g_0, g_2, g_4$ should be positive independently.\\
This construction is easily generalized to the case of conserved higher spin operators. To do so, one finds a symmetric traceless projection operator and acts on a polarization tensor of the form $\ep^{\mu_1} \ep^{\mu_2} \cdots \ep^{\mu_s}$ with
\ba
&\mathcal{Q}_{\mu_1 \mu_2 \cdots \mu_s}^{\nu_1 \nu_2 \cdots \nu_s} \sim \frac{1}{s!} \left( \mathcal{P}_{\mu_1}^{(\nu_1} \mathcal{P}_{\mu_2}^{\nu_2} \cdots \mathcal{P}_{\mu_s}^{\nu_s )} -\text{traces} \right), \nonumber\\
&\mathcal{P}_{\mu_1}^{\nu_1} \ep_\perp^{\mu_1} = {\ep_\perp}^{\nu_1}, \qquad \mathcal{P}_{\mu_1}^{\nu_1} \hat{v}^{\mu_1} = n^\nu_1, \nonumber\\
& \mathcal{Q}_{\mu_1 \mu_2 \cdots \mu_s}^{\nu_1 \nu_2 \cdots \nu_s} \ep^{\mu_1} \ep^{\mu_2} \cdots \ep^{\mu_s} \sim \ep_\perp^{(\nu_1} \cdots \ep_\perp^{\nu_s)} + n^{(\nu_1} \ep_\perp^{\nu_2} \cdots \ep_\perp^{\nu_s)}, \nonumber\\
&+\cdots + \left( n^{\nu_1 } n^{\nu_2} \cdots n^{\nu_s} -\text{traces}\right),
\ea
corresponding to spin $0,1, \cdots, s-1, s$ representations under $SO(d-2)$. 	Each term has a different number of $\ep_\perp$, therefore the coefficients associated to powers of $\lambda$ are independent and should satisfy positivity constraints separately.

 In summary, for conserved operators, polarization vectors defined in \ref{polarization} result in the most general possible bounds in the holographic collider setup described here.

\subsection*{Non-conserved operators} 

For non-conserved operators, the use of longitudinal polarizations will result in more general constraints. The bounds in this paper were obtained using  $ \ep^\mu = (1,-1,\vec{0})$ as the longitudinal polarization tensor. It would interesting to find polarization tensors that result in the most optimal bounds. A more systematic approach would be useful in obtaining bounds in the light-cone limit to ensure the most stringent possible constraints.

\end{spacing}


\begin{thebibliography}{99}%

\bibitem{Afkhami-Jeddi:2017rmx} 
  N.~Afkhami-Jeddi, T.~Hartman, S.~Kundu and A.~Tajdini,
  ``Shockwaves from the Operator Product Expansion,''
  arXiv:1709.03597 [hep-th].
  


\bibitem{Hartman:2015lfa} 
  T.~Hartman, S.~Jain and S.~Kundu,
  ``Causality Constraints in Conformal Field Theory,''
  JHEP {\bf 1605}, 099 (2016)
  doi:10.1007/JHEP05(2016)099
  [arXiv:1509.00014 [hep-th]].
  
\bibitem{Hartman:2016dxc} 
  T.~Hartman, S.~Jain and S.~Kundu,
  ``A New Spin on Causality Constraints,''
  JHEP {\bf 1610}, 141 (2016)
  doi:10.1007/JHEP10(2016)141
  [arXiv:1601.07904 [hep-th]].

\bibitem{Hofman:2016awc} 
  D.~M.~Hofman, D.~Li, D.~Meltzer, D.~Poland and F.~Rejon-Barrera,
  ``A Proof of the Conformal Collider Bounds,''
  JHEP {\bf 1606}, 111 (2016)
  doi:10.1007/JHEP06(2016)111
  [arXiv:1603.03771 [hep-th]].

\bibitem{Hofman:2008ar} 
  D.~M.~Hofman and J.~Maldacena,
  ``Conformal collider physics: Energy and charge correlations,''
  JHEP {\bf 0805}, 012 (2008)
  doi:10.1088/1126-6708/2008/05/012
  [arXiv:0803.1467 [hep-th]].


\bibitem{Hartman:2016lgu} 
  T.~Hartman, S.~Kundu and A.~Tajdini,
  ``Averaged Null Energy Condition from Causality,''
  arXiv:1610.05308 [hep-th].
  
\bibitem{Cordova:2017zej} 
  C.~Cordova, J.~Maldacena and G.~J.~Turiaci,
  ``Bounds on OPE Coefficients from Interference Effects in the Conformal Collider,''
  JHEP {\bf 1711}, 032 (2017)
  doi:10.1007/JHEP11(2017)032
  [arXiv:1710.03199 [hep-th]].

\bibitem{Meltzer:2017rtf} 
  D.~Meltzer and E.~Perlmutter,
  ``Beyond $a=c$: Gravitational Couplings to Matter and the Stress Tensor OPE,''
  arXiv:1712.04861 [hep-th].
  
\bibitem{Chowdhury:2017vel} 
  S.~D.~Chowdhury, J.~R.~David and S.~Prakash,
  ``Constraints on parity violating conformal field theories in $d=3$,''
  JHEP {\bf 1711}, 171 (2017)
  doi:10.1007/JHEP11(2017)171
  [arXiv:1707.03007 [hep-th]].

\bibitem{Brigante:2008gz} 
  M.~Brigante, H.~Liu, R.~C.~Myers, S.~Shenker and S.~Yaida,
  ``The Viscosity Bound and Causality Violation,''
  Phys.\ Rev.\ Lett.\  {\bf 100}, 191601 (2008)
  doi:10.1103/PhysRevLett.100.191601
  [arXiv:0802.3318 [hep-th]].
  
  \bibitem{Hofman:2009ug} 
  D.~M.~Hofman,
  ``Higher Derivative Gravity, Causality and Positivity of Energy in a UV complete QFT,''
  Nucl.\ Phys.\ B {\bf 823}, 174 (2009)
  doi:10.1016/j.nuclphysb.2009.08.001
  [arXiv:0907.1625 [hep-th]].
  
\bibitem{Camanho:2009vw} 
  X.~O.~Camanho and J.~D.~Edelstein,
  ``Causality constraints in AdS/CFT from conformal collider physics and Gauss-Bonnet gravity,''
  JHEP {\bf 1004}, 007 (2010)
  doi:10.1007/JHEP04(2010)007
  [arXiv:0911.3160 [hep-th]].
  
  
  \bibitem{Camanho:2014apa} 
  X.~O.~Camanho, J.~D.~Edelstein, J.~Maldacena and A.~Zhiboedov,
  ``Causality Constraints on Corrections to the Graviton Three-Point Coupling,''
  JHEP {\bf 1602}, 020 (2016)
  doi:10.1007/JHEP02(2016)020
  [arXiv:1407.5597 [hep-th]].

\bibitem{Bellazzini:2015cra} 
  B.~Bellazzini, C.~Cheung and G.~N.~Remmen,
  ``Quantum Gravity Constraints from Unitarity and Analyticity,''
  Phys.\ Rev.\ D {\bf 93}, no. 6, 064076 (2016)
  doi:10.1103/PhysRevD.93.064076
  [arXiv:1509.00851 [hep-th]].
  
  
\bibitem{Maldacena:1997re} 
  J.~M.~Maldacena,
  ``The Large N limit of superconformal field theories and supergravity,''
  Int.\ J.\ Theor.\ Phys.\  {\bf 38}, 1113 (1999)
  [Adv.\ Theor.\ Math.\ Phys.\  {\bf 2}, 231 (1998)]
  doi:10.1023/A:1026654312961, 10.4310/ATMP.1998.v2.n2.a1
  [hep-th/9711200].

\bibitem{Witten:1998qj} 
  E.~Witten,
  ``Anti-de Sitter space and holography,''
  Adv.\ Theor.\ Math.\ Phys.\  {\bf 2}, 253 (1998)
  doi:10.4310/ATMP.1998.v2.n2.a2
  [hep-th/9802150].

\bibitem{Gubser:1998bc} 
  S.~S.~Gubser, I.~R.~Klebanov and A.~M.~Polyakov,
  ``Gauge theory correlators from noncritical string theory,''
  Phys.\ Lett.\ B {\bf 428}, 105 (1998)
  doi:10.1016/S0370-2693(98)00377-3
  [hep-th/9802109].
  
  \bibitem{Strominger:1997eq} 
  A.~Strominger,
  ``Black hole entropy from near horizon microstates,''
  JHEP {\bf 9802}, 009 (1998)
  doi:10.1088/1126-6708/1998/02/009
  [hep-th/9712251].

\bibitem{Keller:2011xi} 
  C.~A.~Keller,
  ``Phase transitions in symmetric orbifold CFTs and universality,''
  JHEP {\bf 1103}, 114 (2011)
  doi:10.1007/JHEP03(2011)114
  [arXiv:1101.4937 [hep-th]].

\bibitem{Hartman:2013mia} 
  T.~Hartman,
  ``Entanglement Entropy at Large Central Charge,''
  arXiv:1303.6955 [hep-th].

\bibitem{Hartman:2014oaa} 
  T.~Hartman, C.~A.~Keller and B.~Stoica,
  ``Universal Spectrum of 2d Conformal Field Theory in the Large c Limit,''
  JHEP {\bf 1409}, 118 (2014)
  doi:10.1007/JHEP09(2014)118
  [arXiv:1405.5137 [hep-th]].
  
\bibitem{Fitzpatrick:2014vua} 
  A.~L.~Fitzpatrick, J.~Kaplan and M.~T.~Walters,
  ``Universality of Long-Distance AdS Physics from the CFT Bootstrap,''
  JHEP {\bf 1408}, 145 (2014)
  doi:10.1007/JHEP08(2014)145
  [arXiv:1403.6829 [hep-th]].

\bibitem{Perlmutter:2016pkf} 
  E.~Perlmutter,
  ``Bounding the Space of Holographic CFTs with Chaos,''
  arXiv:1602.08272 [hep-th].

\bibitem{Cornalba:2006xk} 
  L.~Cornalba, M.~S.~Costa, J.~Penedones and R.~Schiappa,
  ``Eikonal Approximation in AdS/CFT: From Shock Waves to Four-Point Functions,''
  JHEP {\bf 0708}, 019 (2007)
  doi:10.1088/1126-6708/2007/08/019
  [hep-th/0611122].

\bibitem{Cornalba:2006xm} 
  L.~Cornalba, M.~S.~Costa, J.~Penedones and R.~Schiappa,
  ``Eikonal Approximation in AdS/CFT: Conformal Partial Waves and Finite N Four-Point Functions,''
  Nucl.\ Phys.\ B {\bf 767}, 327 (2007)
  doi:10.1016/j.nuclphysb.2007.01.007
  [hep-th/0611123].

\bibitem{Cornalba:2007zb} 
  L.~Cornalba, M.~S.~Costa and J.~Penedones,
  ``Eikonal approximation in AdS/CFT: Resumming the gravitational loop expansion,''
  JHEP {\bf 0709}, 037 (2007)
  doi:10.1088/1126-6708/2007/09/037
  [arXiv:0707.0120 [hep-th]].


\bibitem{Heemskerk:2009pn} 
  I.~Heemskerk, J.~Penedones, J.~Polchinski and J.~Sully,
  ``Holography from Conformal Field Theory,''
  JHEP {\bf 0910}, 079 (2009)
  doi:10.1088/1126-6708/2009/10/079
  [arXiv:0907.0151 [hep-th]].

  
\bibitem{Mack:2009gy} 
  G.~Mack,
  ``D-dimensional Conformal Field Theories with anomalous dimensions as Dual Resonance Models,''
  Bulg.\ J.\ Phys.\  {\bf 36}, 214 (2009)
  [arXiv:0909.1024 [hep-th]].

\bibitem{Mack:2009mi} 
  G.~Mack,
  ``D-independent representation of Conformal Field Theories in D dimensions via transformation to auxiliary Dual Resonance Models. Scalar amplitudes,''
  arXiv:0907.2407 [hep-th].

\bibitem{Fitzpatrick:2010zm} 
  A.~L.~Fitzpatrick, E.~Katz, D.~Poland and D.~Simmons-Duffin,
  ``Effective Conformal Theory and the Flat-Space Limit of AdS,''
  JHEP {\bf 1107}, 023 (2011)
  doi:10.1007/JHEP07(2011)023
  [arXiv:1007.2412 [hep-th]].

\bibitem{Heemskerk:2010ty} 
  I.~Heemskerk and J.~Sully,
  ``More Holography from Conformal Field Theory,''
  JHEP {\bf 1009}, 099 (2010)
  doi:10.1007/JHEP09(2010)099
  [arXiv:1006.0976 [hep-th]].

\bibitem{Fitzpatrick:2011hu} 
  A.~L.~Fitzpatrick and J.~Kaplan,
  ``Analyticity and the Holographic S-Matrix,''
  JHEP {\bf 1210}, 127 (2012)
  doi:10.1007/JHEP10(2012)127
  [arXiv:1111.6972 [hep-th]].

\bibitem{Fitzpatrick:2011ia} 
  A.~L.~Fitzpatrick, J.~Kaplan, J.~Penedones, S.~Raju and B.~C.~van Rees,
  ``A Natural Language for AdS/CFT Correlators,''
  JHEP {\bf 1111}, 095 (2011)
  doi:10.1007/JHEP11(2011)095
  [arXiv:1107.1499 [hep-th]].

\bibitem{ElShowk:2011ag} 
  S.~El-Showk and K.~Papadodimas,
  ``Emergent Spacetime and Holographic CFTs,''
  JHEP {\bf 1210}, 106 (2012)
  doi:10.1007/JHEP10(2012)106
  [arXiv:1101.4163 [hep-th]].


 
  

  
\bibitem{Komargodski:2012ek} 
  Z.~Komargodski and A.~Zhiboedov,
  ``Convexity and Liberation at Large Spin,''
  JHEP {\bf 1311}, 140 (2013)
  doi:10.1007/JHEP11(2013)140
  [arXiv:1212.4103 [hep-th]].
  



\bibitem{Fitzpatrick:2012yx} 
  A.~L.~Fitzpatrick, J.~Kaplan, D.~Poland and D.~Simmons-Duffin,
  ``The Analytic Bootstrap and AdS Superhorizon Locality,''
  JHEP {\bf 1312}, 004 (2013)
  doi:10.1007/JHEP12(2013)004
  [arXiv:1212.3616 [hep-th]].

  
  


\bibitem{Fitzpatrick:2012cg} 
  A.~L.~Fitzpatrick and J.~Kaplan,
  ``AdS Field Theory from Conformal Field Theory,''
  JHEP {\bf 1302}, 054 (2013)
  doi:10.1007/JHEP02(2013)054
  [arXiv:1208.0337 [hep-th]].

\bibitem{Goncalves:2014rfa} 
  V.~Goncalves, J.~Penedones and E.~Trevisani,
  ``Factorization of Mellin amplitudes,''
  JHEP {\bf 1510}, 040 (2015)
  doi:10.1007/JHEP10(2015)040
  [arXiv:1410.4185 [hep-th]].

\bibitem{Hijano:2015zsa} 
  E.~Hijano, P.~Kraus, E.~Perlmutter and R.~Snively,
  ``Witten Diagrams Revisited: The AdS Geometry of Conformal Blocks,''
  JHEP {\bf 1601}, 146 (2016)
  doi:10.1007/JHEP01(2016)146
  [arXiv:1508.00501 [hep-th]].
  
\bibitem{Alday:2016htq} 
  L.~F.~Alday and A.~Bissi,
  ``Unitarity and positivity constraints for CFT at large central charge,''
  arXiv:1606.09593 [hep-th].
  

\bibitem{Costa:2014kfa} 
  M.~S.~Costa, V.~Gonçalves and J.~Penedones,
  ``Spinning AdS Propagators,''
  JHEP {\bf 1409}, 064 (2014)
  doi:10.1007/JHEP09(2014)064
  [arXiv:1404.5625 [hep-th]].
  

\bibitem{Alday:2014tsa} 
  L.~F.~Alday, A.~Bissi and T.~Lukowski,
  ``Lessons from crossing symmetry at large N,''
  JHEP {\bf 1506}, 074 (2015)
  doi:10.1007/JHEP06(2015)074
  [arXiv:1410.4717 [hep-th]].

\bibitem{Caron-Huot:2017vep} 
  S.~Caron-Huot,
  ``Analyticity in Spin in Conformal Theories,''
  JHEP {\bf 1709}, 078 (2017)
  doi:10.1007/JHEP09(2017)078
  [arXiv:1703.00278 [hep-th]].
  
\bibitem{Afkhami-Jeddi:2016ntf} 
  N.~Afkhami-Jeddi, T.~Hartman, S.~Kundu and A.~Tajdini,
  ``Einstein gravity 3-point functions from conformal field theory,''
  arXiv:1610.09378 [hep-th].
 

\bibitem{Kulaxizi:2017ixa} 
  M.~Kulaxizi, A.~Parnachev and A.~Zhiboedov,
  ``Bulk Phase Shift, CFT Regge Limit and Einstein Gravity,''
  arXiv:1705.02934 [hep-th].



\bibitem{Costa:2017twz} 
  M.~S.~Costa, T.~Hansen and J.~Penedones,
  ``Bounds for OPE coefficients on the Regge trajectory,''
  JHEP {\bf 1710}, 197 (2017)
  doi:10.1007/JHEP10(2017)197
  [arXiv:1707.07689 [hep-th]].



\bibitem{Costa:2012cb} 
  M.~S.~Costa, V.~Goncalves and J.~Penedones,
  ``Conformal Regge theory,''
  JHEP {\bf 1212}, 091 (2012)
  doi:10.1007/JHEP12(2012)091
  [arXiv:1209.4355 [hep-th]].

\bibitem{Kravchuk:2018htv} 
  P.~Kravchuk and D.~Simmons-Duffin,
  ``Light-ray operators in conformal field theory,''
  arXiv:1805.00098 [hep-th].
  
\bibitem{Beccaria:2017dmw} 
  M.~Beccaria and A.~A.~Tseytlin,
  ``C$_{T}$ for higher derivative conformal fields and anomalies of (1, 0) superconformal 6d theories,''
  JHEP {\bf 1706}, 002 (2017)
  doi:10.1007/JHEP06(2017)002
  [arXiv:1705.00305 [hep-th]].
  
\bibitem{Ooguri:2016pdq} 
  H.~Ooguri and C.~Vafa,
  ``Non-supersymmetric AdS and the Swampland,''
  Adv.\ Theor.\ Math.\ Phys.\  {\bf 21}, 1787 (2017)
  doi:10.4310/ATMP.2017.v21.n7.a8
  [arXiv:1610.01533 [hep-th]].

\bibitem{Faulkner:2016mzt} 
  T.~Faulkner, R.~G.~Leigh, O.~Parrikar and H.~Wang,
  ``Modular Hamiltonians for Deformed Half-Spaces and the Averaged Null Energy Condition,''
  JHEP {\bf 1609}, 038 (2016)
  doi:10.1007/JHEP09(2016)038
  [arXiv:1605.08072 [hep-th]].

\bibitem{Czech:2016xec} 
  B.~Czech, L.~Lamprou, S.~McCandlish, B.~Mosk and J.~Sully,
  ``A Stereoscopic Look into the Bulk,''
  JHEP {\bf 1607}, 129 (2016)
  doi:10.1007/JHEP07(2016)129
  [arXiv:1604.03110 [hep-th]].
  
  
\bibitem{deBoer:2016pqk} 
  J.~de Boer, F.~M.~Haehl, M.~P.~Heller and R.~C.~Myers,
  ``Entanglement, holography and causal diamonds,''
  JHEP {\bf 1608}, 162 (2016)
  doi:10.1007/JHEP08(2016)162
  [arXiv:1606.03307 [hep-th]].



 



\bibitem{Kabat:2012hp} 
  D.~Kabat, G.~Lifschytz, S.~Roy and D.~Sarkar,
  ``Holographic representation of bulk fields with spin in AdS/CFT,''
  Phys.\ Rev.\ D {\bf 86}, 026004 (2012)
  doi:10.1103/PhysRevD.86.026004, 10.1103/PhysRevD.86.029901
  [arXiv:1204.0126 [hep-th]].

\bibitem{Maldacena:2015waa} 
  J.~Maldacena, S.~H.~Shenker and D.~Stanford,
  ``A bound on chaos,''
  JHEP {\bf 1608}, 106 (2016)
  doi:10.1007/JHEP08(2016)106
  [arXiv:1503.01409 [hep-th]].
  
\bibitem{Engelhardt:2016aoo} 
  N.~Engelhardt and S.~Fischetti,
  ``The Gravity Dual of Boundary Causality,''
  Class.\ Quant.\ Grav.\  {\bf 33}, no. 17, 175004 (2016)
  doi:10.1088/0264-9381/33/17/175004
  [arXiv:1604.03944 [hep-th]].

\bibitem{Gao:2000ga} 
  S.~Gao and R.~M.~Wald,
  ``Theorems on gravitational time delay and related issues,''
  Class.\ Quant.\ Grav.\  {\bf 17}, 4999 (2000)
  doi:10.1088/0264-9381/17/24/305
  [gr-qc/0007021].


\bibitem{Horowitz:1999gf} 
  G.~T.~Horowitz and N.~Itzhaki,
  ``Black holes, shock waves, and causality in the AdS / CFT correspondence,''
  JHEP {\bf 9902}, 010 (1999)
  doi:10.1088/1126-6708/1999/02/010
  [hep-th/9901012].
  
\bibitem{Maldacena:2015iua} 
  J.~Maldacena, D.~Simmons-Duffin and A.~Zhiboedov,
  ``Looking for a bulk point,''
  JHEP {\bf 1701}, 013 (2017)
  doi:10.1007/JHEP01(2017)013
  [arXiv:1509.03612 [hep-th]].
    
\bibitem{Costa:2011mg} 
  M.~S.~Costa, J.~Penedones, D.~Poland and S.~Rychkov,
  ``Spinning Conformal Correlators,''
  JHEP {\bf 1111}, 071 (2011)
  doi:10.1007/JHEP11(2011)071
  [arXiv:1107.3554 [hep-th]].
  
\bibitem{Osborn:1993cr} 
  H.~Osborn and A.~C.~Petkou,
  ``Implications of conformal invariance in field theories for general dimensions,''
  Annals Phys.\  {\bf 231}, 311 (1994)
  doi:10.1006/aphy.1994.1045
  [hep-th/9307010].
    
\bibitem{Costa:2011dw} 
  M.~S.~Costa, J.~Penedones, D.~Poland and S.~Rychkov,
  ``Spinning Conformal Blocks,''
  JHEP {\bf 1111}, 154 (2011)
  doi:10.1007/JHEP11(2011)154
  [arXiv:1109.6321 [hep-th]].
















\bibitem{Bonifacio:2017nnt} 
  J.~Bonifacio, K.~Hinterbichler, A.~Joyce and R.~A.~Rosen,
  ``Massive and Massless Spin-2 Scattering and Asymptotic Superluminality,''
  arXiv:1712.10020 [hep-th].

\bibitem{Hinterbichler:2017qcl} 
  K.~Hinterbichler, A.~Joyce and R.~A.~Rosen,
  ``Eikonal Scattering and Asymptotic Superluminality of Massless Higher Spins,''
  arXiv:1712.10021 [hep-th].



\bibitem{Erdmenger:1996yc} 
  J.~Erdmenger and H.~Osborn,
  ``Conserved currents and the energy momentum tensor in conformally invariant theories for general dimensions,''
  Nucl.\ Phys.\ B {\bf 483}, 431 (1997)
  doi:10.1016/S0550-3213(96)00545-7
  [hep-th/9605009].


\bibitem{Zhiboedov:2012bm} 
  A.~Zhiboedov,
  ``A note on three-point functions of conserved currents,''
  arXiv:1206.6370 [hep-th].

\bibitem{Zhiboedov:2013opa} 
  A.~Zhiboedov,
  ``On Conformal Field Theories With Extremal a/c Values,''
  JHEP {\bf 1404}, 038 (2014)
  doi:10.1007/JHEP04(2014)038
  [arXiv:1304.6075 [hep-th]].

\bibitem{Li:2015itl} 
  D.~Li, D.~Meltzer and D.~Poland,
  ``Conformal Collider Physics from the Lightcone Bootstrap,''
  JHEP {\bf 1602}, 143 (2016)
  doi:10.1007/JHEP02(2016)143
  [arXiv:1511.08025 [hep-th]].
  


\bibitem{Giombi:2011rz} 
  S.~Giombi, S.~Prakash and X.~Yin,
  ``A Note on CFT Correlators in Three Dimensions,''
  JHEP {\bf 1307}, 105 (2013)
  doi:10.1007/JHEP07(2013)105
  [arXiv:1104.4317 [hep-th]].
  
    
\bibitem{Bonora:1985cq} 
  L.~Bonora, P.~Pasti and M.~Bregola,
  ``Weyl Cocycles,''
  Class.\ Quant.\ Grav.\  {\bf 3}, 635 (1986).
  doi:10.1088/0264-9381/3/4/018
  
\bibitem{Deser:1993yx} 
  S.~Deser and A.~Schwimmer,
  ``Geometric classification of conformal anomalies in arbitrary dimensions,''
  Phys.\ Lett.\ B {\bf 309}, 279 (1993)
  doi:10.1016/0370-2693(93)90934-A
  [hep-th/9302047].
  
\bibitem{Bastianelli:2000hi} 
  F.~Bastianelli, S.~Frolov and A.~A.~Tseytlin,
  ``Conformal anomaly of (2,0) tensor multiplet in six-dimensions and AdS / CFT correspondence,''
  JHEP {\bf 0002}, 013 (2000)
  doi:10.1088/1126-6708/2000/02/013
  [hep-th/0001041].
  
\bibitem{Boulanger:2007ab} 
  N.~Boulanger,
  ``Algebraic Classification of Weyl Anomalies in Arbitrary Dimensions,''
  Phys.\ Rev.\ Lett.\  {\bf 98}, 261302 (2007)
  doi:10.1103/PhysRevLett.98.261302
  [arXiv:0706.0340 [hep-th]].
  
\bibitem{Hung:2011xb} 
  L.~Y.~Hung, R.~C.~Myers and M.~Smolkin,
  ``On Holographic Entanglement Entropy and Higher Curvature Gravity,''
  JHEP {\bf 1104}, 025 (2011)
  doi:10.1007/JHEP04(2011)025
  [arXiv:1101.5813 [hep-th]].

  
\bibitem{Beccaria:2015uta} 
  M.~Beccaria and A.~A.~Tseytlin,
  ``Conformal a-anomaly of some non-unitary 6d superconformal theories,''
  JHEP {\bf 1509}, 017 (2015)
  doi:10.1007/JHEP09(2015)017
  [arXiv:1506.08727 [hep-th]].
  
\bibitem{Butter:2016qkx} 
  D.~Butter, S.~M.~Kuzenko, J.~Novak and S.~Theisen,
  ``Invariants for minimal conformal supergravity in six dimensions,''
  JHEP {\bf 1612}, 072 (2016)
  doi:10.1007/JHEP12(2016)072
  [arXiv:1606.02921 [hep-th]].
  
\bibitem{Butter:2017jqu} 
  D.~Butter, J.~Novak and G.~Tartaglino-Mazzucchelli,
  ``The component structure of conformal supergravity invariants in six dimensions,''
  JHEP {\bf 1705}, 133 (2017)
  doi:10.1007/JHEP05(2017)133
  [arXiv:1701.08163 [hep-th]].
  
 
  
\bibitem{Henningson:1998gx} 
  M.~Henningson and K.~Skenderis,
  ``The Holographic Weyl anomaly,''
  JHEP {\bf 9807}, 023 (1998)
  doi:10.1088/1126-6708/1998/07/023
  [hep-th/9806087].
  
\bibitem{Fischler:2015kro} 
  W.~Fischler and S.~Kundu,
  ``Membrane paradigm, gravitational $\Theta$-term and gauge/gravity duality,''
  JHEP {\bf 1604}, 112 (2016)
  doi:10.1007/JHEP04(2016)112
  [arXiv:1512.01238 [hep-th]].
  W.~Fischler and S.~Kundu,
  ``Physical effects of the gravitational $\Theta$ -parameter,''
  Int.\ J.\ Mod.\ Phys.\ D {\bf 25}, no. 12, 1644022 (2016)
  doi:10.1142/S0218271816440223
  [arXiv:1612.06010 [hep-th]].
  
  
  
\bibitem{Lue:1998mq} 
  A.~Lue, L.~M.~Wang and M.~Kamionkowski,
  ``Cosmological signature of new parity violating interactions,''
  Phys.\ Rev.\ Lett.\  {\bf 83}, 1506 (1999)
  doi:10.1103/PhysRevLett.83.1506
  [astro-ph/9812088].

\bibitem{Alexander:2004wk} 
  S.~Alexander and J.~Martin,
  ``Birefringent gravitational waves and the consistency check of inflation,''
  Phys.\ Rev.\ D {\bf 71}, 063526 (2005)
  doi:10.1103/PhysRevD.71.063526
  [hep-th/0410230].
  
\bibitem{Lidsey:1995np} 
  J.~E.~Lidsey, A.~R.~Liddle, E.~W.~Kolb, E.~J.~Copeland, T.~Barreiro and M.~Abney,
  ``Reconstructing the inflation potential : An overview,''
  Rev.\ Mod.\ Phys.\  {\bf 69}, 373 (1997)
  doi:10.1103/RevModPhys.69.373
  [astro-ph/9508078].
  
\bibitem{Baumann:2015xxa} 
  D.~Baumann, H.~Lee and G.~L.~Pimentel,
  ``High-Scale Inflation and the Tensor Tilt,''
  JHEP {\bf 1601}, 101 (2016)
  doi:10.1007/JHEP01(2016)101
  [arXiv:1507.07250 [hep-th]].
  
\bibitem{Maldacena:2011nz} 
  J.~M.~Maldacena and G.~L.~Pimentel,
  ``On graviton non-Gaussianities during inflation,''
  JHEP {\bf 1109}, 045 (2011)
  doi:10.1007/JHEP09(2011)045
  [arXiv:1104.2846 [hep-th]].
  
\bibitem{Soda:2011am} 
  J.~Soda, H.~Kodama and M.~Nozawa,
  ``Parity Violation in Graviton Non-gaussianity,''
  JHEP {\bf 1108}, 067 (2011)
  doi:10.1007/JHEP08(2011)067
  [arXiv:1106.3228 [hep-th]].
  
\bibitem{Shiraishi:2011st} 
  M.~Shiraishi, D.~Nitta and S.~Yokoyama,
  ``Parity Violation of Gravitons in the CMB Bispectrum,''
  Prog.\ Theor.\ Phys.\  {\bf 126}, 937 (2011)
  doi:10.1143/PTP.126.937
  [arXiv:1108.0175 [astro-ph.CO]].
  
  
  
\bibitem{Bartolo:2017szm} 
  N.~Bartolo and G.~Orlando,
  ``Parity breaking signatures from a Chern-Simons coupling during inflation: the case of non-Gaussian gravitational waves,''
  JCAP {\bf 1707}, 034 (2017)
  doi:10.1088/1475-7516/2017/07/034
  [arXiv:1706.04627 [astro-ph.CO]].
  
\end{thebibliography}
\end{document}